\documentclass{theoretics}

\def\fancytoc{1} 
\def\exclaims{1} 
\def\newfigs{1} 

\def\latexVer{1} 
\def\niceN{1} 

\ifnum\latexVer=1
\else
\fi

\ifnum\fancytoc=1 
\usepackage{tocloft} 
\renewcommand{\cftpartpresnum}{Part~} 
\newlength{\mylen} 
\fi

\title{Robustly Self-Ordered Graphs: Constructions and Applications to Property Testing}

\TCSauthor[affil1]{Oded Goldreich}{oded.goldreich@weizmann.ac.il}
\TCSauthor[affil2]{Avi Wigderson}{avi@ias.edu}

\TCSaffil[affil1]{Faculty of Mathematics and Computer Science, Department of Computer Science, Weizmann Institute of Science, Rehovot, Israel}
\TCSaffil[affil2]{School of Mathematics, Institute for Advanced Study, Princeton, USA}

\TCSshorttitle{Robustly Self-Ordered Graphs}

\TCSvolume{2022}
\TCSarticlenum{1}
\TCSreceived{Dec 1, 2021}
\TCSrevised{May 20, 2022}
\TCSaccepted{Jul 30, 2022}
\TCSpublished{Dec 12, 2022}
\TCSkeywords{Asymmetric graphs, expanders, Testing Graph Properties,
two-\linebreak source extractors, non-malleable extractors, coding theory,
tolerant testing, random graphs}
\TCSciteas{Oded Goldreich and Avi Wigderson, Robustly Self-Ordered Graphs: Constructions and Applications to Property Testing. TheoretiCS, 1 (2022), Article 1, 80 pages.}
\TCSdoi{10.46298/theoretics.22.1}
\TCSshortnames{O. Goldreich and A. Wigderson}
\TCSthanks{An extended abstract of this work has appeared in the
  proceedings of the 35th Annual Conference on Computational Complexity, CCC 2021.}

\def\etal{{\it et~al.}}
\newcommand{\prob}{{\rm Pr}}
\newcommand{\Exp}{{\rm E}}
\newcommand{\e}{\epsilon}

\newcommand{\bitset}{\{0,1\}}
\newcommand{\xth}{{\rm th}}

\newcommand{\imu}{{\mu^{-1}}}
\newcommand{\si}{\mu}

\newcommand{\mypar}[1]{\subparagraph{{#1}}}

\newcommand{\ang}[1]{{\langle{#1}\rangle}}

\newcommand{\ceil}[1]{{\lceil#1\rceil}}

\newcommand{\xor}{\oplus}
\newcommand{\eqdef}{\stackrel{\rm def}{=}}

\newcommand{\GF}{{\rm GF}}
\newcommand{\SL}{{\rm SL}}
\newcommand{\GG}{{\hat G}}

\newcommand{\st}{{\rm st}}

\newcommand{\ov}{\overline}

\newcommand{\ee}{\chi}

\newcommand{\BT}{\begin{theorem}}   \newcommand{\ET}{\end{theorem}}
\newcommand{\BD}{\begin{definition}}   \newcommand{\ED}{\end{definition}}
\newcommand{\BCT}{\begin{construction}} \newcommand{\ECT}{\end{construction}}
\newcommand{\BCR}{\begin{corollary}} \newcommand{\ECR}{\end{corollary}}
\newtheorem{conj}[theorem]{Conjecture}      %
\newcommand{\BCJ}{\begin{conj}} \newcommand{\ECJ}{\end{conj}}

\newcommand{\BP}{\begin{proposition}}   \newcommand{\EP}{\end{proposition}}

\newcommand{\BR}{\begin{remark}}   \newcommand{\ER}{\end{remark}}
\newcommand{\BO}{\begin{open}}   \newcommand{\EO}{\end{open}}
\newcommand{\BL}{\begin{lemma}}   \newcommand{\EL}{\end{lemma}}
\newtheorem{tech-lem}{Lemma}[thm]  
\newcommand{\BCM}{\begin{claim}}   \newcommand{\ECM}{\end{claim}}
\newtheorem{techclm}{Claim}[subsection]        %
\newcommand{\Bcm}{\begin{techclm}}   \newcommand{\Ecm}{\end{techclm}}
\newtheorem{tech-obs}{Observation}[theorem]          %
\newcommand{\Bobs}{\begin{tech-obs}}   \newcommand{\Eobs}{\end{tech-obs}}
\newtheorem{tech-def}[tech-lem]{Definition}            %
\newtheorem{tech-const}{Construction}[subsection]     %
\newcommand{\Bct}{\begin{tech-const}} \newcommand{\Ect}{\end{tech-const}}
\newcommand{\BE}{\begin{enumerate}}
\newcommand{\EE}{\end{enumerate}}
\newcommand{\BI}{\begin{itemize}}
\newcommand{\EI}{\end{itemize}}
\newcommand{\BDes}{\begin{description}}
\newcommand{\EDes}{\end{description}}

\newcommand{\BPF}{\begin{proof}} \newcommand{\EPF}{\end{proof}}
\newcommand{\BPFS}{\begin{proof}[Proof Sketch]} \newcommand{\EPFS}{\end{proof}}
\newcommand{\Bpf}{\begin{subproof}} \newcommand{\Epf}{\end{subproof}}

\newcommand{\tildeO}{{\widetilde{O}}}

\ifnum\niceN=1
  \font\tenmsb=msbm10 scaled\magstep1
  \font\sevenmsb=msbm7 scaled\magstep1
  \font\fivemsb=msbm5 scaled\magstep1
  \newfam\msbfam
  \textfont\msbfam=\tenmsb
  \scriptfont\msbfam=\sevenmsb
  \scriptscriptfont\msbfam=\fivemsb
  
  \newcommand{\N}{\mathbb N}
  \newcommand{\R}{\mathbb R}
  
\else
  \newcommand{\lndI}{\makebox[0pt][l]{\hspace*{1pt}I}}
  \newcommand{\N}{\mbox{\bf\lndI N}}
  \newcommand{\R}{\mbox{\bf\lndI R}}
  
\fi

\newcommand{\CG}{{\cal C}}

\newcommand{\poly}{{\rm poly}}

\newcommand{\C}{{\cal C}}

\newcommand{\STOC}{ACM Symposium on the Theory of Computing}
\newcommand{\FOCS}{IEEE Symposium on Foundations of Computer Science}
\newcommand{\JCSS}{Journal of Computer and System Science}

\newcommand{\SICOMP}{SIAM Journal on Computing}

\newcommand{\Sym}{{\rm Sym}}
\newcommand{\nmE}{{\tt nmE}}

\renewcommand{\sf}{\em}

\begin{document}


\maketitle

\begin{abstract}
A graph~$G$ is called {\em self-ordered}\/ (a.k.a asymmetric)
if the identity permutation is its only automorphism.
Equivalently, there is a unique isomorphism from~$G$
to any graph that is isomorphic to~$G$.
We say that $G=(V,E)$ is {\em robustly self-ordered}\/
if the size of the symmetric difference between $E$
and the edge-set of the graph obtained
by permuting $V$ using any permutation $\pi:V\to V$
is proportional to the number of non-fixed-points of $\pi$.
In this work, we initiate the study of the structure,
construction and utility of robustly self-ordered graphs.

We show that
robustly self-ordered bounded-degree graphs exist (in abundance),
and that they can be constructed efficiently, in a strong sense.
Specifically, given the index of a vertex in such a graph,
it is possible to find all its neighbors in polynomial-time
(i.e., in time that is poly-logarithmic in the size of the graph).

We provide two very different constructions, in tools and structure.
The first, a direct construction, is based on proving
a sufficient condition for robust self-ordering,
which requires that an auxiliary graph is expanding.
The second construction is iterative, boosting the property
of robust self-ordering from smaller to larger graphs.
Structurally, the first construction always yields expanding graphs,
while the second construction may produce graphs that have many
tiny (sub-logarithmic) connected components.
%

We also consider graphs of unbounded degree,
seeking correspondingly unbounded robustness parameters.
We again demonstrate that such graphs (of linear degree)
exist (in abundance),
and that they can be constructed efficiently, in a strong sense.
This turns out to require very different tools.
Specifically, we show that the construction of such graphs reduces to
the construction of non-malleable two-source extractors
(with very weak parameters but with some additional natural features).

We demonstrate that robustly self-ordered bounded-degree graphs
are useful towards obtaining lower bounds on the query complexity
of testing graph properties
both in the bounded-degree and the dense graph models.
Indeed, their robustness offers efficient, local
and distance preserving reductions from testing problems
on ordered structures (like sequences)
to the unordered (effectively unlabeled) graphs.
One of the results that we obtain, via such a reduction,
is a subexponential separation between the query complexities
of testing and tolerant testing of graph properties
in the bounded-degree graph model.
\end{abstract}



\section{Introduction}
For a (labeled) graph $G=(V,E)$, and a bijection $\phi:V\to V'$,
we denote by $\phi(G)$ the graph $G'=(V',E')$
such that $E'=\{\{\phi(u),\phi(v)\}:\{u,v\}\!\in\!E\}$,
and say that $G'$ is {\sf isomorphic} to~$G$.
The set of {\sf automorphisms} of the graph $G=(V,E)$,
denoted ${\tt aut}(G)$,
is the set of permutations that preserve the graph~$G$;
that is, $\pi\in{\tt aut}(G)$ if and only if $\pi(G)=G$.
We say that a graph is {\sf asymmetric} (equiv., {\sf self-ordered})
if its set of automorphisms is a singleton,
which consists of the trivial automorphism
(i.e., the identity permutation).
We actually prefer the term {\em self-ordered}, because we take
the perspective that is offered by the following equivalent definition.

\BD[Self-ordered (a.k.a asymmetric) graphs]
\label{asymmetric:def}
The graph $G=([n],E)$ is {\sf self-ordered}
if for every graph $G'=(V',E')$ that is isomorphic to~$G$
there exists a unique bijection $\phi:V'\to[n]$ such that $\phi(G')=G$.
\ED
In other words,
given an isomorphic copy $G'=(V',E')$ of a fixed graph $G=([n],E)$,
there is a unique bijection $\phi:V'\to[n]$ that orders the vertices of $G'$
such that the resulting graph (i.e., $\phi(G')$) is identical to~$G$.
Indeed, if $G'=G$, then this unique bijection is the identity permutation.%
\footnote{Naturally, we are interested in efficient algorithms
that find this unique ordering, whenever it exists;
such algorithms are known
when the degree of the graph is bounded~\cite{L}.}

In this work, we consider a feature,
which we call {\em robust self-ordering},
that is a quantitative version self-ordering.
Loosely speaking, a graph $G=([n],E)$ is robustly self-ordered if,
for every permutation $\pi:[n]\to[n]$,
the size of the symmetric difference between~$G$ and $\pi(G)$
is proportional to the number of non-fixed-points under $\pi$;
that is, $|E\triangle\{\{\pi(u),\pi(v)\}\!:\!\{u,v\}\!\in\!E\}|$
is proportional to $|\{i\!\in\![n]\!:\!\pi(i)\!\neq\!i\}|$.
(In contrast, self-ordering only means that
the size of the symmetric difference is positive
if the number of non-fixed-points is positive.)

\BD[Robustly self-ordered graphs]
\label{robust-asymmetric:def}
A graph $G=(V,E)$ is said to be {\sf $\gamma$-robustly self-ordered}
if for every permutation $\pi:V\to{V}$
it holds that
\begin{equation}\label{robust-asymmetric:eqdef}
\big|E\triangle\left\{\{\pi(u),\pi(v)\}\!:\!\{u,v\}\!\in\!E\right\}\big|
  \;\geq\; \gamma\cdot|\{i\in[n]\!:\!\pi(i)\neq i\}|,
\end{equation}
where $\triangle$ denotes the symmetric difference operation.
An infinite family of graphs $\{G_n=([n],E_n)\}_{n\in\N}$
{\em(such that each $G_n$ has maximum degree $d$)}
is called {\sf robustly self-ordered}
if there exists a constant $\gamma>0$, called the {\sf robustness parameter},
such that for every $n$ the graph $G_n$ is $\gamma$-robustly self-ordered.
\ED
Note that
$|E_n\triangle\{\{\pi(u),\pi(v)\}\!:\!\{u,v\}\!\in\!E_n\}|
 \leq 2d\cdot|\{i\in[n]\!:\!\pi(i)\neq i\}|$
always holds (for families of maximum degree $d$).
The term ``robust'' is inspired by
the property testing literature (cf.~\cite{RS}),
where it indicates that some ``parametrized violation''
is reflected proportionally in some ``detection parameter''.

The second part of Definition~\ref{robust-asymmetric:def}
is tailored for bounded-degree graphs, which will be our focus
in Section~\ref{edge-colored:sec}--\ref{random-graphs:sec}.
Nevertheless,
in Sections~\ref{dense-basics:sec}--\ref{inter-deg:sec}
we consider graphs
of unbounded degree and unbounded robustness parameters.
In this case, for a function $\rho:\N\to\R$,
we say that an infinite family of graphs $\{G_n=([n],E_n)\}_{n\in\N}$
is {\sf $\rho$-robustly self-ordered} if for every $n$
the graph $G_n$ is $\rho(n)$-robustly self-ordered.
Naturally, in this case, the graphs
must have $\Omega(\rho(n)\cdot n)$ edges.%
\footnote{Actually, all but at most one vertex
must have degree at least $\rho(n)/2$.}
In Sections~\ref{dense-basics:sec}--\ref{dense-pt:sec}
we consider the case of $\rho(n)=\Omega(n)$.

\subsection{Robustly self-ordered bounded-degree graphs}
The first part of this paper
(i.e., Section~\ref{edge-colored:sec}--\ref{random-graphs:sec})
focuses on the study of robustly self-ordered bounded-degree graphs.

\subsubsection{Our main results and motivation}
\label{intro:bd-results}
We show that robustly self-ordered ($n$-vertex) graphs of bounded-degree
not only exist (for all $n\in\N$),
but can be efficiently constructed in a strong (or local) sense.
Specifically, we prove the following result.

\BT[Constructing robustly self-ordered bounded-degree graphs]
\label{main:ithm}
For all sufficiently large $d\in\N$,
there exist an infinite family of $d$-regular
robustly self-ordered graphs $\{G_n\}_{n\in\N}$
and a polynomial-time algorithm that,
given $n\in\N$ and a vertex $v\in[n]$ in the $n$-vertex graph $G_n$,
finds all neighbors of $v$ {\rm(in $G_n$)}.
\ET
We stress that the algorithm runs in time that is polynomial
in the description of the vertex; that is, the algorithm runs
in time that is polylogarithmic in the size of the graph.
Theorem~\ref{main:ithm} holds both for graphs that consists
of connected components of logarithmic size
and for ``strongly connected'' graphs (i.e., expanders).

Recall that given an isomorphic copy $G'$ of such a graph $G_n$,
the original graph $G_n$ (i.e., along with its unique ordering)
can be found in polynomial-time~\cite{L}.
Furthermore, we show that the pre-image of each vertex of $G'$
in the graph $G_n$ (i.e., its index in the aforementioned ordering)
can be found in time that is polylogarithmic in the size of the graph
(see discussion in Section~\ref{local-so:sec},
culminating in Theorem~\ref{local-so4strong-construction:thm}).%
\footnote{The algorithm asserted above is said to perform
{\em local self-ordering}\/ of $G'$ according to $G_n$.
For $\phi(G')=G_n$, given a vertex $v$ in $G'$,
this algorithm returns $\phi(v)$ in $\poly(\log n)$-time.
In contrast, a {\em local reversed self-ordering}\/ algorithm
is given a vertex $i\in[n]$ of $G_n$ and returns $\phi^{-1}(i)$.
The second algorithm is also presented in Section~\ref{local-so:sec}
(see Theorem~\ref{local-reversed-so:thm}).\label{intro:local-so+rso:fn}}

We present two proofs of Theorem~\ref{main:ithm}.
Loosely speaking, the first proof reduces to proving
that a $2d$-regular $n$-vertex graph representing the
action of $d$ permutations on $[n]$ is robustly self-ordered
if the $2\cdot\binom{n}{2}$-vertex graph representing the action
of these permutations on (ordered) vertex-pairs is an expander.%
\footnote{Here and throughout this paper, by {\em expander}\/
we mean families of bounded-degree graphs that
have constant expansion (cf., e.g.,~\cite{HLW}).}
The graphs constructed in this proof are expanders,
whereas the graphs constructed via by the second proof
can be either expanders or consist
of connected components of logarithmic size.
More importantly,
the graphs constructed in the second proof are coupled with
{\em local self-ordering and local reversed self-ordering algorithms}\/
(see Section~\ref{local-so:sec}).
The second proof proceeds in three steps,
starting from the mere existence of
robustly self-ordered bounded-degree $\ell$-vertex graphs,
which yields a construction that runs in $\poly(\ell^\ell)$-time.
Next, a $\poly(n)$-time construction of $n$-vertex graphs
is obtained by using the former graphs as small subgraphs
(of $o(\log n)$-size).
Lastly, strong (a.k.a local) constructability
is obtained in an analogous manner.
For more details, see Section~\ref{techniques:sec}.

We demonstrate that robustly self-ordered bounded-degree graphs
are useful towards obtaining lower bounds on the query complexity
of testing graph properties in the bounded-degree graph model.
Specifically, we use these graphs as a key ingredient in
a general methodology of transporting lower bounds regarding
testing binary strings to lower bounds regarding
testing graph properties in the bounded-degree graph model.
In particular, using the methodology,
we prove the following two results.
\BE
\item
A {\em subexponential separation between the query complexities of testing
and tolerant testing of graph properties in the bounded-degree graph model};
that is, for some constant $c>0$, the query complexity
of tolerant testing is at least $\exp(q^c)$,
where $q$ is the query complexity of standard testing.

This result, which appears as Theorem~\ref{pt-tolerant:thm},
is obtained by transporting an analogous result
that was known for testing binary strings~\cite{FF}.
%
\item
A linear query complexity lower bound for testing
an {\em efficiently recognizable}\/ graph property
in the bounded-degree graph model,
where the lower bound holds even if the tested graph is restricted
to consist of connected components of logarithmic size
(see Theorem~\ref{pt-lb:thm}).

As discussed in Section~\ref{pt:sec},
an analogous result was known in the general case
(i.e., without the restriction on the size of the connected components),
and we consider it interesting that the result holds also
in the special case of graphs with small connected components.
\EE

To get a feeling of why robustly self-ordered graphs are relevant to
such transportation, recall that strings are ordered objects,
whereas graphs properties are effectively sets of unlabeled graphs,
which are unordered objects.
Hence, we need to make the graphs (in the property) ordered,
and furthermore make this ordering robust in the very sense
that is reflected in Definition~\ref{robust-asymmetric:def}.
%
We comment that the theme of reducing ordered structures to unordered
structures occurs often in the theory of computation and in logic,
and is often coupled with analogues of query complexity.

Lastly, in Section~\ref{random-graphs:sec},
we prove that random $2d$-regular graphs
are robustly self-ordered; see Theorem~\ref{random-works:thm}.
This extends work in probabilistic graph theory,
which proves a similar result for the weaker notion
of self-ordering~\cite{B1,B2}.

\subsubsection{Techniques}\label{techniques:sec}
As stated above, we present two different constructions
that establish Theorem~\ref{main:ithm}:
A direct construction and a three-step construction.
Both constructions utilize a variant of the notion
of robust self-ordering that refers to edge-colored graphs,
which we review first.

\paragraph{The edge-coloring methodology.}
At several different points, we found it useful to start by
demonstrating the robust self-ordering feature in a relaxed model
in which edges are assigned a constant number of colors,
and the symmetric difference between graphs accounts also
for edges that have different colors in the two graphs
(see Definition~\ref{colored-robust:def}).
This allows us to analyze different sets of edges separately.

For example, we actually analyze the direct construction
in the edge-colored model, while associating each
of the underlying $d$ permutations with a different color.
This association allows for analyzing the effect
of each permutation separately (see below).
%
Another example, which arises in the three-step construction,
occurs when we super-impose a robustly self-ordered graph
with an expander graph
in order to make the robustly self-ordered graph expanding
(as needed for the second and third step of the
aforementioned three-step construction).
In this case, assigning the edges of each of the two graphs
a different color, allows for easily retaining
the robust self-ordering feature (of the first graph).

We obtain robustly self-ordered graphs (in the original sense)
by replacing all edges that are assigned a specific color
with copies of a constant-sized (asymmetric) gadget,
where different (and in fact non-isomorphic) gadgets
are used for different edge colors.
The soundness of this transformation
is proved in Theorem~\ref{colored2std:clm}.

\paragraph{The direct construction.}
For any $d$ permutations, $\pi_1,\ldots,\pi_d:[n]\to[n]$,
we consider the {\em Schreier graph}\/ (see~\cite[Sec.~11.1.2]{HLW})
defined by the action of these permutation on $[n]$;
that is, the edge-set of this graph
is $\{\{v,\pi_i(v)\}\!:\!v\!\in\![n]\,\&\,i\!\in\![d]\}$.
Loosely speaking, we prove that {\em this $2d$-regular $n$-vertex graph
is robustly self-ordered if another Schreier graph is an expander}.
The second Schreier graph represents the action
of the same permutations on {\em pairs}\/ of vertices (in $[n]$);
that is, this graph consisting of the vertex-set $\{(u,v)\!:\!u,v\!\in\![n]\}$
and the edge-set
$\{\{(u,v),(\pi_i(u),\pi_i(v))\}\!:\!u,v\!\in\![n]\,\&\,i\!\in\![d]\}$.%
\footnote{Equivalently, we consider only pairs of distinct vertices;
that is, the vertex-set $\{(u,v)\!:\!u,v\!\in\![n]\,\&\,u\!\neq\!v\}$.}

The argument is actually made with respect to edge-colored
directed graphs (i.e., the edge-set of the first graph
is $\{(v,\pi_i(v))\!:\!v\!\in\![n]\,\&\,i\!\in\![d]\}$
and the directed edge $(v,\pi_i(v))$ is assigned the color~$i$).
Hence, we also present a transformation of robustly self-ordered
edge-colored directed graphs to analogous undirected graphs.
Specifically, we replace the directed edge $(u,v)$ colored $j$
by a 2-path with a designated auxiliary vertex $a_{u,v,j}$,
while coloring the edge $\{u,a_{u,v,j}\}$ by~$2j-1$
and the edge $\{a_{u,v,j},v\}$ by~$2j$.

We comment that permutations satisfying the foregoing condition
can be efficiently constructed;
for example, any set of expanding generators for $\SL_2(p)$
(e.g., the one used by~\cite{LPS})
yield such permutations on $[n]\equiv\{(1,i):i\in\GF(p)\}\cup\{(0,1)\}$
(see Proposition~\ref{direct:LPS:rem}).%
\footnote{In this case, the primary Schreier graph represents the
natural action of the group on the 1-dimensional subspaces of~$\GF(p)^2$.}


\paragraph{The three-step construction.}
Our alternative construction of robustly self-ordered
(bounded-degree) $n$-vertex graphs proceeds in three steps.

\BE
\item
First, we prove the existence of bounded-degree $n$-vertex graphs
that are robustly self-ordered (see Theorem~\ref{existence:thm}),
while observing that this yields
a $\exp(O(n\log n))$-time algorithm for constructing them.
\item
Next (see Theorem~\ref{construction:thm}),
we use the latter algorithm to construct
robustly self-ordered $n$-vertex bounded-degree graphs
that consist of $2\ell$-sized connected components,
where $\ell=\frac{O(\log n)}{\log\log n}$;
these connected components are far from being isomorphic to one another,
and are constructed using robustly self-ordered $\ell$-vertex graphs
as a building block.
This yields an algorithm that constructs the $n$-vertex graph
in $\poly(n)$-time, since $\exp(O(\ell\log\ell))=\poly(n)$.
\item
Lastly, we derive Theorem~\ref{main:ithm}
(restated as Theorem~\ref{strong-construction:thm})
by repeating the same strategy as in Step~2,
but using the construction of Theorem~\ref{construction:thm}
for the construction of the small connected components
(and setting $\ell=O(\log n)$).
This yields an algorithm that finds the neighbors of a vertex
in the $n$-vertex graph in $\poly(\log n)$-time,
since $\poly(\ell)=\poly(\log n)$.
\EE

The foregoing description of Steps~2 and~3 yields graphs
that consists of small connected components.
We obtain analogous results for ``strongly connected'' graphs
(i.e., expanders) by superimposing these graphs with expander graphs
(while distinguishing the two types of edges by using colors
(see the foregoing discussion)).
In fact, it is essential to perform this transformation
(on the result of Step~2) before taking Step~3;
the transformation itself appears
in the proof of Theorem~\ref{make:regular+expanding:thm}.


\paragraph{Using large collections of pairwise far apart permutations.}
One ingredient in the foregoing three-step construction is the use of
a single $\ell$-vertex robustly self-ordered (bounded-degree) graph
towards obtaining a {\em large}\/ collection of $2\ell$-vertex
(bounded-degree) graphs such that every two graphs
are far from being isomorphic to one another,
where ``large'' means $\exp(\Omega(\ell\log\ell))$ in one case
(i.e., in the proof of Theorem~\ref{construction:thm})
and $\exp(\Omega(\ell))$ in another case
(i.e., in the proof of Theorem~\ref{strong-construction:thm}).
Essentially, this is done by constructing a large collection
of permutations of $[\ell]$ that are pairwise far-apart,
and letting the $i^\xth$ graph consists of two copies
of the $\ell$-vertex graph that are matched according to
the $i^\xth$ permutation (see the aforementioned proofs).
(Actually, we use two robustly self-ordered $\ell$-vertex graphs
that are far from being isomorphic (e.g., have different degree).)

A collection of $L=\exp(\Omega(\ell\log\ell))$ pairwise far-apart
permutations over $[\ell]$ can be constructed in $\poly(L)$-time
by selecting the permutations one by one, while relying on the
existence of a permutation that augments the current sequence
(while preserving the distance condition,
see the proof of Theorem~\ref{construction:thm}).
A collection of $L=\exp(\Omega(\ell))$ pairwise far-apart
permutations over $[\ell]$ can be locally constructed such that
the $i^\xth$ permutation is constructed in $\poly(\ell)$-time
by using sequences of disjoint transpositions determined
via a good error correcting code
(see the proof of Theorem~\ref{strong-construction:thm}).

The foregoing discussion begs the challenge of obtaining
a construction of a collection of $L=\exp(\Omega(\ell\log\ell))$
permutations over $[\ell]$ that are pairwise far-apart
along with a polynomial-time algorithm that, on input $i\in[L]$,
returns a description of the $i^\xth$ permutation
(i.e., the algorithm should run in $\poly(\log L)$-time).
%
We meet this challenge in~\cite{GW:perm}.
Note that such a collection constitutes a an asymptotically good code
over the alphabet $[\ell]$, where the permutations are the codewords
(being far-apart corresponds to constant relative distance
and $\log L=\Omega(\log(\ell!))$ corresponds to constant rate).


\paragraph{On the failure of some natural approaches.}
We mention that natural candidates for robustly self-ordered
bounded-degree graphs fail.
In particular, there exist expander graphs that
are not robustly self-ordered.
In fact, any Cayley graph is symmetric
(i.e., has non-trivial automorphisms).%
\footnote{Specifically, multiplying the vertex labels (say, on the right)
by any non-zero group element yields a non-trivial automorphism
(assuming that edges are defined by multiplying with a generator on the left).
Such automorphisms cannot be constructed in general for Schreier graphs,
and some Schreier graphs have no automorphisms
(e.g., the ones we construct here).}
%
%

In light of the above,
it is interesting that expansion {\em can}\/ serve as
a sufficient condition for robust self-ordering
(as explained in the foregoing review of the direct construction);
recall, however, that this works for Schreier graphs,
and expansion needs to hold for the action on vertex-pairs.

\paragraph{On optimization:}
We made no attempt to minimize the degree bound
and maximize the robustness parameter. 
Note that we can obtain 3-regular robustly self-ordered graphs
by applying degree reduction; that is, given a $d$-regular graph,
we replace each vertex by a $d$-cycle and use each
of these vertices to ``hook'' one original edge.
To facilitate the analysis, we may use one color for the edges of
the $d$-cycles and another color for the other (i.e., original) edges.%
\footnote{Needless to say, we later replace all colored edges
by copies of adequate (3-regular) constant-sized gadgets.}
Hence, the issue at hand is actually one of maximizing
the robustness parameter of the resulting 3-regular graphs.

\paragraph{Caveat (tedious):}
Whenever we assert a $d$-regular $n$-vertex graph,
we assume that the trivial conditions hold;
specifically, we assume that $n>d$ and that $nd$ is even
(or, alternatively, allow for one exceptional vertex of degree $d-1$).

\subsection{Robustly self-ordered dense graphs}
In the second part of this paper
(i.e., Sections~\ref{dense-basics:sec}--\ref{inter-deg:sec})
we consider graphs of unbounded degree,
seeking correspondingly unbounded robustness parameters.
In particular, we are interested in $n$-vertex graphs
that are $\Omega(n)$-robustly self-ordered,
which means that they must have $\Omega(n^2)$ edges.

The construction of $\Omega(n)$-robustly self-ordered graphs
offers yet another alternative approach towards the construction
of bounded-degree graphs that are $\Omega(1)$-robustly self-ordered.
%
Specifically, we show that $n$-vertex graphs
that are $\Omega(n)$-robustly self-ordered
can be efficiently transformed into $O(n^2)$-vertex
bounded-degree graphs that are $\Omega(1)$-robustly self-ordered;
see Proposition~\ref{deg-reduction:clm},
which is essentially proved
by the ``degree reduction via expanders'' technique,
while using a different color for the expanders' edges,
and then using gadgets to replace colored edges
(see Theorem~\ref{colored2std:clm}).


\subsubsection{Our main results}
It is quite easy to show that random $n$-vertex graphs
are $\Omega(n)$-robustly self-ordered
(see Proposition~\ref{dense:random:clm});
in fact, the proof is easier than the proof of the analogous result
for bounded-degree graphs (Theorem~\ref{random-works:thm}).
Unfortunately, constructing $n$-vertex graphs
that are $\Omega(n)$-robustly self-ordered seems
to be no easier than constructing robustly self-ordered
bounded-degree graphs. In particular, it seems to
require completely different techniques and tools.

\BT[Constructing $\Omega(n)$-robustly self-ordered graphs]
\label{dense:ithm} 
There exist an infinite family
of dense $\Omega(n)$-robustly self-ordered graphs $\{G_n\}_{n\in\N}$
and a polynomial-time algorithm that,
given $n\in\N$ and a pair of vertices $u,v\in[n]$
in the $n$-vertex graph $G_n$,
determines whether or not $u$ is adjacent to $v$ in $G_n$.
\ET
Unlike in the case of bounded-degree graphs, in general,
we cannot rely on an efficient isomorphism test for
finding the original ordering of $G_n$, when given
an isomorphic copy of it. However, we can obtain
dense $\Omega(n)$-robustly self-ordered graphs
for which this ordering can be found efficiently
(see Theorem~\ref{dense:effective:thm}).
%

Our proof of Theorem~\ref{dense:ithm} is by a reduction to
the construction of non-malleable two-source extractors,
where a suitable construction of the latter was provided
by Chattopadhyay, Goyal, and Li~\cite{CGL}.
We actually present two different reductions
(Theorems~\ref{nmE2RSO:prop} and~\ref{nmE2RSO-tri:thm}),
one simpler than the other but yielding a less efficient construction
when combined with the known constructions of extractors.
We mention that the first reduction (Theorem~\ref{nmE2RSO:prop})
is partially reversible (see Proposition~\ref{reverse-nmE2RSO:prop},
which reverses a special case captured in Remark~\ref{nmE2RSO:rem}).

We show that $\Omega(n)$-robustly self-ordered $n$-vertex graphs
can be used to transport lower bounds
regarding testing binary strings to lower bounds regarding
testing graph properties in the dense graph model.
This general methodology, presented in Section~\ref{dense-pt:sec},
is analogous to the methodology for the bounded-degree graph model,
which is presented in Section~\ref{pt:sec}.

We mention that in a follow-up work~\cite{GW:na-vs-ad},
we employed this methodology in order to resolve several
open problems regarding the relation between adaptive
and non-adaptive testers in the dense graph model.
In particular, we proved that there exist graph properties
for which any non-adaptive tester must have query complexity
that is almost quadratic in the query complexity of
the best general (i.e., adaptive) tester,
whereas it has been known for a couple of decades that
the query complexity of non-adaptive testers is at most
quadratic in the query complexity of adaptive testers.

\paragraph{The case of intermediate degree bounds.}
Lastly, in Section~\ref{inter-deg:sec},
we consider $n$-vertex graphs of degree bound $d(n)$,
for every $d:\N\to\N$ such that $d(n)\in[\Omega(1),n]$.
Indeed, the bounded-degree case
(studied in Section~\ref{edge-colored:sec}--\ref{random-graphs:sec})
and the dense graph case
(studied in Sections~\ref{dense-basics:sec}--\ref{dense-pt:sec})
are special cases (which correspond to $d(n)=O(1)$ and $d(n)=n$).
Using results from these two special cases,
we show how to construct $\Omega(d(n))$-robustly
self-ordered $n$-vertex graphs of maximum degree $d(n)$,
for all $d:\N\to\N$.

\subsubsection{Techniques}\label{techniques-dense:sec}
As evident from the foregoing description, we reduce the construction
of $\Omega(n)$-robustly self-ordered $n$-vertex graphs
to the construction of non-malleable two-source extractors.

Non-malleable two-source extractors were introduced in~\cite{CG:nmE},
as a variant on seeded (one-source) non-malleable extractors,
which were introduced in~\cite{DW}.
Loosely speaking,
we say that $\nmE:\bitset^\ell\times\bitset^\ell\to\bitset^m$
is a non-malleable two-source extractor for a class of sources~$\C$
if for every two independent sources in~$\C$, denoted $X$ and $Y$,
and for every two functions $f,g:\bitset^\ell\to\bitset^\ell$
that have no fixed-point it holds that $(\nmE(X,Y),\nmE(f(X),g(Y)))$
is close to $(U_m,\nmE(f(X),g(Y))$,
where $U_m$ denotes the uniform distribution over $\bitset^m$.
We show that a non-malleable two-source extractor
for the class of $\ell$-bit sources of min-entropy $\ell-O(1)$,
with a single output bit (i.e., $m=1$) and constant error,
suffices for constructing $\Omega(n)$-robustly self-ordered $n$-vertex graphs.
Recall that constructions with much stronger parameters
(e.g., min-entropy $\ell-\ell^{\Omega(1)}$, negligible error,
and $m=\ell^{\Omega(1)}$)
were provided by Chattopadhyay, Goyal, and Li~\cite[Thm.~1]{CGL}.
(These constructions are quite complex.
Interestingly, we are not aware of a simpler way
of obtaining the weaker parameters that we need.)

Actually, we show two reductions of the construction
of $\Omega(n)$-robustly self-ordered $n$-vertex graphs
to the construction of non-malleable two-source extractors.
In both cases we use extractors that operate on pairs of sources
of length $\ell=\log_2n-O(1)$ that have min-entropy $k=\ell-O(1)$,
hereafter called {\sf $(\ell,k)$-sources}.
The extractor is used to define a bipartite graph
with~$2^\ell$ vertices on each side, and a clique
is placed on the vertices of one side so that a permutation
that maps vertices from one side to the other side yields
a proportional symmetric difference
(between the original graph and the resulting graph).

The first reduction, presented in Theorem~\ref{nmE2RSO:prop},
requires the extractor to be {\em quasi-orthogonal},
which means that the residual functions obtained by
any two different fixings of one of the extractor's two arguments
are almost unbiased and uncorrelated.
Using the fact that non-malleable two-source extractors
for $(\ell,k)$-sources can be made quasi-orthogonal in $\exp(\ell)$-time,
we obtain an explicit construction
of $\Omega(n)$-robustly self-ordered $n$-vertex graphs
(i.e., the $n$-vertex graph is constructed in $\poly(n)$-time).

The second reduction, presented in Theorem~\ref{nmE2RSO-tri:thm},
yields a strongly explicit construction as asserted in
Theorem~\ref{dense:ithm}
(i.e., the adjacency predicate of the $n$-vertex graph
is computable in $\poly(\log n)$-time).
This reduction uses an arbitrary non-malleable two-source extractor,
and shifts the quasi-orthogonality condition to
two auxiliary bipartite graphs.

Both reductions are based on the observation that if the number
of non-fixed-points (of the permutation) is very large,
then the non-malleability condition implies a large symmetric difference
(between the original graph and the resulting graph).
This holds as long as there are at least $\Omega(2^{\ell})$
non-fixed-points on each of the two sides of the corresponding
bipartite graph (which corresponds to the extractor).
The complementary case is handled by the quasi-orthogonality condition,
and this is where the two reductions differ.

The simpler case, presented in the first construction
(i.e., Theorem~\ref{nmE2RSO:prop}),
is that the extractor itself is quasi-orthogonal.
In this case we consider the non-fixed-points on the side that
has more of them.
The quasi-orthogonality condition gives us a contribution
of approximately $0.5\cdot2^{\ell}$ units per each non-fixed-point,
whereas the upper-bound on the number of non-fixed-points
on the other side implies that most of these contributions
actually count in the symmetric difference
(between the original graph and the resulting graph).

In the second construction (i.e., Theorem~\ref{nmE2RSO-tri:thm}),
we augment the foregoing $2^\ell$-by-$2^\ell$ bipartite graph,
which is now determined by any non-malleable extractor,
with an additional $4\cdot2^\ell$-vertex clique
that is connected to the two original $2^\ell$-vertex sets
by a bipartite graph that is merely quasi-orthogonal.
The analysis is analogous to the one used in the proof
of Theorem~\ref{nmE2RSO:prop}, but is slightly more complex
because we are dealing with a slightly more complex graph.

\paragraph{Errata regarding the original posting.}
%
We retract the claims made in our initial posting~\cite{GW:rso-vo}
regarding the construction of non-malleable two-source extractors
(which are quasi-orthogonal) as well as the claims
about the construction of relocation-detecting codes
(see Theorems~1.5 and~1.6 in the original version).%
\footnote{In~\cite{GW:rso-vo} quasi-orthogonality is called niceness;
we prefer the current term, which is less generic.}
The source of trouble is a fundamental flaw
in the proof of~\cite[Lem.~9.7]{GW:rso-vo},
which may as well be wrong.

\subsection{Perspective}
Asymmetric graphs were famously studied by Erdos and Renyi~\cite{ER},
who considered the (absolute) distance of asymmetric graphs from
being symmetric (i.e., the number of edges that should
be removed or added to a graph to make it symmetric),
calling this quantity the {\sf degree of asymmetry}.
They studied the extremal question of determining the
largest possible degree of asymmetry of $n$-vertex graphs
(as a function of $n$).
We avoided the term ``robust asymmetry''
because it could be confused with the degree of asymmetry,
which is a very different notion.
In particular, the degree of asymmetry cannot exceed twice
the degree of the graph (e.g., by disconnecting two vertices),
whereas our focus is on robustly self-ordered graphs
of bounded-degree.

We mention that Bollobas proved that,
{\em for every constant $d\geq3$,
almost all $d$-regular graphs are asymmetric}~\cite{B1,B2}.
This result was extended to varying $d\in[3,n-4]$
by Kim, Sudakov, and Vu~\cite{KSV}.
We also mention that their proof of~\cite[Thm.~3.1]{KSV}
implies that a random $n$-vertex Erdos--Renyi graph
with edge probability~$p$ is $2p(1-p)n$-robustly self-ordered.

\subsection{Roadmaps}
This work consists of two parts.
The first part (Sections~\ref{edge-colored:sec}--\ref{random-graphs:sec})
refers to bounded-degree graphs,
and the second part
(Sections~\ref{dense-basics:sec}--\ref{inter-deg:sec})
refers to dense graphs.
These parts are practically independent of one another,
except that Theorem~\ref{small-degree:thm}
builds upon Section~\ref{random-graphs:sec}.
%
Even when focusing on one of these two parts,
its contents may attract attention from diverse perspectives.
Each such perspective may benefit from a different roadmap.


\paragraph{Efficient combinatorial constructions.}
As mentioned above,
in the regime of bounded-degree graphs
we present two different constructions
that establish Theorem~\ref{main:ithm}.
Both constructions make use of the edge-colored model
and the transformations presented in Section~\ref{edge-colored:sec}.
The direct construction is presented in Section~\ref{direct:sec},
and the three-step construction appears in Section~\ref{three-step:sec}.
The three-step construction is augmented by local self-ordering and
local reversed self-ordering algorithms (see Section~\ref{local-so:sec}).%
\footnote{For a locally constructable $G_n$ and $G'=\phi^{-1}(G_n)$,
a {\em local self-ordering}\/ algorithm is given a vertex $v$ in $G'$,
and returns $\phi(v)$.
In contrast, a {\em local reversed self-ordering}\/ algorithm
is given a vertex $i\in[n]$ of $G_n$ and returns $\phi^{-1}(i)$.
Both algorithms run in $\poly(\log n)$-time.}
In the regime of dense graphs,
Sections~\ref{dense-basics:sec} and~\ref{dense+nmE:sec}
refer to the constructability of a couple of combinatorial objects;
see roadmap ``for the dense case'' below.

\paragraph{Potential applications to property testing.}
In Section~\ref{pt:sec} we demonstrate applications
of Theorem~\ref{main:ithm} to proving lower bounds
(on the query complexity) for the bounded-degree graph testing model.
Specifically, we present a methodology of transporting bounds
regarding testing properties of strings to bounds
regarding testing properties of bounded-degree graphs.
%
The specific applications presented in Section~\ref{pt:sec}
rely on Section~\ref{three-step:sec}.
For the first application (Theorem~\ref{pt-lb:thm})
the construction presented in Section~\ref{step2:sec} suffices;
for the second application (i.e., Theorem~\ref{pt-tolerant:thm}, which
establishes a separation between testing and tolerant testing
in the bounded-degree graph model), the local computation tasks studied
in Section~\ref{local-so:sec} are needed.
An analogous methodology for the dense graph testing model
is presented in Section~\ref{dense-pt:sec}.

\paragraph{Properties of random graphs.}
As stated above, it turns out that random $O(1)$-regular graphs
are robustly self-ordered.
This result is presented in Section~\ref{random-graphs:sec},
and this section can be read independently of any other section.
(In addition, Section~\ref{dense-basics:sec} presents a proof
that random (dense) $n$-vertex graphs are $O(n)$-robustly self-ordered.)

\paragraph{The dense case and non-malleable two-source extractors.}
The regime of dense graphs is studied in
Sections~\ref{dense-basics:sec}--\ref{dense-pt:sec},
where the construction of such graphs is undertaken
in Section~\ref{dense+nmE:sec}.
In Section~\ref{dense-basics:sec}, we show
that $\Omega(n)$-robustly self-ordered $n$-vertex graphs
provide yet another way of obtaining $\Omega(1)$-robustly
self-ordered bounded-degree graphs.
In Section~\ref{dense+nmE:sec}, we reduce the construction
of $O(n)$-robustly self-ordered $n$-vertex graphs
to the construction of non-malleable two-source extractors.
As outlined in Section~\ref{techniques-dense:sec},
we actually present two different reductions,
where a key issue is the quasi-orthogonality condition.

Lastly, in Section~\ref{inter-deg:sec},
for every $d:\N\to\N$ such that $d(n)\in[\Omega(1),n]$,
we show how to construct $n$-vertex graphs of maximum degree $d(n)$
that are $\Omega(d(n))$-robustly self-ordered.
Some of the results and techniques presented in this section
are also relevant to the setting of bounded-degree graphs.

\part{The Case of Bounded-Degree Graphs}
As stated in Section~\ref{techniques:sec},
a notion of robust self-ordering of edge-colored graphs
plays a pivotal role in our study of
robustly self-ordered bounded-degree graphs.
This notion as well as a transformation from it to the uncolored version
(of Definition~\ref{robust-asymmetric:def})
is presented in Section~\ref{edge-colored:sec}.

In Section~\ref{direct:sec}, we present a direct construction
of $O(1)$-regular robustly self-ordered edge-colored graphs;
applying the foregoing transformation,
this provides our first proof of Theorem~\ref{main:ithm}.
Our second proof of Theorem~\ref{main:ithm}
is presented in Section~\ref{three-step:sec},
and consists of a three-step process
(as outlined in Section~\ref{techniques:sec}).
Sections~\ref{direct:sec} and~\ref{three-step:sec}
can be read independently of one another,
but both rely on Section~\ref{edge-colored:sec}.

In Section~\ref{pt:sec} we demonstrate the applicability
of robustly self-ordered bounded-degree graphs to property testing;
specifically, to proving lower bounds (on the query complexity)
for the bounded-degree graph testing model.
For these applications, the global notion of constructability,
established in Section~\ref{step2:sec}, suffices.
This construction should be preferred over the direct
construction presented in Section~\ref{direct:sec},
because it can also yields graphs with small connected components.
More importantly, the subexponential separation between
the complexities of testing and tolerant testing of graph properties
(i.e., Theorem~\ref{pt-tolerant:thm})
relies on the construction of Section~\ref{three-step:sec}
and specifically on the local computation tasks studied
in Section~\ref{local-so:sec}.

Lastly, in Section~\ref{random-graphs:sec},
we prove that random $O(1)$-regular graphs are robustly self-ordered.
This section may be read independently of any other section.

\section{The Edge-Colored Variant}
\label{edge-colored:sec}
Many of our arguments are easier to make in a model
of (bounded-degree) graphs in which edges are colored
(by a bounded number of colors),
and where one counts the number of mismatches between colored edges.
Namely, an edge that appears in one (edge-colored) graph
contributes to the count if it either does not appear in the
other (edge-colored) graph or appears in it under a different color.
Hence, we define a notion of robust self-ordering for edge-colored graphs.
We shall then transform robustly self-ordered edge-colored graphs
to robustly self-ordered ordinary (uncolored) graphs,
while preserving the degree, the asymptotic number of vertices,
and other features such as expansion and degree-regularity.
Specifically, the transformation consists of replacing the
colored edges by copies of different connected,
asymmetric (constant-sized) gadgets such that different colors
are reflected by different gadgets.

We start by providing the definition of the edge-colored model.
Actually, for greater flexibility, we will consider multi-graphs;
that is, graphs with possible parallel edges and self-loops.
Hence, we shall consider multi-graphs $G=(V,E)$
coupled with an edge-coloring function $\chi\!:\!E\!\to\!\N$,
where $E$ is a multi-set containing both pairs of vertices
and singletons (representing self-loops).
Actually, it will be more convenient to represent self-loops
as 2-element multi-sets containing two copies of the same vertex.

\BD[Robust self-ordering of edge-colored multi-graphs]
\label{colored-robust:def}
Let $G=(V,E)$ be a multi-graph with colored edges,
where $\chi\!:\!E\!\to\!\N$ denotes this coloring,
and let $E_i$ denote the multi-set of edges colored~$i$
{\rm(i.e., $E_i=\{e\!\in\!E\!:\!\chi(e)\!=\!i\}$)}.
We say that $(G,\chi)$ is {\sf $\gamma$-robustly self-ordered}
if for every permutation $\mu:V\to{V}$ it holds that
\begin{equation}\label{robust4colored:eqdef}
\sum_{i\in\N}
\Big|E_i\;\triangle
       \left\{\{\mu(u),\mu(v)\}\!:\!\{u,v\}\!\in\!E_i\right\}\Big|
  \;\geq\; \gamma\cdot|\{i\!\in\!V\!:\!\mu(i)\neq i\}|,
\end{equation}
where $A\triangle B$ denotes the symmetric difference
between the multi-sets $A$ and $B$; that is $A\triangle B$
contains $t$ occurrences of $e$ if the absolute difference between
the number of occurrences of $e$ in $A$ and $B$ equals~$t$.
\ED
(Definition~\ref{robust-asymmetric:def} is obtained
as a special case when the multi-graph is actually a graph
and all edges are assigned the same color.)

We stress that whenever we consider ``edge-colored graphs''
we actually refer to edge-colored multi-graphs
(i.e., we explicitly allow parallel edges and self-loops).%
\footnote{We comment that a seemingly more appealing definition
can be used for edge-colored (simple) graphs.
Specifically, in that case (i.e., $E\subseteq\binom{V}{2}$),
we can extend $\chi\!:\!E\!\to\!\N$ to non-edges
by defining $\chi(\{u,v\})=0$ if $\{u,v\}\not\in E$,
and say that $(G,\chi)$ is {\sf $\gamma$-robustly self-ordered}
if for every permutation $\mu:V\to{V}$ it holds that\\[-1ex]
$$\left|\left\{\{u,v\}\in \binom{V}{2}:
       \chi(\{\mu(u),\mu(v)\})\!\neq\!\chi(\{u,v\})\right\}\right|
  \;\geq\; \gamma\cdot|\{i\!\in\!V:\mu(i)\neq i\}|.$$}
In contrast, whenever we consider (uncolored) graph,
we refer to simple graphs (with no parallel edges and no self-loops).

Our transformation of robustly self-ordered edge-colored multi-graphs
to robustly self-ordered ordinary graphs depends on the number
of colors used by the multi-graph.
In particular, $\gamma$-robustness of edge-colored multi-graph
that uses $c$ colors gets translated to $(\gamma/f(c))$-robustness
of the resulting graph, where $f:\N\to\N$ is an unbounded function.
Hence, we focus on coloring functions that use a constant number of colors,
denoted $c$. That is, fixing a constant $c\in\N$,
we shall consider multi-graphs $G=(V,E)$ coupled with an edge-coloring
function $\chi\!:\!E\!\to\![c]$.

\subsection{Transformation to standard (uncolored) version}
\label{edge-colored:trans:sec}
As a preliminary step for the transformation,
we add self-loops to all vertices
and make sure that parallel edges are assigned different colors.
The self-loops make it easy to distinguish
the original vertices from auxiliary vertices
that are parts of gadgets introduced in the main transformation.
Different colors assigned to parallel edges are essential
to the mere asymmetry of the resulting graph,
since we are going to replace edges of the same color
by copies of the same gadget.

\BCT[Preliminary step towards Construction~\ref{colored2standard:ct}]
\label{pre-colored2standard:ct}
For a fixed $d\geq3$,
given a multi-graph $G=(V,E)$ of maximum degree $d$
and an edge-coloring function $\chi\!:\!E\!\to\![c]$,
we define a multi-graph $G=(V,E')$ and
an edge-coloring function $\chi'\!:\!E'\!\to\![d\cdot c+1]$ as follows.
\BE
\item For every pair of vertices $u$ and~$v$ that are connected
by few parallel edges, denoted $e^{(1)}_{u,v},\ldots,e^{(d')}_{u,v}$,
we change, for each $i\in[d']$, the color of $e^{(i)}_{u,v}$
to $\chi'(e^{(i)}_{u,v})\gets(i-1)\cdot c+\chi(e^{(i)}_{u,v})$.
This includes also the case $u=v$.
\item We augment the multi-graph with self-loops colored $d\cdot c+1$;
that is, $E'$ is the multi-set $E\cup\{e_v:v\!\in\!V\}$,
where $e_v$ is a self-loop added to $v$, and $\chi'(e_v)=dc+1$.
\EE%
{\rm(Other edges $e\!\in\!E$ maintain their color;
that is, for them $\chi'(e)=\chi(e)$ holds.)}
\ECT

(For simplicity, we re-color all parallel edges, save the first one,
rather than re-coloring only parallel edges that have the same color.)
Note that refining the coloring may only increase
the robustness parameter of an edge-colored multi-graph.
Clearly, $G'$ preserves many features of~$G$.
In particular, it preserves $\gamma$-robust self-ordering,
expansion, degree-regularity, and the number of vertices.

As stated above, our transformation of edge-colored multi-graphs
to ordinary graphs uses gadgets, which are constant-size graphs.
Specifically, when handling a multi-graph of maximum degree $d$
with edges that are colored by $c$ colors,
we shall use $c$ different {\em connected and asymmetric}\/ graphs.
Furthermore, in order to maintain $d$-regularity,
we shall use $d$-regular graphs as gadgets;
and in order to have better control on the number
of vertices in the resulting graph,
each of these gadgets will contain $k=k(d,c)$ vertices.
The existence of such ($d$-regular) asymmetric (and connected) graphs
is well-known, let alone that it is known that
a random $d$-regular $k$-vertex graph is asymmetric
(for any constant $d\geq3$)~\cite{B1,B2}.

We stress that the different gadgets are
each connected and asymmetric,
and it follows that they are not isomorphic to one another.
We designate in each gadget an edge $\{p,q\}$,
called the {\sf designated edge},
such that omitting this edge does not disconnect the gadget.
The endpoints of this edge
will be used to connect two vertices of the original multi-graph.
Specifically, we replace each edge $\{u,v\}$ (of the original multi-graph)
that is colored $i$ by a copy of the $i^\xth$ gadget,
while omitting its designated edge $\{p,q\}$,
and connecting $u$ to $p$ and $v$ to $q$.
The construction is spelled out below.

We say that a (non-simple) multi-graph $G=(V,E)$
coupled with an edge-coloring $\chi$ is {\sf eligible}
if each of its vertices contains a self-loop,
and parallel edges are assigned different colors.
Recall that eligibility comes almost for free
(by applying Construction~\ref{pre-colored2standard:ct}).
We shall apply the following construction
only to eligible edge-colored multi-graphs.

\BCT[The main transformation]
\label{colored2standard:ct}
For a fixed $d\geq3$ and $c$, let $k=k(d,c)$
and $G_1,\ldots,G_c$ be different asymmetric
and connected $d$-regular graphs over the vertex-set $[k]$.
Given a multi-graph $G=(V,E)$ of maximum degree $d$
and an edge-coloring function $\chi\!:\!E\!\to\![c]$,
we construct a graph $G'=(V',E')$ as follows.
\begin{quote}
Suppose that the multi-set $E$ has size $m$.
Then, for each $j\in[m]$, if the $j^\xth$ edge of $E$
connects vertices $u$ and~$v$, and is colored $i$,
then we replace it by a copy of $G_i$, 
while omitting its designated edge and connecting
one of its endpoints to $u$ and the other endpoint to $v$.

Specifically, assuming that $V=[n]$ and
recalling that $j$ is the index of the edge
{\rm(colored~$i$)} that connects $u$ and~$v$,
let $G_i^{u,v}$ be an isomorphic copy of $G_i$
that uses the vertex set $\{n+(j-1)\cdot k+i:i\!\in\![k]\}$.
Let $\{p,q\}$ be the designated edge in $G_i^{u,v}$,
and $\GG_i^{u,v}$ be the graph that results from $G_i^{u,v}$
by omitting $\{p,q\}$.
Then, we replace the edge $\{u,v\}$ by $\GG_i^{u,v}$,
and add the edges $\{u,p\}$ and $\{v,q\}$.
\end{quote}
Hence, $V'=[n+m\cdot k]$
and $E'$ consists of the edges of all $\GG_i^{u,v}$'s
as well as the edges connecting the endpoint of the corresponding
designated edges to the corresponding vertices~$u$ and~$v$.
\ECT
We stress that, although~$G$ may have parallel edges and self-loops,
the graph $G'$ has neither parallel edges nor self-loops.
Also note that $G'$ preserve various properties of~$G$
such as degree-regularity, number of connected components,
and expansion (up to a constant factor).


We shall show that if the edge-colored multi-graph $G=(V,E)$
is robustly self-ordered (in the edge-colored sense),
then the resulting graph $G'=(V',E')$ is robustly self-ordered
(in the ordinary sense). The proof of this fact relies on
a correspondence between the colored edges of~$G$ and the gadgets in $G'$.
For starters, suppose that the permutation $\mu':V'\to V'$
maps $V$ to $V$ (i.e., $\mu'(V)=V$),
and gadgets to the corresponding gadgets; that is,
if $\mu'$ maps the vertex-pair $(u,v)\in V^2$
to $(\mu'(u),\mu'(v))\in V^2$,
then $\mu'$ maps the vertices in the possible gadget that
connects~$u$ and~$v$ to the vertices in the gadget that
connects $\mu'(u)$ and $\mu'(v)$.
In such a case, letting $\mu$ be the restriction of $\mu'$ to $V$,
a difference of $D$ colored edges between~$G$ and $\mu(G)$
translates to a difference of
at least $D$ edges between~$G'$ and $\mu'(G')$,
due to the difference between the gadgets
that replace the corresponding (colored) edges of $G'$,
%
whereas the number of non-fixed-point vertices in $\mu'$ is $k$ times
larger than the number of non-fixed-point vertices in $\mu$.
Assuming that~$G$ is $\gamma$-robustly self-ordered,
it follows that $\mu$ has at most $D/\gamma$ non-fixed-points.
Hence, in this case we have
 \begin{equation*}
\frac{|G'\triangle\,\mu'(G')|}{|\{v\in V':\mu'(v)\!\neq\!v\}|}
  \geq \frac{D}{k\cdot|\{v\in V:\mu(v)\!\neq\!v\}|}
   \geq \frac{D}{k\cdot D/\gamma}
 \end{equation*}
which equals $\gamma/k$.
However, in general, $\mu'$ needs not satisfy the foregoing condition.
Nevertheless, if $\mu'$ splits some gadget
or maps some gadget in a manner that is inconsistent with
the vertices of $V$ connected by it,
then this gadget contributes at least one unit
to the difference between~$G'$ and $\mu'(G')$,
whereas the number of non-fixed-point vertices in this gadget
is at most~$k$.
Lastly, if $\mu'$ maps vertices of a gadget to other
vertices in the same gadget, then we get a contribution
of at least one unit due to the asymmetry of the gadget.
The foregoing argument is made rigorous in the proof of the following theorem.

\BT[From edge-colored robustness to standard robustness]
\label{colored2std:clm}
For constant $d\geq3$ and $c\in\N$,
suppose that the multi-graph $G=(V,E)$ coupled with $\chi\!:\!E\!\to\![c]$
is eligible and $\gamma$-robustly self-ordered.
Then, the graph $G'=(V',E')$ resulting from
Construction~\ref{colored2standard:ct}
is $(\gamma/3k)$-robustly self-ordered,
where $k=k(d,c)>d$ is the number of vertices in a gadget
{\rm(as determined above)} and $\gamma\leq1$.
\ET

\BPF
As a warm-up, let us verify that $G'$ is asymmetric.
We first observe that the vertices of~$G$ are uniquely identified
(in $G'$), since they are the only vertices that are incident at copies
of the gadget that replaces the self-loops.%
\footnote{Indeed, a vertex of $G'$ is in $V$ if and only if
omitting it from $G'$ yields several connected components
such that (at least) one of them is a copy of the gadget
that replaces the self-loops (with the designated edge missing).}
Hence, any automorphism of $G'$ must map $V$ to $V$.
Consequently, for any $i$, such an automorphism $\mu'$
must map each copy of $G_i$ to a copy of $G_i$,
which means that when permuting $V$ according to $\mu'$
the edges of~$G$ as well as their colors are preserved.
By the ``colored asymmetry'' of~$G$,
this implies that $\mu'$ maps each $v\in V$ to itself,
and consequently each copy of $G_i$ must be mapped (by $\mu'$) to itself.
Finally, using the asymmetry of the $G_i$'s,
it follows that each vertex of each copy of $G_i$ is mapped to itself.
Hence, $\mu'$ must be the identity permutation.

We now turn to proving that $G'$ is actually robustly self-ordered.
Considering an arbitrary permutation $\mu':V'\to V'$,
we lower-bound the distance between~$G'$ and $\mu'(G')$
as a function of the number of non-fixed-points under $\mu'$
(i.e., of $v\in V'$ such that $\mu'(v')\neq v'$).
We do so by considering the contribution
of each non-fixed-point to the distance between~$G'$ and $\mu'(G')$.
We first recall the fact that the vertices of $V$
(resp., of gadgets) are uniquely identified in $\mu'(G')$
by virtue of the gadgets that replace self-loops
(see the foregoing warm-up).

\BDes
\item[{\em Case 1}:] {\em Vertices of some copy of $G_i$
that are not mapped by $\mu'$ to a single copy of $G_i$;
that is, vertices in some $G_i^{u,v}$ that are not
mapped by $\mu'$ to some $G_i^{u',v'}$.}

(This includes the case of vertices $w'$ and $w''$ of some $G_i^{u,v}$
such that $\mu'(w')$ is in $G_{i'}^{u',v'}$
and~$\mu'(w'')$ is in $G_{i''}^{u'',v''}$,
but $(i',u',v')\neq(i'',u'',v'')$.
It also includes the case of a copy of $G_i$ that is mapped by $\mu'$
to a copy of $G_j$ for $j\neq i$,
and the case that a vertex $w$ in some $G_i^{u,v}$
that is mapped by $\mu'$ to a vertex in $V$.)

The set of vertices $S_i^{u,v}$ of each such copy (i.e., $G_i^{u,v}$)
contribute at least one unit to the difference between~$G'$ and $\mu'(G')$,
since $\mu'(S_i^{u,v})$ induces a copy of $\GG_i$ in $\mu(G')$
but not in $G'$,
where here we also use the fact that the $\GG_i$'s are connected
(and not isomorphic (for the case of $i'=i''\neq i$)).
%
%
Note that the total contribution of all vertices of the current case
equals at least the number of gadgets in which they reside.
Hence, if the current case contains $n_1$ vertices,
then their contribution to the distance between~$G'$ and $\mu'(G')$
is at least $n_1/k$.

Ditto for vertices that do not belong to a single copy of $G_i$
and are mapped by $\mu'$ to a single copy of $G_i$.
This also includes $v\in V$ being mapped to some copy of some $G_i$,
but in this case we get a contribution of one unit
(rather than $1/k$ amortized units)
per each such vertex (i.e., $v\in V$ such that $\mu(v)\not\in V$).
%
\item[{\em Case 2}:] {\em Vertices of some copy of $G_i$
that are mapped by $\mu'$ to a single copy of $G_i$,
while not preserving their indices inside $G_i$.}

(This refers to vertices of some $G_i^{u,v}$
that are mapped by $\mu'$ to vertices of $G_i^{u',v'}$,
where $(u',v')$ may but need not equal $(u,v)$,
such that for some $j\in[k]$ the $j^\xth$ vertex of $G_i^{u,v}$
is not mapped by $\mu'$ to the $j^\xth$ vertex of $G_i^{u',v'}$.)%
\footnote{Recall that $G_i^{u,v}$ and $G_i^{u',v'}$ are both
copies of the $k$-vertex graph $G_i$, which is an asymmetric graph,
and so the notion of the $j^\xth$ vertex in them is well-defined.
Formally, the $j^\xth$ vertex of $G_i^{u,v}$ is $\phi^{-1}(j)$ such
that $\phi$ is the (unique) bijection satisfying $\phi(G_i^{u,v})=G_i$.}

By the fact that $G_i$ is asymmetric,
it follows that each such copy contributes at least one unit
to the difference between~$G'$ and $\mu'(G')$,
and so (again) the total contribution of all these vertices
is proportional to their number; that is,
if the number of vertices in this case is $n_2$,
then their contribution is at least $n_2/k$.

\item[{\em Case 3}:]
{\em Vertices $v\in V$ such that $\mu'(v)\in V\setminus\{v\}$.}

(This is the main case, and here we use the hypothesis that
the edge-colored multi-graph~$G$ is robustly self-ordered.

Intuitively,
the hypothesis that the edge-colored~$G$ is robustly self-ordered
implies that such vertices contribute proportionally
to the difference between the colored versions of
the multi-graphs~$G$ and $\mu(G)$,
where $\mu$ is the restriction of $\mu'$ to $V$.
Indeed, we first assume, for simplicity, that $\mu'(V)=V$,
an assumption we shall have to dispose of later.
In this case, the number of tuples $(\{u,w\},i)$
such that $\{u,w\}$ is colored $i$ in exactly one of these multi-graph
(i.e., either in~$G$ or in $\mu(G)$ but not in both)
is at least $\gamma\cdot|\{v\!\in\!V:\mu(v)\neq v\}|$.
Assuming, without loss of generality that $\chi(\{u,w\})=i$
but either $\{\imu(u),\imu(w)\}\not\in E$
or $\chi(\{\imu(u),\imu(w)\})=j\neq i$,
we observe that $\imu(u)$ and $\imu(w)$ cannot
be connected in $G'$ via a copy of $G_i$.
We consider two sub-cases:
\BE
\item $\mu'$ maps a copy of $G_i$ to $G_i^{u,w}$,
but either $\imu(u)$ or $\imu(w)$ is not connected to this copy in $G'$.
In this sub-case we get a contribution of at least one unit,
since $u$ and $w$ are connected to $G_i^{u,w}$ in $G'$.
\item $\mu'$ does not map a copy of $G_i$ to $G_i^{u,w}$.
In this sub-case, it follows that some vertices that do not belong
to a copy of $G_i$ are mapped by $\mu'$ to $G_i^{u,w}$
which means that Case~1 applies for each such a tuple.
\EE
Hence, if the number of vertices in the current case is $n_3$,
then the number of tuples (handled by the two sub-cases)
is at least $\gamma\cdot n_3$, and we get a contribution
of at least $\gamma\cdot{n_3}/{k}$
(since the second sub-case is handled via Case~1).

The foregoing description is based on the assumption that $\mu(V)=V$.
If this does not hold, then we redefine $\mu$ such that $\mu(v)\not\in V$
is modified such that $\mu(v)=r$ if $r\in V$ has no preimage under $\mu'$.
(Of course, each such $r$ is only used once.)
Indeed, the modified $\mu$ may be ficticiously charged with $d$ edges
per each modification, but each such modification arises due to $v\in V$
that contributes at least one unit via (the last part of) Case~1.
Hence, the amortized over-counting of $d\cdot \gamma/k$ units
is compansated by the unit contributed in Case~1.

\item[{\em Case 4}:] {\em Vertices of some copy of $G_i$
that are mapped by $\mu'$ to a different copy of $G_i$.}

This refers to the case that $\mu'$ maps $G_i^{u,v}$
to $G_i^{u',v'}$ such that $(u',v')\neq(u,v)$,
which corresponds to mapping the gadget to a gadget
connecting a different pair of vertices
(but by an edge of the same color).

For $u,v,u',v'$ and $i$ as above, if $\mu'(u)=u'$ and $\mu'(v)=v'$,
then a gadget that connects $u$ and~$v$ in $G'$
is mapped to a gadget that does not connects them in $\mu'(G')$
(but rather connects the vertices $u'$ and $v'$,
whereas either $u'\neq u$ or $v'\neq v$).
So, due to the gadget-edge incident at either $u$ or $v$,
we get a contribution of at least one unit to the
difference between~$G'$ and $\mu'(G')$,
whereas the number of vertices in this gadget is $k$.
Hence, the contribution is proportional to the
number of non-fixed-points of the current type.
Otherwise (i.e., $(\mu'(u),\mu'(v))\neq(u',v')$),
we get a vertex as in Case~3,
and get a proportional contribution again.
%
\EDes
Hence, the contribution of each of these cases
to the difference between~$G'$ and $\mu'(G')$
is proportional to the number of vertices involved.
Specifically, if there are $n_i$ vertices in Case~$i$,
then we get a contribution-count of
at least $\gamma\cdot\sum_{i\in[4]}n_i/k$,
where some of these contributions were possibly counted thrice.
The claim follows.
\EPF

\BR[Fitting any desired number of vertices]
\label{colored2std:rem}
Assuming that the hypothesis of Theorem~\ref{colored2std:clm}
can be met for any sufficiently large $n\in S\subseteq\N$,
Construction~\ref{colored2standard:ct} yields robustly
self-ordered $n'$-vertex graphs for any $n'\in\{k\cdot n:n\!\in\!S\}$,
where $k=k(d,c)$ is as in Theorem~\ref{colored2std:clm}.
To obtain such graphs also for $n'$ that is not a multiple of $k$,
we may use two gadgets with a different number of vertices
for replacing at least one of the sets of colored edges.
\ER

\subsection{Application: Making the graph regular and expanding}
\label{edge-colored:applic:sec}
We view the edge-colored model as an intermediate locus
in a two-step methodology for constructing
robustly self-ordered graphs of bounded-degree.
First, one constructs edge-colored multi-graphs that
are robustly self-ordered
in the sense of Definition~\ref{colored-robust:def},
and then converts them to ordinary robustly self-ordered graphs
(in the sense of Definition~\ref{robust-asymmetric:def}),
by using Construction~\ref{colored2standard:ct}
(while relying on Theorem~\ref{colored2std:clm}).

We demonstrate the usefulness of this methodology
by showing that it yields a simple way of
making robustly self-ordered graphs be also expanding
as well as regular, while maintaining a bounded degree.
We just augment the original graph by super-imposing
an expander (on the same vertex set),
while using one color for the edges of the original graph
and another color for the edges of the expander.
Note that we do not have to worry about the possibility of
creating parallel edges (since they are assigned different colors).
The same method applies in order to make the graph regular.
We combine both transformations in the following result,
which we shall use in the sequel.

\BT[Making the graph regular and expanding]
\label{make:regular+expanding:thm}
For constant $d\geq3$ and~$\gamma$,
there exists an efficient algorithm
that given a $\gamma$-robustly self-ordered
graph $G=(V,E)$ of maximum degree $d$,
returns a $(d+O(1))$-regular multi-graph expander
coupled with a 2-coloring of its edges
such that the edge-colored multi-graph is $\gamma$-robustly self-ordered
{\rm(in the sense of Definition~\ref{colored-robust:def})}.
\ET
The same idea can be applied to edge-colored multi-graphs;
in this case, we use one color more than given.
We could have avoided the creation of parallel edges
with the same color by using more colors,
but preferred to relegate this task to
Construction~\ref{pre-colored2standard:ct},
while recalling that it preserves both the expansion
and the degree-regularity.
Either way, applying Theorem~\ref{colored2std:clm}
to the resulting edge-colored multi-graph,
we obtain robustly self-ordered (uncolored) graphs.
\medskip

\BPF
For any $d''\geq d+d'$,
given a graph $G=(V,E)$ of maximum degree $d$ that
is $\gamma$-robustly self-ordered and a $d'$-regular expander
graph $G'=(V,E')$,
we construct the desired $d''$-regular multi-graph $G''$
by super-imposing the two graphs on the same vertex set,
while assigning the edges of each of these graphs a different color.
In addition, we add edges to make the graph regular,
and color them using the same color as used for the expander.%
\footnote{We assume for simplicity that $|V'|$ is even.
Alternatively, assuming that~$G$ contains no isolated vertex,
we first augment it with an isolated vertex
and apply the transformation on the resulting graph.
Yet another alternative is to consider only even $d''$.}
%
Details follow.

\BI
\item
We superimpose~$G$ and $G'$ (i.e., create a multi-graph $(V,E\cup E')$),
while coloring the edges of~$G$ (resp., $G'$)
with color~1 (resp., color~2).

Note that this may create parallel edges,
but with different colors.
\item
Let $d_v\leq d+d'$ denote the degree of vertex $v$
in the resulting multi-graph.
Then, we add edges to this multi-graph so that each vertex has degree~$d''$.
These edges will also be colored~2.

(Here, unless we are a bit careful, we may introduce
parallel edges that are assigned the same color.
This can be avoided by using more colors for these added edges,
but in light of Construction~\ref{pre-colored2standard:ct}
(which does essentially the same)
there is no reason to worry about this aspect.)
\EI
(Recall that the resulting edge-colored multi-graph is denoted $G''$.)

The crucial observation is that, since the edges of~$G$
are given a distinct color in $G''$, the added edges
do not harm the robust self-ordering feature of~$G$.
Hence, for any permutation $\mu:V\to V$,
any vertex-pair that contributes to the symmetric difference
between~$G$ and $\mu(G)$, also contributes to an inequality
between colored edges of $G''$ and $\mu(G'')$
(by virtue of the edges colored~1).
\EPF

\subsection{Local computability of the transformations}
\label{edge-colored:local:sec}
In this subsection, we merely point out that the transformation
presented in Constructions~\ref{pre-colored2standard:ct}
and~\ref{colored2standard:ct} as well as the one underlying
the proof of Theorem~\ref{make:regular+expanding:thm}
preserve efficient local computability
(e.g., one can determine the neighborhood of a vertex in
the resulting multi-graph by making a polylogarithmic number
of neighbor-queries to the original multi-graph).
Actually, this holds provided that we augment the
(local) representation of graphs, in a natural manner.

Recall that the standard representation of bounded-degree graphs
is by their incidence functions.
Specifically, a graph $G=([n],E)$ of maximum degree $d$ is represented
by the {\sf incident function} $g:[n]\times[d]\to[n]\cup\{0\}$
such that $g(v,i)=u\in[n]$ if $u$ is the $i^\xth$ neighbor of $v$,
and $g(v,i)=0$ if $v$ has less than $i$ neighbors.
This does not allow us to determined the identity of the $j^\xth$
edge in~$G$, nor even to determine the number of edges in~$G$,
by making a polylogarithmic number of queries to $g$,
whereas this determination is needed for
a local implementation of Construction~\ref{colored2standard:ct}.
Nevertheless, efficient local computability is preserved
if we use the following local representation
(presented for edge-colored multi-graphs).

\BD[Local representation]
For $d,c\in\N$, a {\sf local representation}
of a multi-graph $G=([n],E)$ of maximum degree $d$
that is coupled with a coloring $\chi\!:\!E\!\to\![c]$
is provided by the following three functions:
\BE
\item
An incidence function $g_1:[n]\times[d]\to\N\cup\{0\}$
such that $g_1(v,i)=j\in\N$
if $j$ is the index of the $i^\xth$ edge that incident at vertex $v$,
and $g_1(v,i)=0$ if $v$ has less than $i$ incident edges.
\item
An edge enumeration function $g_2:\N\to(\binom{[n]}{2}\times[c])\cup\{0\}$
such that $g_2(j)=(\{u,v\},\chi(e_j))$ if the $j^\xth$ edge,
denoted $e_j$, connects the vertices $u$ and~$v$,
and $g_2(j)=0$ if the multi-graph has less than $j$ edges.
\item
An vertex enumeration {\rm(by degree)}
function $g_3:[d]\times[n]\to[n]\cup\{0\}$
such that $g_3(i,j)=v\in[n]$
if $v$ is the $j^\xth$ vertex of degree $i$ in the multi-graph,
and $g_3(i,j)=0$ if the multi-graph has less than $j$ vertices of degree $i$.
\EE
\ED

The aforementioned incident function $g:[n]\times[d]\to[n]\cup\{0\}$
can be computed by composing $g_1$ and $g_2$;
in particular, $g(v,i)=u\in[n]$ if $g_2(g_1(v,i))=(\{u,v\},k)$
for some $k\in[c]$.
Needless to say, the function $g_3$ is redundant in the case
that we are guaranteed that the multi-graph is regular.
One may augment the foregoing representation by providing also
the total number of edges, but this number can be determined
by binary search.

\BT[The foregoing transformations preserve local computability]
\label{local-computability:thm}
The local representation of the multi-graph that
result from Construction~\ref{pre-colored2standard:ct}
can be computed by making a polylogarithmic number of queries
to the given multi-graph.
The same holds for Construction~\ref{colored2standard:ct}
and for the transformation underlying
the proof of Theorem~\ref{make:regular+expanding:thm}.
\ET

\BPF
For Construction~\ref{pre-colored2standard:ct},
we mostly need to enumerate all parallel edges
that connect~$u$ and~$v$. This can be done easily
by querying the incidence function on $(u,1),\ldots,(u,d)$
and querying the edge enumeration function on the non-zero answers.
(In addition, when adding a self-loop on vertex $v\in[n]$,
we need to determine the degree of $v$
as well as the number of edges in the multi-graph
(in order to know how to index the self-loop
in the incidence and edge enumeration functions, respectively).)

For Construction~\ref{colored2standard:ct},
we merely need to determine the color of the $j^\xth$ edge,
its endpoint
(and its index in the incidence list of each of its endpoints),
in order to replace this colored edge by the relevant gadget.
Recall that the relevant gadget
uses the vertices $n+(j-1)\cdot k+1,\ldots,n+j\cdot k$
and its edges are determined by the color of the edge that it replaces.

For the transformation underlying
the proof of Theorem~\ref{make:regular+expanding:thm},
adding edges to make the multi-graph regular requires
determining the index of a vertex in the list of all vertices
of the same degree (in order to properly index the added edges).
Here is where we use the vertex enumeration (by degree) function.
(We also need a local procedure $I$ for transforming
a sorted $n$-long sequence $(d_1,\ldots,d_n)\in[d'']$
into an all-$d''$ sequence by making pairs of increments;
that is, given $j\in[D]$ such that $D=(d''n-\sum_{i\in[n]}d_i)/2$,
we should determine a pair $I(j)=(I_1(j),I_2(j))\in[n]^2$
such that for every $i\in[n]$ it holds
that $d_i+|I_1^{-1}(i)|+|I_2^{-1}(i)|=d''$.)
\EPF

\section{The Direct Construction}
\label{direct:sec}
We shall make use of the edge-colored variant presented
in Section~\ref{edge-colored:sec}, while relying on the fact
that robustly self-ordered colored multi-graphs can be efficiently
transformed into robustly self-ordered (uncolored) graphs.
Actually, it will be easier to present the construction
as a directed edge-colored multi-graph.
Hence, we first define a variant of robust self-ordering
for directed edge-colored multi-graph
(see Definition~\ref{directed-robust:def}),
then show how to construct such multi-graphs
(see Section~\ref{direct-main:sec}),
and finally show how to transform the directed variant
into an undirected one (see Section~\ref{direct-trans:sec}).

The construction is based on $d$ permutations,
denoted $\pi_1,\ldots,\pi_d:[n]\to[n]$,
and consists of the directed edge-colored multi-graph
that is naturally defined by them.
Specifically, for every $v\in[n]$ and $i\in[d]$,
this multi-graph contains a directed edge, denoted $(v,\pi_i(v))$,
that goes from vertex $v$ to vertex $\pi_i(v)$, and is colored~$i$.

We prove that a sufficient condition for this edge-colored directed
multi-graph, denoted~$G_1$, to be robustly self-ordered is that
a related multi-graph is an expander.
Specifically, we refer to the multi-graph $G_2=(V_2,E_2)$
that represents the actions of these permutations on pairs of vertices
of~$G_1$; that is, $V_2=\{(u,v)\!\in\![n]^2:u\!\neq\!v\}$
and
$E_2=\{\{(u,v),(\pi_i(u),\pi_i(v))\}:(u,v)\!\in\!V_2\,\&\,i\!\in\![d]\}$.

The foregoing requires extending the notion of robustly self-ordered
(edge-colored) multi-graphs to the directed case.
The extension is straightforward and is spelled-out next,
for sake of good order.

\BD[Robust self-ordering of edge-colored directed multi-graphs]
\label{directed-robust:def}
Let $G=(V,E)$ be a directed multi-graph with colored edges,
where $\chi\!:\!E\!\to\!\N$ denotes this coloring,
and let $E_i$ denote the multi-set of directed edges colored~$i$.
We say that $(G,\chi)$ is {\sf $\gamma$-robustly self-ordered}
if for every permutation $\mu:V\to{V}$ it holds that
\begin{equation}\label{robust4directed:eqdef}
\sum_{i\in\N}
\Big|E_i\;\triangle
       \left\{(\mu(u),\mu(v))\!:\!(u,v)\!\in\!E_i\right\}\Big|
  \;\geq\; \gamma\cdot|\{i\!\in\!V\!:\!\mu(i)\neq i\}|,
\end{equation}
where $A\triangle B$ denotes the symmetric difference
between the multi-sets $A$ and $B$
{\rm(as in Definition~\ref{colored-robust:def})}.
\ED
(The only difference between Definition~\ref{directed-robust:def}
and Definition~\ref{colored-robust:def}
is that \eqref{robust4directed:eqdef} refers to
the directed edges of the directed multi-graph,
whereas \eqref{robust4colored:eqdef} refers
to the undirected edges of the undirected multi-graph.)

In Section~\ref{direct-main:sec} we present a construction
of a directed edge-colored $O(1)$-regular multi-graph
that is $\Omega(1)$-robustly self-ordered.
We shall actually present a sufficient condition
and a specific instantiation that satisfies it.
In Section~\ref{direct-trans:sec} we show how to transform
any directed edge-colored multi-graph into
an undirected one while preserving all relevant features;
that is, bounded robustness, bounded degree, regularity, expansion,
and local computability.

\subsection{A sufficient condition for robust self-ordering
of directed colored graphs}
\label{direct-main:sec}
For any $d$ permutations, $\pi_1,\ldots,\pi_d:[n]\to[n]$,
we consider two multi-graphs.
\BE
\item The {\sf primary multi-graph (of $\pi_1,\ldots,\pi_d$)}
is a {\em directed}\/ multi-graph, denoted $G_1=([n],E_1)$,
such that $E_1=\{(v,\pi_i(v)):v\!\in\![n]\,\&\,i\!\in\![d]\}$.
This directed multi-graph is coupled with an edge-coloring in
which the directed edge from $v$ to $\pi_i(v)$ is colored~$i$.
\item The {\sf secondary multi-graph (of $\pi_1,\ldots,\pi_d$)}
is an undirected multi-graph, denoted $G_2=(V_2,E_2)$,
such that $V_2=\{(u,v)\!\in\![n]^2:u\!\neq\!v\}$
and $E_2=\{\{(u,v),(\pi_i(u),\pi_i(v))\}:(u,v)\!\in\!V_2\,\&\,i\!\in\![d]\}$.
\EE
Recalling that we wish the secondary multi-graph to be an expander,
we mention that an archetypical case is when each of the foregoing
multi-graphs is a {\em Schreier graph}\/ that correspond to
the action of the permutation $\pi_1,\ldots,\pi_d$
on the corresponding vertex sets (i.e., $[n]$ and $V_2$, respectively).
See Proposition~\ref{direct:LPS:rem} and a
wider perspective at the (paragraph at the) end of this subsection.

We now state the main result of this section,
which asserts that the primary multi-graph~$G_1$ is robustly self-ordered
if the secondary multi-graph $G_2$ is an expander.
We use the combinatorial definition of expansion:
{\em A multi-graph $G=(V,E)$ is {\sf $\gamma$-expanding}
if, for every subset~$S$ of size at most $|V|/2$,
there are at least $\gamma\cdot|S|$ vertices in $V\setminus S$
that neighbor some vertex in~$S$}.

\BT[Expansion of $G_2$ implies robust self-ordering of~$G_1$]
\label{primary-secondary:thm}
For any $d\geq2$ permutations, $\pi_1,\ldots,\pi_d:[n]\to[n]$,
if the secondary multi-graph $G_2$ of $\pi_1,\ldots,\pi_d$
is $\gamma$-expanding,
then the primary directed multi-graph~$G_1$ of $\pi_1,\ldots,\pi_d$ coupled
with the foregoing edge-coloring is $(\gamma/2)$-robustly self-ordered.
Furthermore, $G_1$
{\em(or rather the undirected multi-graph underlying~$G_1$)}
is $\min(0.25,\gamma/3)$-expanding.
\ET

\BPF
Let $\mu:[n]\to[n]$ be an arbitrary permutation,
and let $T=\{v\!\in\![n]:\mu(v)\!\neq\!v\}$
be its set of non-fixed-points.
Then, the size of the symmetric difference between~$G_1$ and $\mu(G_1)$
equals $2\cdot\sum_{i\in[d]}|D_i|$ such that $v\in D_i$
if $(\mu(v),\mu(\pi_i(v)))$ is either not an edge in~$G_1$
or is not colored~$i$ in it,
whereas $(v,\pi_i(v))$ is an edge colored~$i$ in~$G_1$.
Note that
if $(\mu(v),\mu(\pi_i(v)))$ is not an $i$-colored edge in~$G_1$,
then $\pi_i(\mu(v))\neq\mu(\pi_i(v))$.
Hence, $D_i=\{v\!\in\![n]:\mu(\pi_i(v))\neq\pi_i(\mu(v))\}$.

The key observation (proved next) is that
{\em if $v\in T\setminus D_i$,
then $(\pi_i(v),\pi_i(\mu(v)))\in T_2$,
where $T_2=\{(v,\mu(v)):v\!\in\!T\}$ represents
the set of replacements performed by $\mu$}.
This fact implies that if $\sum_{i\in[d]}|D_i|$
is small in comparison to $|T|$,
then the set $T_2$ (which is a set of vertices in $G_2$)
does not expand much, in contradiction to the hypothesis.
Details follow.

\Bobs[Key observation] 
\label{key-observation}
For $T,~D_i$ and $T_2$ as defined above, if $v\in T\setminus D_i$, 
then $(\pi_i(v),\pi_i(\mu(v))\in T_2$.
\Eobs
Recall that $v\in T$ implies $(v,\mu(v))\in T_2$.
Observation~\ref{key-observation} asserts that
if (in addition to $v\in T$) it holds that $v\not\in D_i$,
then $(\pi_i(v),\pi_i(\mu(v))$ is also in $T_2$.
This means that the vertices in
$\{(\pi_i(v),\pi_i(\mu(v))):v\!\in\!T\setminus D_i\}$
do not contribute to the expansion of the set $T_2$ in $G_2$.
\medskip

\Bpf
Since $v\not\in D_i$ we have $\pi_i(\mu(v))=\mu(\pi_i(v))$,
and $\mu(\pi_i(v))\neq\pi_i(v)$ follows,
because otherwise $\pi_i(\mu(v))=\pi_i(v)$,
which implies $\mu(v)=v$ in contradiction to $v\in T$.
However, $\mu(\pi_i(v))\neq\pi_i(v)$ means that $\pi_i(v)\in T$,
and $(\pi_i(v),\mu(\pi_i(v)))\in T_2$ follows.
Using $\mu(\pi_i(v))=\pi_i(\mu(v))$ again,
we get $(\pi_i(v),\pi_i(\mu(v)))\in T_2$.
\Epf


\mypar{Establishing the main claim (i.e., robustness of~$G_1$).}
Recall that Observation~\ref{key-observation} implies that
$\{(\pi_i(v),\pi_i(\mu(v))):v\!\in\!T\setminus D_i\}\subseteq T_2$.
On the other hand, assuming that the sequence of $\pi_i$'s
contains its own inverses
(i.e., $\forall i\!\in\![d]\exists j\!\in\![d]$ such that $\pi^{-1}_i=\pi_j$),
we observe that $\bigcup_{i\in[d]}\{(\pi_i(v),\pi_i(\mu(v))):v\!\in\!T\}$
is the neighborhood of $T_2$ in the multi-graph $G_2$
(since $\{(\pi_i(v),\pi_i(\mu(v))):i\!\in\![d]\}$
is the neighbor-set of $(v,\mu(v))$ in $G_2$).
Using the $\gamma$-expansion of the set $T_2$ in $G_2$
(while relying on $|T_2|\leq n<|V_2|/2$),
it follows that
$$\sum_{i\in[d]}
  |\{(\pi_i(v),\pi_i(\mu(v))):v\!\in\!T\cap D_i\}|\geq\gamma\cdot|T_2|.$$
Hence, $\sum_{i\in[d]}|D_i|\geq\gamma\cdot|T|$, and
the main claim follows in this case.
We reduce the general case to this special case
by augmenting the sequence of $\pi_i$'s by their inverses
(i.e., we add the permutations $\pi_1^{-1},\ldots,\pi_d^{-1}$,
which are associated colors $d+1,\ldots,2d$).
Observing that the corresponding primary graph is $\gamma$-robustly
self-ordered and that it is twice more robust than the original~$G_1$,
the claim follows.

\mypar{Establishing the furthermore claim (i.e., expansion of~$G_1$).}
The expansion of~$G_1$ is shown by relating sets of vertices of~$G_1$
to the corresponding sets of pairs in $G_2$.
Specifically, for and $S\subset[n]$ of size at most $n/2$,
we consider the set $T=\{(u,v)\!\in\!V_2:u,v\!\in\!S\}$,
which has size
$|S|\cdot(|S|-1)\leq\frac{n}{2}\cdot(\frac{n}{2}-1)<\frac{|V_2|}{2}$.
Letting $T'$ denote the set of neighbors of $T$ in $G_2$,
and $S'$ denote the set of neighbors of~$S$ in~$G_1$,
on the one hand we have $|T'\setminus T|\geq\gamma\cdot|T|$
(by expansion of $T$ in $G_2$), and on the other hand
$|T'\setminus T|\leq2\cdot|S|\cdot|S'\setminus S|
 +|S'\setminus S|\cdot(|S'\setminus S|-1)$.
This implies $|S'\setminus S| \geq (\gamma/3)\cdot|S|$
(unless $|S|\leq4$, which can be handled by using $|S'\setminus S|\geq1$).
\EPF

\paragraph{Primary and secondary multi-graphs based on $\SL_2(p)$.}
Recall that $\SL_2(p)$ is the multiplicative group of 2-by-2 matrices
over $\GF(p)$ that have determinant~1.
There are several different explicit constructions of constant-size
expanding generating sets for $\SL_2(p)$,
namely making the associated Cayley graph an expander
(see, e.g.,~\cite{LPS}, \cite[Thm.~4.4.2(i)]{Lub}, and~\cite{BG}).
We use any such generating set to define
a directed (edge-colored) multi-graph~$G_1$ on $p+1$ vertices,
and show that the associated multi-graph on pairs, $G_2$, is an expander.

\BP[Expanding generators for $\SL_2(p)$
yield an expanding secondary multi-graph]
\label{direct:LPS:rem}
For any prime $p>2$,
let $V=\{(1,i)^\top:i\in\GF(p)\}\cup\{(0,1)^\top\}$,
and $M_1,\ldots,M_d\in\SL_2(p)$.
For every $i\in[d]$, define $\pi_i:V\to V$
such that $\pi_i(u)=v$ if $v\in V$
is a non-zero multiple of $M_iu$.
Then:
\BE
\item Each $\pi_i$ is a bijection.
\item If the Cayley multi-graph
$\CG=\CG(\SL_2(p),\{M_1,\ldots,M_d\})
  =(\SL_2(p),\{\{M,M_iM\}:M\!\in\SL_2(p)\,\&\,i\!\in\![d]\})$
is an expander,
then the {\rm(Schreier)} multi-graph $G_2$ with
vertex-set $P=\{(v,v'):v\!\in\!V\,\&\,v'\!\in\!V\setminus\{v\}\}$
and edge-set $\{\{(v,v'),(\pi_i(v),\pi_i(v'))\}:(v,v')\!\in\!P\}$
is an expander.
\EE
\EP
Part~1 implies that these permutations yield a primary
(directed edge-colored) multi-graph on the vertex-set $V$,
whereas Part~2 asserts that the corresponding secondary graph
is an expander (if the corresponding Cayley graph is expanding).
Note that $|V|=p+1$ and $|P|=(p+1)p$,
whereas $|\SL_2(p)| = p^3-p = (p-1)\cdot|P|$.
\medskip

\BPF
Part~1 follows by observing that for every $M\in\SL_2(p)$
and every vector $v\in\GF(p)^2$ and scalar $\alpha\in\GF(p)$
it holds that $M\alpha v=\alpha Mv$.
Consequently, if for some non-zero $\alpha,\alpha'\in\GF(p)$
it holds that $\alpha Mv=\alpha'Mv'$,
then $Mv=M\alpha''v'$ for $\alpha''=\alpha'/\alpha$,
which implies $v=\alpha''v'$ (since~$M$ is invertible).
Hence, $\pi_i(v)=\pi_i(v')$, for $v,v'\in V$,
implies that $M_iv$ and $M_iv'$ are non-zero multiples
of the same $w\in V$, which implies $v=v'$
(since $V$ contains a single non-zero multiple of each vector in it).

Part~2 follows by observing that the vertices of $G_2$
correspond to equivalence classes of the vertices of $\CG$
{\em that are preserved by $\SL_2(p)$},
where $A,B\in\SL_2(p)$ are equivalent if the columns of~$A$
are non-zero multiples of the corresponding columns of $B$.
That is, we consider an equivalence relation, denoted $\equiv$,
such that for $A=[A_1|A_2]$ and $B=[B_1|B_2]$ in $\SL_2(p)$
it holds that $A\equiv B$ if $A_i=\alpha_iB_i$ for both $i\in\{1,2\}$,
where $\alpha_1,\alpha_2\in[p-1]$ (and, in fact, $\alpha_2=1/\alpha_1$).%
\footnote{Recall that ${\rm det}(A)=1={\rm det}(B)$, whereas
${\rm det}([\alpha_1B_1|\alpha_2B_2])=\alpha_1\alpha_2\cdot{\rm det}(B)$.
Note that each equivalence class contains a single element of $P$.}
%
%
%
By saying that these {\em equivalence classes are preserved by $\SL_2(p)$},
we mean that, for every $A,B,M\in\SL_2(p)$,
if $A\equiv B$, then $MA \equiv MB$.
Hence, the (combinatorial) expansion of $G_2$
follows from the expansion of $\CG$,
because the neighbors of a vertex-set $S\subseteq P$ in $G_2$
are the vertices of $G_2$ that are equivalent to $T'$
such that $T'$ is the set of vertices of $\CG$ that neighbor
(in $\CG$) vertices that are equivalent to vertices in~$S$.%
%
\footnote{Specifically, let~$S$ have density at most half in $P$,
and let $T$ be the set of vertices of $\CG$ that are equivalent to~$S$.
Note that $|T|=(p-1)\cdot|S|$,
since each equivalence class contains a single element of $P$.
By the foregoing, the set of neighbors of $T$ in $\CG$, denoted $T'$,
is a collection of equivalence classes of vertices of $G_2$,
and $|T'\setminus T|=\Omega(|T|)$ by the expansion of $\CG$.
It follows that the set of neighbors of~$S$ in $G_2$, denoted $S'$,
is the set of vertices that are equivalent to $T'$,
which implies that
$|S'\setminus S| = \frac{|T'\setminus T|}{p-1}
 = \frac{\Omega(|T|)}{p-1} = \Omega(|S|)$.}
\EPF


\paragraph{A simple construction.}
Combining Theorem~\ref{primary-secondary:thm}
with Proposition~\ref{direct:LPS:rem},
while using a simple pair of expanding generators
(which does not yield a Ramanujan graph),
we get

\BCR[A simple robustly self-ordered primary multi-graph]
\label{direct:SL2-four:cor}
For any prime $p>2$,
let $V=\{(1,i)^\top:i\in\GF(p)\}\cup\{(0,1)^\top\}$,
and consider the matrices
\begin{equation}
M_1\eqdef\left(\begin{array}{cc} 1 & 1 \\ 0 & 1 \end{array}\right)
\;\;\;\;\;\mbox{\rm and}\; \;\;\;\;
M_2\eqdef\left(\begin{array}{cc} 0 & 1 \\ -1 & 0 \end{array}\right)
\end{equation}
Then,
for $\pi_1$ and $\pi_2$ defined as in Proposition~\ref{direct:LPS:rem},
the corresponding primary {\rm(directed edge-colored)} multi-graph
is robustly self-ordered.
\ECR
This follows from the fact that the corresponding
Cayley graph $\CG(\SL_2(p),\{M_1,M_2\})$
is an expander~\cite[Thm.~4.4.2(i)]{Lub}.

\paragraph{Perspective.}
The foregoing construction using the group $\SL_2(p)$
is a special case of a much more general family of constructions,
and the elements of the proof of Proposition~\ref{direct:LPS:rem}
follow an established theory (explained, e.g., in~\cite[Sec.~11.1.2]{HLW}),
which we briefly describe.

Let $H$ be any finite group,
and~$S$ an expanding generating set of $H$
(i.e., the Cayley graph $\CG(H,S)$ is an expander).
Assume that $H$ acts on a finite set $V$
(i.e., each $h\in H$ is associated with a permutation on $V$,
and $h'h(v)=h'(h(v))$ for every $h,h'\in H$  and $v\in V$).
Then, the primary (directed edge-colored) multi-graph~$G_1$ on vertices $V$
can be constructed from the permutations defined by members of~$S$.
The secondary multi-graph $G_2$ is naturally defined by
the action of~$S$ on pairs of elements in $V$.
Finally, the expansion of $\CG(H,S)$ implies that
every connected component of $G_2$ is an expander.%
\footnote{Indeed, this was easy to demonstrate directly
in the case of Proposition~\ref{direct:LPS:rem}.}
Thus, whenever this (Schreier) graph $G_2$ is connected
(as it is in Proposition~\ref{direct:LPS:rem}),
one may conclude that~$G_1$ is a directed edge-colored
robustly self-ordered multi-graph.


\subsection{From the directed variant to the undirected one}
\label{direct-trans:sec}
In this section we show how to transform
directed (edge-colored) multi-graphs,
of the type constructed in Section~\ref{direct-main:sec},
into undirected ones, while preserving all relevant features
(i.e., bounded robustness, bounded degree, regularity, expansion,
and local computability).
The transformation is extremely simple and natural:
We replace the directed edge $(u,v)$ colored $j$
by a 2-path with a designated auxiliary vertex $a_{u,v,j}$,
while coloring the edge $\{u,a_{u,v,j}\}$ by~$2j-1$
and the edge $\{a_{u,v,j},v\}$ by~$2j$.
Evidently, this colored 2-path encodes the direction
of the original edge (as well as the original color).

Note that the foregoing transformation works well
provided that there are no parallel edges that are
colored with the same color, a condition which is satisfied
by the construction presented in Section~\ref{direct-main:sec}.
Furthermore, since the latter construction has no vertices
of (in+out) degree less that $2d\geq4$,
there is no need to mark the original vertices by self-loops.
Hence, a preliminary step
akin to Construction~\ref{pre-colored2standard:ct}
is unnecessary here, although it can be performed in general.

\BP[From directed robust self-ordering to undirected robust self-ordering]
\label{directed2undirected:clm}
For constants $d\geq3$ and $c\in\N$,
let $G=(V,E)$ be a directed multi-graph in which each vertex
has between three and $d$ incident edges {\rm(in both directions)},
and suppose that~$G$ is coupled with
an edge-coloring function $\chi\!:\!E\!\to\![c]$
such that no parallel edges {\em(in the same direction)}
are assigned the same color.
Letting $E_i=\{e\!\in\!E:\chi(e)=i\}$
denote the set of directed edges colored~$i$ in~$G$,
consider the undirected multi-graph $G'=(V',E')$
such that $V'=V\cup\{a_{u,v,i}:(u,v)\!\in\!E_i\}$
and $E'=\bigcup_{j\in[2c]}E'_j$ where
\begin{eqnarray*}
E'_{2i-1} &=& \{\{u,a_{u,v,i}\}:(u,v)\!\in\!E_i\}, \\
E'_{2i} &=& \{\{a_{u,v,i},v\}:(u,v)\!\in\!E_i\},
\end{eqnarray*}
and the edge-coloring function $\chi'\!:\!E'\!\to\![2c]$
that assigns the edges of $E'_j$ the color~$j$
{\rm(i.e., $\chi'(e)=j$ for every $e\!\in\!E'_j$)}.
Then, if $(G,\chi)$ is $\gamma$-robustly self-ordered
{\rm(in the sense of Definition~\ref{directed-robust:def})},
then $(G',\chi')$ is $(\gamma/2d)$-robustly self-ordered
{\rm(in the sense of Definition~\ref{colored-robust:def})},
provided $\gamma\leq1$.
\EP
We comment that the transformation of $(G,\chi)$ to $(G',\chi')$
preserves bounded robustness, bounded degree, regularity, expansion,
and local computability (cf.\ Theorem~\ref{local-computability:thm}).
\medskip

\BPF
The proof is analogous to the proof of Theorem~\ref{colored2std:clm},
but it is much simpler because the gadgets used in the current
transformation (i.e., the auxiliary vertices $a_{u,v,i}$)
are much simpler.

Considering an arbitrary permutation $\mu':V'\to V'$,
we lower-bound the distance between~$G'$ and $\mu'(G')$
as a function of the number of non-fixed-points under $\mu'$.
We do so by considering the contribution
of each non-fixed-point to the distance between~$G'$ and $\mu'(G')$.
We first recall the fact that the vertices of $V$
(resp., the auxiliary vertices) are uniquely identified in $\mu'(G')$
by virtue of the their degree, since each vertex of $V$ has degree
at least three (in $G'$) whereas the auxiliary vertices have degree~2.

\BDes
\item[{\em Case 1}:]
{\em Auxiliary vertices of the form $a_{u,v,i}$ that are not mapped by $\mu'$
to auxiliary vertices of the form~$a_{u',v',i}$; that is,
$\mu'(a_{u,v,i})\in(V\cup\bigcup_{j\neq i}\{a_{u',v',j}:(u',v')\!\in\!E\})$.}

Each such vertex $a_{u,v,i}$ contributes at least one unit
to the difference between~$G'$ and $\mu'(G')$,
since the two edges incident at $a_{u,v,i}$
(in $G'$) are colored~$2i-1$ and $2i$ respectively,
whereas $\mu(a_{u,v,i})$ has either more than two edges (in $G'$)
or its two edges are colored~$2j-1$ and $2j$, respectively,
where for $j\neq i$.
Hence, if the current case contains $n_1$ vertices,
then their contribution to the distance between~$G'$ and $\mu'(G')$
is at least $n_1$.

Ditto for vertices of $V$ that are mapped by $\mu'$ to
an auxiliary vertex.

\item[{\em Case 2}:]
{\em Vertices $v\in V$ such that $\mu'(v)\in V\setminus\{v\}$.}


Intuitively,
the hypothesis that the edge-colored directed~$G$
is robustly self-ordered,
implies that such vertices contribute proportionally
to the difference between the colored versions of
the directed multi-graphs~$G$ and $\mu(G)$,
where $\mu$ is the restriction of $\mu'$ to $V$.
Indeed, we first assume, for simplicity, that $\mu'(V)=V$,
an assumption we shall have to dispose of later.
In this case, the number of tuples $((u,w),i)$
such that $(u,w)$ is colored $i$ in exactly one of these multi-graph
(i.e., either in~$G$ or in $\mu(G)$ but not in both)
is at least $\gamma\cdot|\{v\!\in\!V:\mu(v)\neq v\}|$.
Assume, without loss of generality that $(u,w)\in E_i$
but either $(\imu(u),\imu(w))\not\in E$
(which includes the case that $\imu(u),\imu(w)\in V$ does not hold)
or $(\imu(u),\imu(w))\in E_j$ for $j\neq i$.
We consider two sub-cases:
\BE
\item $\mu'$ maps some $a_{u',w',i}$ to $a_{u,w,i}$,
but either $u$ or $w$ is not connected in $G'$ to this $a_{u',w',i}$
via an edge with the relevant color (i.e., either $2i-1$ or $2i$).
In this sub-case we get a contribution of at least one unit,
since $u$ and $w$ are connected to $a_{u,w,i}$ in $G'$.
\item A vertex not in $\{a_{u',w',i}:(u',w')\!\in\!E_i\}$
is mapped by $\mu'$ to $a_{u,w,i}$,
which means that Case~1 applies for each such a tuple.
\EE
Hence, if the number of vertices in the current case is $n_2$,
then the number of tuples (handled by the two sub-cases)
is at least $\gamma\cdot n_2$,
and we get a contribution of at least $\gamma\cdot n_2$.

The foregoing description is based on the assumption that $\mu(V)=V$.
If this does not hold, then we redefine $\mu$ such that $\mu(v)\not\in V$
is modified such that $\mu(v)=r$ if $r\in V$ has no preimage under $\mu'$.
(Of course, each such $r$ is only used once.)
Indeed, the modified $\mu$ may be ficticiously charged with $d$ edges
per each modification, but each such modification arises due to $v\in V$
that contributes at least one unit in Case~1.
Hence, the amortized over-counting of $d\cdot \gamma$ units
is partially compansated by the unit contributed in Case~1.

\item[{\em Case 3}:]
{\em Auxiliary vertices of the form $a_{u,v,i}$ that are mapped by $\mu'$
to auxiliary vertices of the form~$a_{u',v',i}$ for $(u'v')\neq(u,v)$;
that is,
$\mu'(a_{u,v,i})\in\{a_{u',v',i}:(u',v')\!\in\!E_i\setminus\{(u,v)\}\}$.}

For $u,v,u',v'$ and $i$ as above, if $\mu'(u)=u'$ and $\mu'(v)=v'$,
then an auxiliary vertex that connects $u$ and~$v$ in $G'$
is mapped to an auxiliary vertex that does not connects them in~$\mu'(G')$
(but rather connects the vertices $u'$ and $v'$,
whereas either $u'\neq u$ or $v'\neq v$).
So we get a contribution of at least one unit to the
difference between~$G'$ and $\mu'(G')$
(i.e., the edge incident at either $u$ or $v$).
Hence, the contribution is proportional to the
number of non-fixed-points of the current type.
Otherwise (i.e., $(\mu'(u),\mu'(v))\neq(u',v')$),
we get a vertex as in either Case~1 or Case~2,
and get a proportional contribution again.
\EDes
Hence, the contribution of each of these cases
to the difference between~$G'$ and $\mu'(G')$
is proportional to the number of vertices involved.
Specifically, if there are $n_i$ vertices in Case~$i$,
then we get a contribution-count of
at least $\gamma\cdot\sum_{i\in[3]}n_1$,
where some of these contributions were possibly counted twice.
The claim follows.
\EPF

\section{The Three-Step Construction}
\label{three-step:sec}
In this section we present a different construction
of bounded-degree graphs that are robustly self-ordered.
It uses totally different techniques than the ones utilized
in the construction presented in Section~\ref{direct:sec}.
Furthermore, the current construction offers the flexibility
of obtaining either graphs that have small connected components
(i.e., of logarithmic size)
or graphs that are highly connected (i.e., are expanders).
Actually, one can obtain anything in-between
(i.e., $n$-vertex graphs that consist of $s(n)$-sized
connected components that are each an expander,
for any $s(n)=\Omega((\log n)/\log\log n)$).
We mention that robustly self-ordered bounded-degree graphs
with small connected components are used in the proof
of Theorem~\ref{pt-lb:thm}.

As stated in Section~\ref{techniques:sec},
the current construction proceeds in three steps.
First, in Section~\ref{step1:sec}, we prove the existence
of robustly self-ordered bounded-degree graphs,
and observe that such $\ell$-vertex graphs can actually
be found in $\poly(\ell!)$-time (i.e., $\exp(\tildeO(\ell))$-time).
Next, setting $\ell=\Omega((\log n)/\log\log n)$,
we use these graphs as part of $2\ell$-vertex
connected components in an $n$-vertex
(robustly self-ordered bounded-degree)
graph that is constructed in $\poly(n)$-time
(see Section~\ref{step2:sec}).
Lastly, in Section~\ref{step3:sec},
we repeat this strategy using the graphs
constructed in Section~\ref{step2:sec},
and obtain exponentially larger graphs
that are locally constructible.

In addition, in Section~\ref{local-so:sec},
we show that the foregoing graphs can be locally self-ordered.
That is, given a vertex $v$ in any graph $G'=(V',E')$
that is isomorphic to the foregoing $n$-vertex graph
and oracle access to the incidence function of $G'$,
we can find in $\poly(\log n))$-time the vertex
to which this unique isomorphism maps $v$.


\subsection{Existence}
\label{step1:sec}
As stated above, we start with establishing the mere existence
of bounded-degree graphs that are robustly self-ordered.

\BT[Robustly self-ordered graphs exist]
\label{existence:thm}
For any sufficiently large constant~$d$,
there exists a family $\{G_n\}_{n\in\N}$
of robustly self-ordered $d$-regular graphs.
Furthermore, these graphs are expanders.
\ET
Actually, it turns out that random $d$-regular graphs
are robustly self-ordered; see Theorem~\ref{random-works:thm}.
Either way, given the existence of such $n$-vertex graphs,
they can actually be found in $\poly(n!)$-time,
by an exhaustive search. Specifically,
for each of the possible $o((dn)!)$ graphs,
we check the robust self-ordering condition
by checking all $n!-1$ relevant permutation.
(The expansion condition can be checked similarly,
by trying all $(0.5+o(1))\cdot2^n$ relevant subsets of~$[n]$.)

The proof of Theorem~\ref{existence:thm} utilizes
a simpler probabilistic argument than the one used
in the proof of Theorem~\ref{random-works:thm}.
This argument (captured by Claim~\ref{random-colored:clm})
refers to the auxiliary model of edge-colored multi-graphs
(see Definition~\ref{colored-robust:def})
and is combined with a transformation of
this model to the original model of uncolored graphs
(provided in Construction~\ref{colored2standard:ct}
and analyzed in Theorem~\ref{colored2std:clm}).
Indeed, the relative simplicity of Claim~\ref{random-colored:clm}
is mainly due to using the edge-colored model
(see digest at the end of Section~\ref{random-graphs:sec}).
\medskip

\BPF
To facilitate the proof, we present the construction
while referring to the edge-colored model presented in
Section~\ref{edge-colored:sec}.
We shall then apply Theorem~\ref{colored2std:clm}
and obtain a result for the original model
(of uncolored simple graphs).

For $m=n/O(1)$,
we shall consider $2m$-vertex multi-graphs
that consists of two $m$-vertex cycles,
using a different color for the edges of each cycle,
that are connected by $d'=O(1)$ random perfect matching,
which are also each assigned a different color.
(Hence, we use $2+d'$ colors in total.)
We shall show that (w.h.p.) a random multi-graph constructed in this way
is robustly self-ordered (in the colored sense).
(Note that parallel edges, if they exist,
will be assigned different colors.)
Specifically,
we consider a generic $2m$-vertex multi-graph that is determined
by $d'$ perfect matchings of $[m]$ with $\{m+1,\ldots,2m\}$.
Denoting this sequence of perfect matchings by ${\ov M}=(M_1,\ldots,M_{d'})$,
we consider the (edge-colored) multi-graph $G_{\ov M}([2m],E_{\ov M})$
given by
\begin{eqnarray*}
E_{\ov M} &=& C_1\cup C_2 \cup\bigcup_{j\in[d']} M_j \\
& &\mbox{\rm where $C_1=\{\{i,i+1\}:i\in[m-1]\}\;\cup\;\{\{m,1\}\}$} \\
& &\mbox{\rm and $C_2=\{\{m+i,m+i+1\}:i\in[m-1]\}\;\cup\;\{\{2m,m+1\}\}$}
\end{eqnarray*}
and a coloring $\chi$ in which the edges of $C_j$ are colored $j$
and the edges of $M_j$ are colored $j+2$.
(That is, for $i\in\{1,2\}$, the set $C_i$ forms a cycle
of the form $((i-1)m+1,(i-1)m+2,\ldots,\allowbreak(i-1)m+m,(i-1)m+1)$
and its edges are colored~$i$.)
Note that the $d'+1$ edges incident at each vertex are assigned $d'+1$
different colors.

\Bcm{\em(W.h.p., $G_{\ov M}$ is robustly self-ordered):}
\label{random-colored:clm}
For some constant $\gamma>0$,
with high probability over the choice of~$\ov M$,
the edge-colored multi-graph $G_{\ov M}$ is $\gamma$-robustly self-ordered.
Furthermore, it is also an expander.
\Ecm

\Bpf
Consider an arbitrary permutation $\mu:[2m]\to[2m]$,
and let $t=|\{i\!\in\![2m]\!:\!\mu(i)\neq i\}|$.
We shall show that, with probability $1-\exp(-\Omega(dt\log m))$
over the choice of $\ov M$, the difference between the colored versions
of $G_{\ov M}$ and $\mu(G_{\ov M})$ is $\Omega(t)$.
Towards this end, we consider two cases.
\BDes
\item[{\em Case 1}:] $|\{i\in[m]:\mu(i)\not\in[m]\}|>t/4$.
Equivalently, $|\{i\in[2m]:\ceil{\mu(i)/m}\neq\ceil{i/m}\}|>t/2$.

The vertices in the set $\{i\in[m]:\mu(i)\not\in[m]\}$
are mapped from the first cycle to the second cycle,
and so rather than having two incident edges
that are colored~1 they have two incident edges colored~2.
Hence, each such vertex contributes two units to the difference
(between the colored versions of $G_{\ov M}$ and $\mu(G_{\ov M})$),
and the total contribution is greater than $2\cdot(t/4)\cdot 2$,
where the first factor of~2 accounts also for vertices
that are mapped from $C_2$ to $C_1$.

\item[{\em Case 2}:] $|\{i\in[m]:\mu(i)\not\in[m]\}|\leq t/4$.
Equivalently, $|\{i\in[2m]:\ceil{\mu(i)/m}\neq\ceil{i/m}\}|\leq t/2$.

We focus on the non-fixed-points of $\mu$ that stay on their
original cycle (i.e., those not considered in Case~1).
Let $A\eqdef\{i\!\in\![m]:\mu(i)\!\neq\!i\wedge\mu(i)\!\in\![m]\}$
and
$B\eqdef\{i\!\in\!\{m+1,....,2m\}:
   \mu(i)\!\neq\!i\wedge\mu(i)\!\in\!\{m+1,\ldots,2m\}\}$.
By the case hypothesis, $|A|+|B|\geq t/2$,
and we may assume (without loss of generality) that $|A|\geq t/4$.
As a warm-up, we first show that
{\em each element of~$A$ contributes
a non-zero number of units to the difference}\/
(between the colored versions of $G_{\ov M}$ and $\mu(G_{\ov M})$)
{\em with probability $1-O(1/m)^{d'}$}, over the choice of $\ov M$.

To see this, let $\pi_j:[m]\to\{m+1,\ldots,2m\}$
be the mapping used in the $j^\xth$ matching;
that is, $M_j=\{\{i,\pi_j(i)\}:i\!\in\![m]\}$,
which means that $\pi_j(i)$ is the $j^\xth$ match of $i$ in $G_{\ov M}$
(i.e., the vertex matched to $i$ by $M_j$).
Then, we consider the event that {\em for some $j\in[d']$,
the $j^\xth$ match of $i\in[m]$ in $\mu(G_{\ov M})$
is different from the $j^\xth$ match of $i$ in $G_{\ov M}$},
and note that when this event occurs $i$ contributes to the difference
(between the colored versions of $G_{\ov M}$ and $\mu(G_{\ov M})$).
Note that $x$ is the $j^\xth$ match of $i$ in $\mu(G_{\ov M})$
if and only if $\imu(x)$ is the $j^\xth$ match of $\imu(i)$ in $G_{\ov M}$,
which holds if and only if $\imu(x)=\pi_j(\imu(i))$
(equiv., $x=\mu(\pi_j(\imu(i)))$).
Hence, $i\in[m]$ contributes to the difference if and only if
for some $j$ it holds that $\pi_j(i)\neq\mu(\pi_j(\imu(i)))$,
because $\pi_j(i)\neq\mu(\pi_j(\imu(i)))$ means that
the edge $\{i,\pi_j(i)\}$ is colored~$j+2$ in $G_{\ov M}$
but is not colored~$j+2$ in $\mu(G_{\ov M})$
(since a different edge incident at~$i$ in $\mu(G_{\ov M})$
is colored~$j+2$).
Letting ${\ov\pi}=(\pi_1,\ldots,\pi_{d'})$,
the probability of the complementary event
(i.e., $i$ does not contribute to the difference)
is given by
\begin{eqnarray*}
\prob_{\ov\pi}
  \left[(\forall j\!\in\![d'])\;\;\pi_j(i)=\mu(\pi_j(\imu(i)))\right]
&=& \prod_{j\in[d']}
    \prob_{\pi_j}\left[\pi_j(i)=\mu(\pi_j(\imu(i)))\right] \\
&\leq& (m-1)^{-d'},
\end{eqnarray*}
where the inequality uses the hypothesis
that $\mu(i)\neq i$ and $i,\mu(i)\!\in\![m]$;
specifically, fixing the value of $\pi_j(\imu(i))$,
leaves $\pi_j(i)$ uniformly distributed
in $S\eqdef\{m+1,\ldots,2m\}\setminus\{\pi_j(\imu(i))\}$,
which means that
$\prob_{\pi_j}[\pi_j(i)\!=\!\mu(v)\,|\,v\!=\!\pi_j(\imu(i))]\leq1/|S|$
(where equality holds if $\mu(v)\in S$).

The same argument generalises to any set $I\subseteq A$
such that $I\cap\mu(I)=\emptyset$.
In such a case, letting $I=\{i_1,\ldots,i_{t'}\}$, we get
\begin{eqnarray*}
\lefteqn{\prob_{\ov\pi}\left[(\forall i\!\in\!I)(\forall j\!\in\![d'])\;\;
                     \pi_j(i)=\mu(\pi_j(\imu(i)))\right]} \\
&=& \prod_{k\in[t']}\prod_{j\in[d']}
    \prob_{\pi_j}\left[\pi_j(i_k)=\mu(\pi_j(\imu(i_k)))
          \left|(\forall k'\!\in\![k-1])\;
                   \pi_j(i_{k'})=\mu(\pi_j(\imu(i_{k'})))\right.\right] \\
&\leq& (m-2t'+1)^{-t'd'},
\end{eqnarray*}
where the inequality uses the hypothesis that $I\cap\mu(I)=\emptyset$;
specifically, for each $k\in[t']$, we use the fact
that $i_k\not\in\{i_1,\ldots,i_{k-1},\imu(i_1),\ldots,\imu(i_{k})\}$.
Hence, fixing the values of $\pi_j(i_{k'})$ for all $k'\in[k-1]$
and the values of $\pi_j(\imu(i_{k'}))$ for all $k'\in[k]$,
and denoting these values
by $u_1,\ldots,u_{k-1}$ and $v_1,\ldots,v_{k}$ respectively,
leaves $\pi_j(i_k)$ uniformly distributed
in $S\eqdef\{m+1,\ldots,2m\}\setminus\{u_1,\ldots,u_{k-1},v_1,\ldots,v_k\}$,
which means that
$\prob_{\pi_j}[\pi_j(i)\!=\!\mu(v_k)\,|\,\mbox{\rm foregoing fixing}]\leq1/|S|$
(where equality holds if $\mu(v_k)\in S$).

Recalling that $|A|\geq t/4$ and $t\leq2m$,
we upper-bound the probability (over the choice of $\ov M$)
that $A$ contains a $t/8$-subset $A'$ such that
$(\forall i\!\in\!A')(\forall j\!\in\![d'])\;\pi_j(i)=\mu(\pi_j(\imu(i)))$,
by taking a union bound over all possible $A'$
and using for each such $A'$ a subset $I\subset A'$
such that $I\cap\mu(I)=\emptyset$.
(So we actually take a union bound over the $I$'s
and derive a conclusion regarding the $t/8$-subsets $A'$.)
Observing that $|I|\geq|A'|/3\geq t/24$,
we conclude that, with probability at most
$\binom{t}{t/24}\cdot(m/2)^{d'\cdot t/24}=\exp(-\Omega(d' t\log m))$
over the choice of~$\ov M$,
the set $A$ contains no $t/8$-subset $A'$ as above.
This means that,
with probability at most $\exp(-\Omega(d' t\log m))$,
less than $t/8$ of the indices $i\in A$ contribute
a non-zero number of units to the difference
(between the colored versions of $G_{\ov M}$ and $\mu(G_{\ov M})$).
\EDes
Hence, we have shown that, for every permutations $\mu:[2m]\to[2m]$,
the probability (over the choice of $\ov M$) that the size
of the symmetric difference between
the colored versions of $G_{\ov M}$ and $\mu(G_{\ov M})$
is smaller than $t/8$ is $\exp(-\Omega(d' t\log m))$,
where $t$ is the number of non-fixed-points of $\mu$.
Letting $\gamma=1/8$ and taking a union bound over
all (non-trivial) permutations $\mu:[2m]\to[2m]$,
we conclude that the probability, over the choice of $\ov M$,
that $G_{\ov M}$ is not $\gamma$-robustly self-ordered is at most
\begin{eqnarray*}
\sum_{t\in[2m]}\binom{2m}{t}\cdot\exp(-\Omega(d' t\log m))
&=& \sum_{t\in[2m]}\exp(-\Omega((d'-O(1))\cdot t\log m)) \\
&=& \exp(-\Omega((d'-O(1))\cdot \log m)),
\end{eqnarray*}
and the claim follows (for any sufficiently large $d'$),
while observing that, with very high probability,
these multi-graphs are expanders.
\Epf

\mypar{Back to the non-colored version.}
We now convert the edge-colored multi-graphs $G=G_{\ov M}$
that are $\gamma$-robustly self-ordered into standard graphs $G'$
that are robustly self-ordered in the original sense.
This is done by using Construction~\ref{colored2standard:ct}
(while relying on Theorem~\ref{colored2std:clm}).
Recall that this transformation also preserves expansion.
Actually, before invoking Construction~\ref{colored2standard:ct},
we augment the multi-graph~$G$ by adding a self-loop to each vertex,
and color all these self-loops using a special color.
Combining Claim~\ref{random-colored:clm}
and Theorem~\ref{colored2std:clm}, the current theorem follows.
\EPF

\subsection{Constructions}
\label{step2:sec}
Having established the existence of bounded-degree graphs that
are robustly self-ordered, we now turn to actually construct them.
We shall use the fact that the proof of existence yields
a construction that runs in time that is polynomial in the
number of possible graphs.
Specifically, for $\ell=\frac{O(\log n)}{\log\log n}$,
we shall construct $\ell$-vertex graphs in $\poly(\ell^\ell)$-time
and use them in our construction of $n$-vertex graphs,
while noting that $\poly(\ell^\ell)=\poly(n)$.

\BT[Constructing robustly self-ordered graphs]
\label{construction:thm}  \label{CONSTRUCTION:THM}   
For any sufficiently large constant~$d$,
there exists an efficiently constructable family $\{G_n\}_{n\in\N}$
of robustly self-ordered graphs of maximum degree $d$.
That is, there exists a polynomial-time algorithm
that on input $1^n$ outputs the $n$-vertex graph $G_n=([n],E_n)$.
Furthermore, $G_n$ consists of connected components
of size $\frac{O(\log n)}{\log\log n}=o(\log n)$.
\ET
Note that the connected components of $G_n$ cannot be any smaller
(than $\frac{O(\log n)}{\log\log n}$).
This is the case because an asymmetric $n$-vertex bounded-degree graph,
let alone a robustly self-ordered one, cannot have connected components
of size $\frac{o(\log n)}{\log\log n}$ (because the number of $t$-vertex
graphs of bounded-degree is $t^{O(t)}$).
\medskip

\BPF
The proof proceeds in two steps.
We first use the existence of $\ell$-vertex ($d'$-regular)
expander graphs that are robustly self-ordered
towards constructing a sequence of $m=\exp(\Omega(\ell\log\ell))$
bounded-degree $2\ell$-vertex graphs that are robustly self-ordered,
expanding, and far from being isomorphic to one another.
We construct this sequence of $2\ell$-vertex graphs in $\poly(m)$-time,
using the fact that $(\ell!)^{O(1)}=\poly(m)$.
In the second step, we show that the $(m\cdot2\ell)$-vertex graph
that consists of these $2\ell$-vertex graphs (as its connected components)
is robustly self-ordered.
Note that this graph is constructed in time that is
polynomial in its size, since its size is $\Omega(m)$,
whereas it is constructed in $\poly(m)$-time.%
\footnote{We mention that a slightly different construction can be
based on the fact that random $\ell$-vertex ($d'$-regular) graphs
are robustly self-ordered expanders (see Theorem~\ref{random-works:thm}).
In this alternative construction we find a sequence of $m$ such
graphs that are pairwise far from being isomorphic to one another.
As further detailed in Remark~\ref{random-works-alt:rem},
the analysis of the alternative construction is somewhat easier
than the analysis of the construction presented below,
but we need the current construction for the proof
of Theorem~\ref{strong-construction:thm}.}

Given a generic $n$, let $\ell=\frac{O(\log n)}{\log\log n}$,
which implies that $\ell^\ell=\poly(n)$.
By Theorem~\ref{existence:thm}, for all sufficiently large $d'$,
there exist $\ell$-vertex $d'$-regular expander graphs
that are robustly self-ordered
(with respect to the robustness parameter $c'$).
Furthermore, we can find such a graph, denoted $G'_\ell$,
in time $\poly(\ell^\ell)=\poly(n)$,
by scanning all $\ell$-vertex $d'$-regular graphs and checking
both the expansion and the robustness (w.r.t parameter $c'$)
conditions for each of them.
Actually, for $d''=d'+1$, we shall also find
an $\ell$-vertex $d''$-regular expander, denoted $G''_\ell$,
that is robustly self-ordered.

\mypar{The construction of $G_n$.}
Using $G'_\ell$ and $G''_\ell$, we construct an $n$-vertex
robustly self-ordered graph, denoted $G_n$,
that consists of $n/2\ell$ connected components that
are pairwise far from being isomorphic to one another.
This is done by picking $m=n/2\ell$ permutations,
denoted $\pi_1,\ldots,\pi_m:[\ell]\to[\ell]$,
that are pairwise far-apart
and constructing $2\ell$-vertex graphs
such that the $i^\xth$ such graph consist of
a copy of $G'_\ell$ and a copy of $G''_\ell$ that are connected
by a matching as determined by the permutation $\pi_i$.
Specifically,
for $G'_\ell=([\ell],E'_\ell)$ and $G''_\ell=([\ell],E''_\ell)$,
the $i^\xth$ connected component is isomorphic
to a graph with the vertex set $[2\ell]$
and the edge set
\begin{equation}\label{edges-of-cc:eq1}
E'_\ell \;\cup\; \{\{\ell+u,\ell+v\}:\{u,v\}\in E''_\ell\}
         \;\cup\; \{\{v,\ell+\pi_i(v)\}:v\!\in\![\ell]\}.
\end{equation}
(The first two sets correspond to the copies of~$G'_\ell$ and $G''_\ell$,
and the third set corresponds to the matching between these copies.
Note that the vertices in $[\ell]$ have degree $d'+1$,
whereas vertices in $\{\ell+1,\ldots,2\ell\}$ have degree $d''+1\neq d'+1$.)

To see that this construction can be carried out in $\poly(n)$-time,
we need to show that the sequence of $m$ pairwise far-apart permutations
can be determined in $\poly(n)$-time, let alone that such a sequence exists.
This is the case, because we can pick the permutation sequentially
(one after the other) by scanning the symmetric group on $[\ell]$
and relying on the fact that for ($i<n$ and) any fixed sequence
of permutations $\pi_1,\ldots,\pi_{i-1}:[\ell]\to[\ell]$
it holds that a random permutation $\pi_i$ is far-apart from
each of the fixed $i-1$ permutations; that is,
$\prob_{\pi_i}[|\{v\in[\ell]:\pi_i(v)\neq\pi_j(v)\}|=\Omega(\ell)]=1-o(1/n)$
for every $j\in[i-1]$.%
\footnote{Specifically, for some $\ell'=\Omega(\ell)$,
we upper-bound $\prob_{\pi}[|\{v\in[\ell]:\pi(v)=v)\}|\geq\ell-\ell']$,
where $\pi:[\ell]\to[\ell]$ is a random permutation.
We do so by observing that the number of permutations
that have at least $\ell-\ell'$ fixed-points is at most
$\binom{\ell}{\ell'}\cdot(\ell'!)=\frac{\ell!}{(\ell-\ell')!}$,
whereas $(\ell-\ell')!=\exp(\Omega(\ell\log\ell))=\omega(n)$
for any $\ell'$ such that $\ell-\ell'=\Omega(\ell)$.}


\mypar{Towards proving that $G_n$ is robustly self-ordered.}
We now prove that the resulting graph~$G_n$,
which consists of these $m$ connected components,
is $c$-robustly self-ordered, where $c$ is a universal constant
(which is independent of the generic $n$).
For starters, let's verify that $G_n$ is self-ordered.
We first note that any automorphism of $G_n$ must
map the vertices of copies of $G'_\ell$ (resp., $G''_\ell$)
to vertices of copies of $G'_\ell$ (resp., $G''_\ell$),
since these are the only vertices of degree~$d'+1$ (resp., $d''+1$).
The connectivity of these copies implies that
the automorphism must map each connected component
to some connected component,
which  determines the $m$ connected components.
The self-ordered feature of $G'_\ell$ and $G''_\ell$
determines a unique ordering on each copy,
whereas the fact the permutations (i.e., $\pi_i$'s)
are different imposes that each connected
component is mapped to itself
(i.e., the order of the connected components is preserved).
Hence, the automorphism must be trivial
(and it follows that $G_n$ is self-ordered).


An analogous argument establishes the robust self-ordering of $G_n$,
where we use the hypothesis that $G'_\ell$ and $G''_\ell$ are expanders
(rather than merely connected),
the choice of the $\pi_i$'s as being far-apart
(rather than merely different),
and the robust self-ordering of $G'_\ell$ and $G''_\ell$
(rather than their mere self-ordering).
Considering an arbitrary permutation $\mu:[n]\to[n]$,
these stronger features are used to establish a lower bound on the size
of the symmetric difference between $G_n$ and $\mu(G_n)$ as follows:
\BI
\item
The fact that $G'_\ell$ is an expander implies that if $\mu$ splits
the vertices of a copy of $G'_\ell$ such that $\ell'$ vertices are mapped
to copies that are different than the other $\ell-\ell'\geq\ell'$ vertices,
then this contributes $\Omega(\ell')$ units
to the difference between $G_n$ and $\mu(G_n)$.
Ditto for $G''_\ell$, whereas mapping a copy of $G'_\ell$
to a copy of $G''_\ell$ contributes $\Omega(\ell)$ units
(per the difference in the degrees).
\item
The robust self-ordering of $G'_\ell$ and $G''_\ell$ implies that
if $\mu$ changes the index of vertices inside a component,
then this yields a proportional difference between $G_n$ and $\mu(G_n)$.
\item
The distance between the $\pi_i$'s
(along with the aforementioned robustness)
implies that if~$\mu$ changes the indices of the connected components,
then each such change contributes $\Omega(\ell)$ units
to the difference between $G_n$ and $\mu(G_n)$.
\EI
The actual implementation of this sketch requires
a careful accounting of the various contributions.
As a first step in this direction we provide
a more explicit description of $G_n$.
We denote the set of vertices
of the copy of $G'_\ell$ (resp., $G''_\ell$)
in the $i^\xth$ connected component of $G_n$
by $F_i=\{2(i-1)\ell+j:j\in[\ell]\}$
(resp., $S_i=\{2(i-1)\ell+\ell+j:j\in[\ell]\}$).
Recall that $F_i$ and $S_i$ are connected by the edge-set
\begin{equation}\label{edges-of-cc:eq2}
\left\{\{2(i-1)\ell+j,2(i-1)\ell+\ell+\pi_i(j)\}:j\!\in\![\ell]\right\}
\end{equation}
whereas the subgraph of $G_n$ induced by $F_i$ (resp., $S_i$)
has the edge-set $\{\{2(i-1)\ell+u,2(i-1)+v\}:\{u,v\}\!\in\!E'_\ell\}$
(resp., $\{\{2(i-1)\ell+\ell+u,2(i-1)+\ell+v\}:\{u,v\}\!\in\!E''_\ell\}$).
In addition, let $F=\bigcup_{i\in[m]}F_i$
(resp., $S=\bigcup_{i\in[m]}S_i$).

\mypar{The actual proof (that $G_n$ is robustly self-ordered).}
Considering an arbitrary permutation $\mu:[n]\to[n]$,
we lower-bound the distance
(i.e., size of the symmetric difference)
between $G_n$ and $\mu(G_n)$
as a function of the number of non-fixed-points under $\mu$
(i.e., the number of $v\in[n]$ such that $\mu(v)\neq v$).
We do so by considering the (average) contribution
of every non-fixed-point to the distance between $G_n$ and $\mu(G_n)$
(i.e., number of pairs of vertices that form an edge in one graph
but not in the other). We may include the same contribution in
few of the following (seven) cases, but this only means that
we are double-counting the contribution by a constant factor.
We first consider the set of vertices that switch sides
(under $\mu$) in the connected components.

\BDes
\item[{\em Case 1}:] {\em Vertices $v\in F$ such that $\imu(v)\in S$}.
Ditto for $v\in S$ such that $\imu(v)\in F$.

Each such vertex contributes at least one unit to the distance
(between $G_n$ and $\mu(G)$)
by virtue of $v$ having degree~$d'+1$ in $G_n$
and strictly higher degree in $\mu(G_n)$,
since vertices in $F$ have degree~$d'+1$ (in $G_n$)
whereas vertices in~$S$ have higher degree (in $G_n$).%
\footnote{Note that $v$ neighbors $u$ in $\mu(G_n)$
if and only if $\imu(v)$ neighbors $\imu(u)$ in $G_n$.}
\EDes
In light of Case~1, we may focus on vertices whose ``type''
(or ``side'') is preserved by $\imu$. Actually, it will be more convenient
to consider the set of vertices whose ``type'' is preserved by $\mu$;
that is, the set
$\{v\!\in\!F\!:\!\mu(v)\!\in\!F\}\cup\{v\!\in\!S\!:\!\mu(v)\!\in\!S\}$.
Next, for each $i\in[m]$, we define $\si'(i)$ to be the index
of the connected component that takes the plurality of $\mu(F_i)$;
that is, $\si'(i)\eqdef j$
if $|\{v\in F_i:\mu(v)\in F_j\}|\geq|\{v\in F_i:\mu(v)\in F_k\}|$
for all $k\in[m]$ (breaking ties arbitrarily).
Next, we consider the set of vertices that are not mapped
according to the plurality vote.

\BDes
\item[{\em Case 2}:]
{\em Vertices $v\in F_i$ such that $\mu(v)\in F\setminus F_{\si'(i)}$}.

For starters, suppose that
$|\{v\!\in\!F_i\!:\!\mu(v)\!\in\!F_{\si'(i)}\}|\geq\ell/2$;
that is, a majority of the vertices of $F_i$ are mapped by~$\mu$
to $F_{\si'(i)}$. In this case, by the expansion of $G'_\ell$,
we get a contribution that is proportional to the size
of the set $F'_i\eqdef\{v\!\in\!F_i\!:\!\mu(v)\!\not\in\!F_{\si'(i)}\}$,
because (in $G_n$) there are $\Omega(|F'_i|)$ edges
between $F'_i$ and the rest of $F_i$
but there are no edges between $F'_i$ and $F_i\setminus F'_i$ in $\mu(G_n)$.
In the general case, we have to be more careful since expansion
is guaranteed only for sets that have size at most $\ell/2$.
In such a case we use an adequate subset of $F'_i$.
Details follow.

Let $R_i\eqdef\{v\!\in\!F_i\!:\!\mu(v)\!\in\!F\setminus F_{\si'(i)}\}$.
We first note that there exists a set $J\subseteq[m]\setminus\{\si'(i)\}$
such that $F'_i\eqdef\bigcup_{j\in J}\{v\!\in\!F_i\!:\!\mu(v)\!\in\!F_j\}$
satisfies $|R_i|/4\leq|F'_i|\leq\ell/2$.
Recall that the subgraph of $G_n$ induced by $F_i$ is an expander,
and consider the edges in $G_n$ that cross the cut
between $F'_i$ and the rest of $F_i$. 
Then, this cut has $\Omega(|F'_i|)$ edges in $G_n$,
but there are no edges between $F'_i$ and $F_i\setminus F'_i$ in $\mu(G_n)$,
because $\imu(F'_i)\subseteq\bigcup_{j\in J}F_j$
and $\imu(F_i\setminus F'_i)\subseteq\bigcup_{j\in[m]\setminus J}F_j$
are not connected in $G_n$.
Hence, the total contribution of the vertices
in $\{v\!\in\!F_i\!:\!\mu(v)\!\in\!F\setminus F_{\si'(i)}\}=R_i$
to the distance (between $G_n$ and $\mu(G)$) is $\Omega(|F'_i|)$,
which is proportional to their number
(i.e., it is $\Omega(|R_i|)$).
\EDes
Defining $\si''(i)$ in an analogous manner with respect to $\mu(S_i)$,
we get an analogous contribution by the expander induced by $S_i$.
Specifically, for each $i\in[m]$, we define $\si''(i)$ to be the index
of the connected component that takes the plurality of $\mu(S_i)$;
that is, $\si''(i)\eqdef j$ if
$|\{v\!\in\!S_i\!:\!\mu(v)\!\in\!S_j\}|
  \geq|\{v\!\in\!S_i\!:\!\mu(v)\!\in\!S_k\}|$
for all $k\in[m]$ (breaking ties arbitrarily).
Analogously to Case~2, we consider the following vertices.

\BDes
\item[{\em Case 3}:]
{\em Vertices $v\in S_i$ such that $\mu(v)\in S\setminus S_{\si''(i)}$}.

Here we get a contribution of
$\Omega(|\{v\!\in\!S_i\!:\!\mu(v)\!\in\!S\setminus S_{\si''(i)}\}|)$,
where the analysis is analogous to Case~2.
\EDes
Recall that if $v\in F_i$ then it holds that $v=2(i-1)\ell+j$
for some $j\in[\ell]$, and that (in $G_n$) vertex $v$ has
a unique neighbor in~$S$, which is $2(i-1)\ell+\ell+\pi_i(j)\in S_i$.
It will be convinient to denote this neighbor by $\phi_i(v)$;
that is, for $v\in F_i$ such that $v=2(i-1)\ell+j$,
we have $\phi_i(v)=2(i-1)\ell+\ell+\pi_i(j)\in S_i$.
The next two cases refer to vertices that are mapped by $\mu$
according to the plurality vote
(e.g., $v\in F_i$ is mapped to $\mu(v)\in F_{\si'(i)}$),
but their match is not mapped accordingly
(i.e., $\phi_i(v)\in S_i$ is not mapped to $S_{\si'(i)}$).

\BDes
\item[{\em Case 4}:]
{\em Vertices $v\in F_i$ such that $\mu(v)\in F_{\si'(i)}$
but $\mu(\phi_i(v))\not\in S_{\si'(i)}$}.

(Note that the condition $v\in F_i$ and $\mu(v)\in F_{\pi'(i)}$
means that vertex $v$ is not covered in Case~2.
If $\si''(i)=\si'(i)$, then $\mu(\phi_i(v))\not\in S_{\si'(i)}$
means that $v$ is covered in Case~3, since $\phi_i(v)\in S_{i}$.
Hence, the current case is of interest only when $\si''(i)\neq\si'(i)$.
In particular, it is of interest when referring to vertices
in the $i^\xth$ connected component of $G_n$ that reside
in the copies of~$G'_\ell$ and $G''_\ell$ and are mapped
according to the plurality votes of these copies,
whereas these two plurality votes are inconsistent.)

We focus on the case that a vast majority of the vertices in
both $F_i$ and $S_i$ are mapped according to the plurality votes
(i.e., $\si'(i)$ and $\si''(i)$), since the complementary cases
are covered by Cases~2 and~3, respectively.
Specifically, if
either $|\{v\!\in\!F_i\!:\!\mu(v)\!\in\![n]\setminus F_{\si'(i)}\}|>\ell/3$
or $|\{u\!\in\!S_i\!:\!\mu(u)\!\in\![n]\setminus S_{\si''(i)}\}|>\ell/3$,
then we get a contribution of $\Omega(\ell)$
either by Cases~1\&2 or by Cases~1\&3.
Otherwise, it follows that
$${|\{v\!\in\!F_i\!:\!\mu(v)\!\in\!F_{\si'(i)}
    \;\wedge\; \mu(\phi_i(v))\!\in\!S_{\si''(i)}\}|}
    \geq \ell -2\cdot\ell/3$$
which implies that, if $\si'(i)\neq\si''(i)$,
then the $i^\xth$ connected component of $G_n$
contributes $\ell/3$ units to the difference
(between $G_n$ and $\mu(G_n)$),
since $v$ and $\phi_i(v)$ are connected in $G_n$,
but $\mu(v)\in F_{\si'(i)}$ and $\mu(\phi_i(v))\in S_{\si''(i)}$
reside in different connected components of $\mu(G_n)$.
(That is, the contribution is due to vertices $v$ of $F_i$
that are mapped by $\mu$ to $F_{\si'(i)}$,
while the corresponding vertices $\phi_i(v)$ of $S_i$
(which are connected to them in $G_n$) are mapped by $\mu$
to $S_{\si''(i)}\subset S\setminus S_{\si'(i)}$,
whereas $F_{\si'(i)}$ and $S_{\si''(i)}$ are not connected in $G_n$,
assuming $\si'(i)\neq\si''(i)$.)

To conclude: The contribution of the vertices of Case~4
(to the difference between $G_n$ and $\mu(G_n)$)
is proportional to the number of these vertices
(where this contribution {\em might}\/ have been counted
already in Cases~1,~2 and~3).

\item[{\em Case 5}:]
{\em Vertices $v\in F_i$ such that $\mu(v)\not\in F_{\si''(i)}$
but $\mu(\phi_i(v))\in S_{\si''(i)}$}.

(Equiv., vertices $v\in S_i$ such that $\mu(v)\in S_{\si''(i)}$
but $\mu(\phi_i^{-1}(v))\not\in F_{\si''(i)}$.)

Analogously to Case~4, the contribution of these vertices
is proportional to their number.
(Analogously, this augments Case~2 only in case $\si''(i)\neq\si'(i)$.)
\EDes
In light of Cases~2--5, we may focus on indices $i\in[m]$
such that $\si'(i)=\si''(i)$ and on vertices in the $i^\xth$
connected component that are mapped by $\mu$ to
the $\si'(i)^\xth$ connected component
(and the same ''type'' per Case~1).
The following case refers to such vertices that do not
maintain their position in this connected component.

\BDes
\item[{\em Case 6}:]
{\em Vertices $v\!=\!2(i-1)\ell+j\in F_i$
such that $\mu(v)\in F_{\si'(i)}\setminus\{2(\si'(i)-1)\ell+j\}$.}

Ditto for $v\!=\!2(i-1)\ell+\ell+j\in S_i$
such that $\mu(v)\in S_{\si''(i)}\setminus\{2(\si''(i)-1)\ell+\ell+j\}$.

(This case refers to vertices in $F_i$ that are mapped to $F_{\si'(i)}$
but do not maintain their index in the relevant copy of $G'_\ell$;
indeed, $v\!=\!2(i-1)\ell+j$ is the $j^\xth$ vertex of $F_i$,
but it is mapped by~$\mu$ to the $k^\xth$ vertex of $F_{\si'(i)}$
(i.e., $\mu(v)\!=\!2(\si'(i)-1)\ell+k$) such that $k\neq j$.)

Fixing $i$, let
$C\eqdef\{v\!=\!2(i-1)\ell+j\in F_i:
  \mu(v)\in F_{\si'(i)}\setminus\{2(\si'(i)-1)\ell+j\}\}$
denote the set of vertices considered in this case,
and $D=\{v\in F_i:\mu(v)\not\in F_{\si'(i)}\}$
denote the set of vertices that we are going to discount for.
As a warm-up, consider first the case that $D=\emptyset$.
In this case, by the robust self-ordering of $G'_\ell$,
the contribution of the vertices in $C$
to the difference between $G_n$ and $\mu(G_n)$
is $\Omega(|C|)$.

In the general case (i.e., where $D$ may not be empty),
we get a contribution of $\Omega(|C|)-d'\cdot|D|$,
where the second term compensates for the fact that the vertices
of $D$ were moved outside of this copy of $G'_\ell$ and replaced
by different vertices that may have different incidences.
Letting~$c'$ be the constant hidden in the $\Omega$-notation,
we get a contribution of
at least $c'\cdot|C|-d'\cdot|D|$,
which is at least $c'\cdot|C|/2$ if $|D|\leq c'\cdot|C|/2d'$.
On the other hand, if $|D|>c'\cdot|C|/2d'$,
then we get a contribution of $\Omega(|D|)=\Omega(|C|)$ by Cases~1--2.

Hence, in both sub-cases we have a contribution of $\Omega(|C|)$
to the difference between $G_n$ and $\mu(G_n)$.

The same analysis applies to
$\{v\!=\!2(i-1)\ell+\ell+j\in S_i:
  \mu(v)\in S_{\si''(i)}\setminus\{2(\si''(i)-1)\ell+\ell+j\}\}$,
where we use the robust self-ordering of $G''_\ell$
and Cases~1\&3.
\EDes
Lastly, we consider vertices that do not fall into any of
the prior cases. Such vertices maintain their type,
are mapped with the plurality vote of their connected component,
which is consistent among its two parts (i.e., $\si'$ and $\si''$),
and maintain their position in that component.
Hence, {\em the hypothesis that they are not fixed-points of $\mu$
can only be attributed to the fact that these vertices are
mapped to a connected component with a different index}.

\BDes
\item[{\em Case 7}:]
{\em Vertices $v\in F_i$ such that
both $\mu(v)\in F_{\si'(i)}\setminus F_{i}$
and $\mu(\phi_i(v))\in S_{\si''(i)}\setminus S_{i}$ hold.}

(We may assume that $\si'(i)\neq i$ and $\si''(i)\neq i$,
since otherwise this set is empty.
We may also assume that $\si'(i)=\si''(i)$,
since the complementary case was covered by Cases~4 and~5.
Hence, we focus on pairs of vertices that are
matched in the $i^\xth$ connected component of $G_n$
and are mapped by $\mu$ to the $k^\xth$ component of $G_n$
such that $k\neq i$.)

For every $i\neq k$,
let $\Delta_{i,k}=\{j\in[\ell]:\pi_i(j)\neq\pi_{k}(j)\}$
be the sets on which $\pi_i$ and $\pi_k$ differ.
(Note that if for every $v=2(i-1)\ell+j\in F_i$
it holds that $\mu(v)=2(k-1)\ell+j$
and $\mu(\phi_i(v))=2(k-1)\ell+\pi_i(j)$
(equiv., $\mu(2(i-1)\ell+\ell+\pi_i(j))=2(k-1)\ell+\pi_i(j)$),
then we get a contribution of $|\Delta_{i,k}|$
to the difference between $G_n$ and $\mu(G_n)$.)

Fixing $i$, let $D=D_1\cup D_2$ such that
\begin{eqnarray*}
D_1&=& \{v\in F_i:\mu(v)\not\in F_{\si'(i)}\;\vee\;
                 \mu(v+\ell)\not\in S_{\si''(i)}\} \\[3pt]
D_2&=& \left\{v\!=\!2(i-1)\ell+j\in F_i:
        \begin{array}{l}
         \mu(v)\in F_{\si'(i)}\setminus\{2(\si'(i)-1)\ell+j\} \\
         \vee\;\;
         \mu(\phi_i(v))
              \in S_{\si''(i)}\setminus\{2(\si''(i)-1)\ell+\ell+\pi_i(j)\}
        \end{array}\right\}
\end{eqnarray*}
(Recall that $\phi_i(2(i-1)\ell+j)=2(i-1)\ell+\ell+\pi_i(j)$.
The set $D_1$ accounts for the vertices covered in Cases~2\&3,
whereas $D_2$ accounts for the vertices covered in
(the two sub-cases of) Case~6.)

As a warm-up, consider first the case that $D=\emptyset$.
In this case, assuming $\si'(i)=\si''(i)\neq i$,
we get a contribution of $|\Delta_{i,\si'(i)}|=\Omega(\ell)$
(to the difference between $G_n$ and $\mu(G_n)$).
This contribution is due to the difference in the edges
that match $F_{\si'(i)}$ and $S_{\si'(i)}$ in $G_n$
and the edges that match $F_i$ and $S_i$ in $G_n$,
where $|\Delta_{i,\si'(i)}|=\Omega(\ell)$ is due to
the fact that the permutations (i.e., $\pi_k$'s) are far-apart.
The hypothesis $D_1=\emptyset$ means that all vertices
of $F_i$ (resp., of $S_i$) are mapped to $F_{\si'(i)}$
(resp., to $S_{\si''(i)}=S_{\si'(i)}$),
whereas $D_2=\emptyset$ means that these vertices preserves
their order within the two parts of the connected component.

The general case (i.e., where $D$ may not be empty)
requires a bit more care.
Suppose that the $\pi_k$'s are $\gamma$-apart;
that is, $|\Delta_{k',k}|>\gamma\cdot\ell$ for every $k'\neq k$.
We focus on the case that a vast majority of the vertices in
both $F_i$ and $S_i$ are mapped according to the plurality votes
(i.e., $\si'(i)$ and $\si''(i)$), since the complementary cases
are covered by Cases~2 and~3, respectively.
Specifically, if $|D_1|>\gamma\ell/3$,
then we get a contribution of $\Omega(\ell)$ by either Case~2 or Case~3.
Likewise, if $|D_2|>\gamma\ell/3$,
then we get a contribution of $\Omega(\ell)$ by Case~6.
So, assuming $\si'(i)=\si''(i)\neq i$, we are left with the case that
$$|\{v\!=\!2(i-1)\ell+j\in F_i\setminus D: j\in\Delta_{i,\si'(i)}\}|
  \geq\gamma\ell-2\gamma\ell/3.$$
In this case,
we get a contribution of at least $\gamma\ell/3$
to the difference between $G_n$ and $\mu(G_n)$.
This contribution is due to the difference in the edges
that match $F_{\si'(i)}$ and $S_{\si'(i)}$ in $G_n$
and the edges that match $F_i$ and $S_i$ in $G_n$,
where edges that have an endpoint (or its $\phi_i$-mate)
in $D$ were discarded.
Specifically, letting $k=\si'(i)=\si''(i)\neq i$,
the pair $(v,w)=(2(i-1)\ell+j,2(i-1)\ell+\ell+\pi_i(j))\in F_i\times S_i$
contributes to the difference if $j\in\Delta_{i,k}$ and
both $\mu(v)=2(k-1)\ell+j\in F_{k}$
and $\mu(w)=2(k-1)\ell+\ell+\pi_i(j)\in S_{k}$ hold
(i.e., $v\not\in D_1$ and $v,\phi_i^{-1}(w)\not\in D_2$).%
\footnote{Recall
that $\phi_i^{-1}(w)=\phi^{-1}((2(i-1)\ell+\ell+\pi_i(j)))=2(i-1)\ell+j=v$.}
Indeed, in this case $\{v,w\}$ is an edge in $G_n$
but $\{v,w\}$ is not an edge in $\imu(G_n)$.
(Hence, if the number of vertices of this case
is $\Omega(|\{u\in[n]:\mu(u)\neq u\}|)$,
then the difference between $G_n$ and $\imu(G_n)$
is $\Omega(|\{u\in[n]:\mu(u)\neq u\}|)$, and the same holds
with respect to the difference between $\mu(G_n)$ and $G_n$.)
\EDes
Combining all these cases, we get a total contribution
that is proportional to $|\{v\in[n]:\mu(v)\neq v\}|$,
where we might have counted the same contribution
in several different cases.
Since the number of cases is a constant, the theorem follows.
\EPF

\paragraph{Digest: Using large collections of pairwise far apart permutations.}
The construction presented in the proof of Theorem~\ref{construction:thm}
utilizes a collection of $(\ell!)^{\Omega(1)}$ permutations over $[\ell]$
that are pairwise far-apart (i.e., every two permutations differ
on $\Omega(\ell)$ inputs). Such a collection is constructed
in $\tildeO(\ell!)$-time by an iterative exhaustive search,
where the permutations are selected iteratively such that
in each iteration we find a permutation that is far from
permutations that were included in previous iterations.
We mention that in Section~\ref{step3:sec} we shall use
a collection of $\exp(\Omega(\ell))$ such permutations that is
locally computable (i.e., given the index of a permutation
we find its explicit description in polynomial time).
We also mention that, in follow-up work~\cite{GW:perm},
we provided a locally computable collection of $(\ell!)^{\Omega(1)}$
that are pairwise far-apart.

\paragraph{Digest: Combining two robustly self-ordered graphs.}
One ingredient in the proof of Theorem~\ref{construction:thm}
is forming connected components that consist of two
robustly self-ordered graphs that have different vertex degrees
and are connected by a bounded-degree bipartite graph.
Implicit in the proof is the fact that
the resulting graph is robustly self-ordered graph.

\BCM[Combining two $\Omega(1)$-robustly self-ordered graphs]
\label{bd:bipartite:clm}
For $i\in\{1,2\}$ and constant \mbox{$\gamma>0$},
let $G_i=(V_i,E_i)$ be an $\gamma$-robustly self-ordered graph,
and consider a graph $G=(V_1\cup V_2,E_1\cup E_2\cup E)$
of maximum degree $d$
such that $E$ contain edges with a single vertex in each $V_i$;
that is, $G$ consists of~$G_1$ and $G_2$
and an arbitrary bipartite graph that connects them.
If the maximum degree in~$G$ of each vertex in $V_1$
is strictly smaller than the minimum degree of each vertex in $V_2$,
then~$G$ is $\gamma/(4d+1)$-robustly self-ordered.
\ECM

\BPFS
For an arbitrary permutation $\mu:V\to V$,
let $T$ denote the set of its non-fixed-points,
and consider the following two cases.

\BDes
\item[{\em Case 1}:] More than $t=\gamma'\cdot|T|$ vertices are
mapped by $\mu$ from~$G_1$ to $G_2$, where $\gamma'=\gamma/(4d+1)$.

In this case,
we get a contribution of at least one unit per each such vertex,
due to the difference in the degrees between $V_1$ and $V_2$.

\item[{\em Case 2}:] at most $t$ vertices are
mapped by $\mu$ from~$G_1$ to $G_2$.

In this case, letting $T_i$ denote the set of non-fixed
vertices in $G_i$ that are mapped by $\mu$ to $G_i$,
we get a contribution of
at least $\sum_{i=1,2}(\gamma\cdot|T_i|-d\cdot t)$ units,
where the negative term is due to possible change in the incidence
with vertices in $T\setminus T_i$.
Hence, the total contribution in this case is at least
$\gamma\cdot(|T|-2t)-2d\cdot t \geq \gamma\cdot|T|-4dt= \gamma'\cdot|T|$.
\EDes
The claim follows.
\EPFS

\paragraph{Regaining regularity and expansion.}
While Theorem~\ref{construction:thm} achieves our main objective,
it useful towards some applications
(see, e.g., the proof of Theorem~\ref{strong-construction:thm})
to obtain this objective with graphs that are both regular and expanding.
This is achieved by applying Theorem~\ref{make:regular+expanding:thm}.
Hence, we have.

\BT[Theorem~\ref{construction:thm}, revised]
\label{construction-add:thm}
For any sufficiently large constant~$d$,
there exists an efficiently constructable family $\{G_n\}_{n\in\N}$
of robustly self-ordered $d$-regular expander graphs.
That is, there exists a polynomial-time algorithm
that on input $1^n$ outputs the $n$-vertex graph $G_n$.
\ET

\subsection{Strong (i.e., local) constructions}
\label{step3:sec}
While Theorem~\ref{construction-add:thm} provides
an efficient construction of
robustly self-ordered $d$-regular expander graphs,
we seek a stronger notion of constructability.
Specifically, rather than requiring that the graph be constructed
in time that is polynomial in its size, we require that the neighbors
of each given vertex can be found in time that is polynomial
in the vertex's name
(i.e., time that is polylogarithmic in the size of the graph).
We call such graphs {\sf locally constructable}
(and comment that the term ``strongly explicit''
is often used in the literature).

\BT[Locally constructing robustly self-ordered graphs]
\label{strong-construction:thm}
For any sufficiently large constant~$d$, there exists
a locally constructable family $\{G_n=([n],E_n)\}_{n\in\N}$
of robustly self-ordered $d$-regular graphs.
That is, there exists a polynomial-time algorithm
that on input $n$ and $v\in[n]$ outputs the list of neighbours
of vertex $v$ in $G_n$.
Furthermore, the graphs are either expanders
or consist of connected components of logarithmic size.
\ET
(Indeed, this establishes Theorem~\ref{main:ithm}.)
We comment that using the result of~\cite{GW:perm},
we can also get connected components of sub-logarithmic size,
as in Theorem~\ref{construction:thm}.%
\footnote{Specifically, the result of~\cite{GW:perm} provides
a construction of a collection of $L=\exp(\Omega(\ell\log\ell))$
permutations over $[\ell]$ that are pairwise far-apart
along with a polynomial-time algorithm that, on input $i\in[L]$,
returns a description of the $i^\xth$ permutation
(i.e., the algorithm should run in $\poly(\log L)$-time).
Using this algorithm,
we can afford to set $\ell=\frac{O(\log n)}{\log\log n}$
as in Theorem~\ref{construction:thm}.}
\medskip

\BPF
We employ the idea that underlies the proof of
Theorem~\ref{construction:thm},
while starting with an {\em efficiently constructable family}\/
of robustly self-ordered graphs
(as provided by Theorem~\ref{construction-add:thm})
rather than with the mere existence of a family of such graphs
(equiv., with $\ell$-vertex graphs that can be constructed
in $\poly(\ell!)$-time).
We use a slightly larger setting of $\ell$,
which allows us to use a collection of $\exp(\Omega(\ell))$
pairwise-far-apart permutations (rather than a collection
of $\exp(\Omega(\ell\log\ell))$ such permutations).
Lastly, we apply the same transformation as in the
proof of Theorem~\ref{construction-add:thm}
(so to regain regularity and expansion).
Details follow.

Given a generic $n$, let $\ell=O(\log n)$,
which implies that $\exp(\ell)=\poly(n)$.
By Theorem~\ref{construction-add:thm},
for all sufficiently large $d'$,
we can construct $\ell$-vertex $d'$-regular expander graphs that
are robustly self-ordered (with respect to the robustness parameter $c$)
in $\poly(\ell)$-time.
Again, we shall use two such graphs:
a $d'$-regular graph, denoted $G'_\ell=([\ell],E'_\ell)$,
and a $d''$-regular graph, denoted $G''_\ell=([\ell],E''_\ell)$,
where $d''=d'+1$.

Using $G'_\ell$ and $G''_\ell$, we construct an $n$-vertex
robustly self-ordered graph, denoted $G_n$,
that consists of $n/2\ell$ connected components that
are pairwise far from being isomorphic to one another.
This is done by picking $m=n/2\ell$ permutations,
denoted $\pi_1,\ldots,\pi_m:[\ell]\to[\ell]$,
that are pairwise far-apart, and constructing $2\ell$-vertex graphs
such that the $i^\xth$ such graph consist of
a copy of $G'_\ell$ and a copy of $G''_\ell$ that are connected
by a matching as determined by the permutation $\pi_i$
(as detailed in \eqref{edges-of-cc:eq}).

Using the fact that $m<2^\ell$
(rather that $m=\exp(\Theta(\ell\log\ell))$),
we can construct each of these permutations in $\poly(\ell)$-time
by using sequences of disjoint traspositions determined
via a good error correcting code.
Specifically, for $k=\log_2m<\log_2n$,
we use an error correcting code $C:\bitset^k\to\bitset^\ell$
of constant rate (i.e., $\ell=O(k)$) and linear distance
(i.e., the codewords are $\Omega(\ell)$ bits apart from each other),
and let $\pi_i(2j-1)=2j-1+C(i)_j$ and $\pi_i(2j)=2j-C(i)_j$,
where $i\in[m]=[2^k]\equiv\bitset^k$ and $j\in[\ell/2]$.
(That is, the $i^\xth$ permutation switches the pair $(2j-1,2j)\in[\ell]^2$
if and only if the $j^\xth$ bit in the $i^\xth$ codeword is~1,
where $C(i)$ is considered the~$i^\xth$ codeword.)

Like in the proof of Theorem~\ref{construction:thm},
the $i^\xth$ connected component of $G_n$ is isomorphic
to a graph with the vertex set $[2\ell]$
and the edge set
\begin{equation}\label{edges-of-cc:eq}
E'_\ell \;\cup\; \{\{\ell+u,\ell+v\}:\{u,v\}\in E''_\ell\}
         \;\cup\; \{\{v,\ell+\pi_i(v)\}:v\!\in\![\ell]\}.
\end{equation}
The key observation is that, for every $i\in[m]$ and $j\in[\ell]$,
the neighborhood of the $j^\xth$ (resp., $(\ell+j)^\xth$) vertex
in the $i^\xth$ connected component of the $n$-vertex graph $G_n$
is determined by $G'_\ell$ and~$\pi_i(j)$
(resp., by $G''_\ell$ and $\pi_i^{-1}(j)$),
which in turn are constructible in $\poly(\ell)$-time.
Hence, the neighborhood of each vertex in $G_n$
can be found in $\poly(\ell)$-time.
This implies local constructability, since $\ell=O(\log n)$.

The fact that $G_n$ is robustly self-ordered was already
established in the proof of Theorem~\ref{construction:thm},
which is oblivious of the permutations used as long as any
pair of permutations disagrees on $\Omega(\ell)$ points.
Lastly, we may obtain regularity and expansion
by applying Theorem~\ref{make:regular+expanding:thm}.
\EPF

\subsection{Local self-ordering}
\label{local-so:sec}
Recall that, by Definition~\ref{asymmetric:def},
a graph $G=([n],E)$ is called {\sf self-ordered}
if for every graph $G'=(V',E')$ that is isomorphic to~$G$
there exists a unique bijection $\phi:V'\to[n]$ such that $\phi(G')=G$.
One reason for our preferring the term ``self-ordered'' over
the classical term ``asymmetric'' is that we envision being
given such an isomorphic copy $G'=(V',E')$ and asked to find its
unique isomorphism to~$G$, which may be viewed as ordering
the vertices of $G'$ according to (their index in) $G$.
The task of finding this unique isomorphism
will be called {\em self-ordering $G'$ according to~$G$}\/
or {\em self-ordering $G'$}\/ (when~$G$ is clear from the context).

Evidently, the task of self-ordering a given graph $G'$
according to a (self-ordered) graph~$G$ that can
be efficiently constructed reduces to testing isomorphism.
When the graphs have bounded-degree the latter task
can be performed in polynomial-time~\cite{L}.
These are general facts that do apply also to
the robustly self-ordered graph $G_n$ constructed
in the proof of Theorem~\ref{strong-construction:thm}.
However, in light of the fact that the graph $G_n$
is {\em locally}\/ constructable, we can hope for more.
Specifically, it is natural to ask if we can perform
self-ordering of a graph $G'$ that is isomorphic to $G_n$
in a {\em local}\/ manner; that is, given a vertex in $G'$
(and oracle access to the incidence function of $G'$),
can we find the corresponding vertex in $G_n$ in $\poly(\log n)$-time?
Let us define this notion formally.

\BD[Locally self-ordering a self-ordered graph]
\label{local-so:def}
We say that a self-ordered graph $G=([n],E)$
is {\sf locally self-ordered}
if there exists a polynomial-time algorithm that,
given a vertex $v$ in any graph $G'=(V',E')$ that is isomorphic to~$G$
and oracle access to the incidence function of $G'$,
finds $\phi(v)\in[n]$ for the unique bijection  $\phi:V'\to[n]$
such that $\phi(G')=G$
{\rm(i.e., the unique isomorphism of $G'$ to~$G$)}.
\ED
Indeed, the isomorphism $\phi$ orders the vertices of $G'$
in accordance with the original (or target) graph~$G$.
We stress that the foregoing algorithm works in time that is polynomial
in the description of a vertex (i.e., $\poly(\log n))$-time),
which is polylogarithmic in the size of the graph (i.e., $n$).
We show that such algorithms exist for the graphs constructed
in the proof of Theorem~\ref{strong-construction:thm}.

\BT[Locally self-ordering the graphs
of Theorem~\ref{strong-construction:thm}]
\label{local-so4strong-construction:thm}
For any sufficiently large constant~$d$, there exists
a locally constructable family $\{G_n=([n],E_n)\}_{n\in\N}$
of robustly self-ordered $d$-regular graphs
that are locally self-ordered.
Furthermore, the graphs are either expanders
or consist of connected components of logarithmic size.
\ET
As in Theorem~\ref{strong-construction:thm},
we can obtain connected components of sub-logarithmic size
by using~\cite{GW:perm}.
\medskip

\BPF
We first consider the version that yields $n$-vertex graphs
that consist of connected components of logarithmic size.
The basic idea is that it we can afford reconstructing
the connected component in which the input vertex reside,
and this allows us both to determine the index of the vertex
in this connected component as well as the index of the component
in the graph. Specifically,
on input a vertex $v$ in a graph $G'$ that is isomorphic to $G_n$,
we proceed as follows.

\BE
\item
Using queries to the incidence function of $G'$,
we explore and retrieve the entire $2\ell$-vertex
connected component in which $v$ resides, where $\ell=\log_2n$.

Recall that this connected component consists of (copies of)
two $\ell$-vertex regular graphs, denoted $G'_\ell$ and $G''_\ell$,
that are connected by a matching.
Furthermore, these graphs have different degrees
and are each (robustly) self-ordered.
\item
Relying on the different degrees,
we identify the foregoing partition of this $2\ell$-vertex component
into two $\ell$-vertex (self-ordered) graphs,
denoted $A_v$ and $B_v$, where $A_v$ (resp., $B_v$)
is isomorphic to $G'_\ell$ (resp., $G''_\ell$).
\item
Relying on the self-ordering of $G'_\ell$ (resp., $G''_\ell$),
we order the vertices of $A_v$ (resp., $B_v$).
This is done by constructing $G'_\ell$ (resp., $G''_\ell$),
and using an isomorphism tester.
The order of the vertices in $A_v$ and $B_v$ also determines
the permutation that defines the matching between the two graphs.
\item
Relying on the correspondence between the permutations used
in the construction and codewords of a good error-correcting code,
we decode the relevant codeword (i.e., this is decoding without error).
This yields the index of the permutation in the collection,
which equals the index of the connected component.
\EE
Note that this refers to the basic construction that was presented
in the proof of Theorem~\ref{strong-construction:thm},
before it was transformed to an expander and made regular.
Recall that both transformations are performed by augmenting
the graph with auxiliary edges that are assigned a different color
than the original edges, and that edges with different colors
are later replaced by copies of different (constant-size) gadgets.
These transformations do not hinder the local self-ordering procedure
described above, since it may identify the original graph
(and ignore the gadgets that replace other edges).
The claim follows.
\EPF

\paragraph{Local reversed self-ordering.}
While {\em local self-ordering}\/ a (self-ordered) graph
seems {\em the natural local version}\/ of self-ordering the graph,
an alternative notion called {\em local reversed self-ordering}\/
will be defined and studied next
(and used in Section~\ref{pt:sec}).
Both notions refer to a self-ordered graph, denoted $G=([n],E)$,
and to an isomorphic copy of it, denoted $G'=(V',E')$;
that is, $G=\phi(G')$ for a (unique) bijection  $\phi:V'\to[n]$.
While local self-ordering is the task of
finding the index of a given vertex of $G'$ according to~$G$
(i.e., given $v\in V'$, find $\phi(v)\in[n]$),
local reversed self-ordering is the task of
finding the vertex of $G'$ that has a given index in~$G$
(i.e., given $i\in[n]$, find $\phi^{-1}(i)\in V'$).
In both cases, the graph~$G$ is locally constructible
and we are given oracle access to the incidence function of $G'$.
In addition, in the reversed task, we assume that the algorithm
is given an arbitrary vertex in $G'$,
since otherwise there is no hope to hit any element of $V'$.%
\footnote{Needless to say, this is not needed in case $V'=[n]$,
which is the case that is used in Section~\ref{pt:sec}.}

\BD[Locally reversed self-ordering]
\label{local-reversed-so:def}
We say that a self-ordered graph $G=([n],E)$
is {\sf locally reversed self-ordered}
if there exists a polynomial-time algorithm that,
given $i\in[n]$
and oracle access to the incidence function of
a graph $G'=(V',E')$ that is isomorphic to~$G$
and an arbitrary vertex $s\in V'$,
finds $\phi^{-1}(i)\in V'$ for the unique bijection  $\phi:V'\to[n]$
such that $\phi(G')=G$
{\rm(i.e., the unique isomorphism of $G'$ to~$G$)}.
\ED
We stress that the foregoing algorithm works in time that is polynomial
in the description of a vertex (i.e., $\poly(\log n))$-time),
which is polylogarithmic in the size of the graph (i.e., $n$).
We show that such algorithms exist for variants of the graphs
constructed in the proof of Theorem~\ref{strong-construction:thm}.
In fact, we show a more general result that refers
to any graph that is locally self-ordered and for which
short paths can be locally found between any given pair of vertices.

\BT[Sufficient conditions for
locally reversed self-ordering of graphs]
\label{local-reversed-so:thm}
Suppose that $\{G_n=([n],E_n)\}_{n\in\N}$ is a family
of bounded degree graphs that is locally self-ordered.
Further suppose that given $v,u\in[n]$,
one can find in polynomial-time a path from $u$
to $v$ in $G_n$.
Then, $\{G_n=([n],E_n)\}_{n\in\N}$
is locally reversed self-ordered.
\ET
We mention that any family of robustly self-ordered graphs
that is locally self-ordered can be transformed
into one that also supports locally finding short paths.
This is done by superimposing the graphs of this family
with graphs that supports locally finding short paths,
while using different colors for the edges of the two graphs
and later replacing these colored edges by gadgets
(as done in Section~\ref{edge-colored:trans:sec}).
We also mention that applying degree reduction to the hyper-cube
(i.e., replacing the original vertices with simple cycles)
yields a graph that supports locally finding short paths.%
\footnote{For any $\ell\in\N$, the resulting graph consists of the
vertex-set $\{\ang{x,i}\!:\!x\!\in\!\bitset^\ell\,\&\,i\!\!\in\![\ell]\}$
and edges that connect $\ang{x,i}$ to $\ang{x\xor0^{i-1}10^{\ell-i},i}$
and to $\ang{x,i+1}$, where $\ell+1$ stands for~1.
For simplicity of exposition, we also add self-loops on all vertices.
Then, given $\ang{x,i}$ and $\ang{y,j}$, we can combine the $2\ell$-path
that goes from $\ang{x,i}$ to $\ang{y,i}$ with the $|j-i|$-path
that goes from $\ang{y,i}$ to $\ang{y,j}$,
where the odd steps on the first path move from $\ang{z,k}$
to $\ang{z\xor0^{i-1}10^{\ell-i},k}$ (or stay in place)
and the even steps (on this path) move from $\ang{z,k}$ to $\ang{z,k+1}$.}
\medskip

\BPF
On input $i\in[n]$ and $s\in V'$,
and oracle access to the incidence function
of a graph $G'=(V',E')$ that is isomorphic to $G_n$,
we proceeds as follows.
\BE
\item
Using the local self-ordering algorithm,
we find $i_0=\phi(s)$, where $\phi:V'\to[n]$
is the unique bijection satisfying $\phi(G')=G$.

\item
Using the path-finding algorithm for~$G$,
we find a $\poly(\log n)$-long path from $i_0$ to $i$ in~$G$.

Let $\ell$ denote the length of the path,
and denote its intermediate vertices by $i_1,\ldots,i_{\ell-1}$;
that is, the full path is $i_0,i_1,\ldots,i_{\ell-1},i_\ell=i$.
\item
Using the local self-ordering algorithm (and our oracle access to $G'$),
we iteratively find the corresponding vertices in $G'$,
denoted $v_1,....,v_\ell$, where $v_j=\phi^{-1}(i_j)$.

For $j=1,\ldots,\ell$, we find $v_j\eqdef\phi^{-1}(i_j)$ as follows.
First, using queries to the incidence function of $G'$,
we find all neighbors (in $G'$) of $v_{j-1}$,
where $v_0\eqdef s$ (and, indeed, $v_0=\phi^{-1}(i_0)$).
Next, using the local self-ordering algorithm,
we find the indices of all these vertices in~$G$; that is,
for every vertex $w$ that neighbors $v_{j-1}$, we find $\phi(w)$.
Last, we set $v_j$ to be the neighbor that has index $i_j$ in~$G$;
that is, $v_j$ satisfies $\phi(v_j)=i_j$.
\EE
Hence, $v_\ell$ is the desired vertex;
that is, $v_\ell$ satisfies $\phi(v_\ell)=i_\ell=i$.

Assuming that
the local self-ordering algorithm has query complexity $q(n)$,
that the paths found in~$G$ have length at most $\ell(n)$,
and that $d$ is the degree bound,
the query complexity of our reversed self-ordering algorithm
is $(1+\ell(n)\cdot d)\cdot(q(n)+1)$,
where we count both our direct queries to the incidence function of~$G$
and the queries performed by the local self-ordering algorithm.
Similar considerations apply to the time complexity.
\EPF

\BCR[A version of Theorem~\ref{local-so4strong-construction:thm}
supporting local reversed self-ordering]
\label{local-reversed-so4rso:cor}
For any sufficiently large constant~$d$, there exists
a locally constructable family $\{G_n=([n],E_n)\}_{n\in\N}$
of robustly self-ordered graphs of maximum degree $d$ that are
both locally self-ordered and locally reversed self-ordered.
\ECR
The corollary follows by combining
Theorem~\ref{local-so4strong-construction:thm}
with Theorem~\ref{local-reversed-so:thm},
while using the augmentation outlined following
the statement of Theorem~\ref{local-reversed-so:thm}.
We mention that Corollary~\ref{local-reversed-so4rso:cor}
will be used in Section~\ref{pt:sec}.


\section{Application to Testing Bounded-Degree Graph Properties}
\label{pt:sec}
Our interest in efficiently constructable bounded-degree graphs
that are robustly self-ordered was triggered by an application
to property testing. Specifically, we observed that such constructions
can be used for proving a linear lower bound on the query complexity
of testing an {\em efficiently recognizable}\/ graph property
in the {\em bounded-degree graph model}.

It is well known that 3-Colorability has such a lower bound~\cite{BOT},
but this set is NP-complete.
On the other hand, linear lower bounds on the query complexity of testing
efficiently recognizable properties of {\em functions}\/ (equiv., sequences)
are well known (see~\cite[Sec.~10.2.3]{GGR}).
So the idea was to transport the latter lower bounds from
the domain of functions to the domain of bounded-degree graphs,
and this is where efficient constructions
of robustly self-ordered bounded-degree graphs come into play.
(We mention that an alternative way of obtaining the desired
lower bound was outlined in~\cite[Sec.~1]{G:ham}, see details below.)

More generally, the foregoing transportation demonstrates
a general methodology of transporting lower bounds that refer to
testing binary strings to lower bounds regarding testing
graph properties in the bounded-degree graph model.
The point is that strings are ordered objects,
whereas graphs properties are effectively sets of unlabeled graphs,
which are unordered objects.
Hence, we need to make the graphs (in the property) ordered,
and furthermore make this ordering robust in the very sense
that is reflected in Definition~\ref{robust-asymmetric:def}.
Essentially, we provide a reduction of testing a property
of strings to testing a (related) property of graphs.

We apply this methodology to obtain a subexponential separation
between the complexities of testing and tolerant testing
of graph properties in the bounded-degree graph model.
This result is obtained by transporting an analogous
result that was known for testing binary strings~\cite{FF}.
In addition to using a reduction from tolerantly testing
a property of strings to tolerantly testing a property of graphs,
this transportation also uses a reduction in the opposite direction,
which relies on the local computation features asserted
in Corollary~\ref{local-reversed-so4rso:cor}.

\paragraph{Organization of this section.}
We start with a brief review of the bounded-degree graph model
for testing graph properties. Next, we prove the aforementioned
linear lower bound on the query complexity of testing
an efficiently recognizable property,
and later we abstract the reduction that underlies this proof.
Observing that this reduction applies also to tolerant testing,
and presenting a reduction in the opposite direction,
we derive the aforementioned separation between testing
and tolerant testing.

\subsection{Background}
Property testing refers to algorithms of sublinear query complexity
for {\em approximate decision}; that is, given oracle access to
an object, these algorithms (called testers) distinguish objects
that have a predetermined property from objects that
are far from the property. Different models of property testing
arise from different query access and different distance measures.

In the last couple of decades, the area of property testing
has attracted significant attention (see, e.g.,~\cite{G:pt}).
Much of this attention was devoted to testing graph properties
in a variety of models including the dense graph model~\cite{GGR},
and the bounded-degree graph model~\cite{GR:bdg}
(surveyed in~\cite[Chap.~8]{G:pt} and~\cite[Chap.~9]{G:pt}, resp.).
In this section, we refer to the bounded-degree graph model,
in which graphs are represented by their incidence function
and distances are measured as the ratio of the number of
differing incidences to the maximal number of edges.

Specifically, for a degree bound $d\in\N$,
we represent a graph $G=([n],E)$ of maximum degree $d$
by the incidence function $g:[n]\times[d]\to[n]\cup\{0\}$
such that $g(v,i)$ indicates the $i^\xth$ neighbor of $v$
(where $g(v,i)=0$ indicates that $v$ has less than $i$ neighbors).
The distance between the graphs $G=([n],E)$ and $G'=([n],E')$
is defined as the size of the symmetric difference
between $E$ and $E'$ over $dn/2$.

A tester for a property $\Pi$ is given oracle access to
the tested object, where here oracle access to a graph
means oracle access to its incidence function.
In addition, such a tester is given a size parameter $n$
(i.e., the number of vertices in the graph),
and a {\sf proximity parameter}, denoted $\e>0$.
Tolerant testers, introduced in~\cite{PRR}
(and briefly surveyed in~\cite[Sec.~12.1]{G:pt}),
are given an additional parameter, $\eta<\e$,
which is called the {\sf tolerance parameter}.

\BD[Testing and tolerant testing graph properties
in the bounded-degree graph model]
\label{bdg:test-bd.def}
For a fixed {\sf degree bound} $d$,
a {\sf tester} for a graph property $\Pi$
is a probabilistic oracle machine that,
on input parameters $n$ and $\e$,
and oracle access to an $n$-vertex graph $G=([n],E)$ of maximum degree $d$,
outputs a binary verdict that satisfies the following two conditions.
\BE
\item
If $G\in\Pi$, then the tester accepts with probability at least~$2/3$.
\item
If~$G$ is $\e$-far from $\Pi$,
then the tester accepts with probability at most~$1/3$,
where~$G$ is {\sf $\e$-far} from $\Pi$ if for every $n$-vertex
graph $G'=([n],E')\in\Pi$ of maximum degree $d$ it holds that
the size of the symmetric difference between $E$ and $E'$
has cardinality that is greater than $\e\cdot dn/2$.
%
\EE
A {\sf tolerant tester} is also given a {\sf tolerance parameter} $\eta$,
and is required to accept with probability at least~$2/3$
any graph that is $\eta$-close to $\Pi$
{\em(i.e., not $\eta$-far from $\Pi$)}.%
\footnote{Of course, a tolerant tester is also required to reject
with probability at least~$2/3$ any graph that is $\e$-far from $\Pi$.}
\ED
We stress that a {\sf graph property} is defined as a property
that is preserved under isomorphism; that is, if $G=([n],E)$
is in the graph property $\Pi$, then all its isomorphic copies
are in the property
(i.e., $\pi(G)\in\Pi$ for every permutation $\pi:[n]\to[n]$).
The fact that we deal with graph properties
(rather than with properties of functions)
is the source of the difficulty
(of transporting results from the domain of functions
to the domain of graphs)
and the reason that robust self-ordering is relevant.%
\footnote{As noted in Section~\ref{intro:bd-results},
this is a special case of the general phenomenon pivoted at
the difference between ordered and unordered structures,
which arises in many contexts (in complexity and logic).}

%
The {\sf query complexity} of a tester for $\Pi$ is a function
(of the parameters $d,n$ and $\e$)
that represents the number of queries made by the tester
on the worst-case $n$-vertex graph of maximum degree $d$,
when given the proximity parameter $\e$.
Fixing $d$, we typically ignore its effect on the complexity
(equiv., treat $d$ as a hidden constant).
Also, when stating that the query complexity is $\Omega(q(n))$,
we mean that this bound holds for all sufficiently small $\e>0$;
that is, there exists a constant $\e_0>0$ such that distinguishing
between $n$-vertex graphs in $\Pi$ and $n$-vertex graphs
that are $\e_0$-far from $\Pi$ requires $\Omega(q(n))$ queries.

\subsection{Our first result and the general methodology}
With the foregoing preliminaries in place,
we state the first result of this section,
which is proved using Theorem~\ref{construction:thm}.

\BT[Linear query complexity lower bound
for testing an efficiently recognizable graph property
in the bounded-degree graph model]
\label{pt-lb:thm}
For any sufficiently large constant~$d$,
there exists an efficiently recognizable graph property $\Pi$
such that testing $\Pi$ in the bounded-degree graph model
{\rm(with degree bound $d$)} has query complexity $\Omega(n)$.
Furthermore, each $n$-vertex graph in $\Pi$
consists of connected components of size $o(\log n)$.
\ET
The main part of the theorem was known before:
As observed in~\cite[Sec.~1]{G:ham},
{\em there exists graph properties that are recognizable in polynomial-time
and yet are extremely hard to test in the bounded-degree graph model}.
This follows from the fact that the local reduction from
testing 3LIN (mod~2) to testing 3-Colorability
used by Bogdanov, Obata, and Trevisan~\cite{BOT}
is invertible in polynomial-time (which is a common feature
of reductions used in the context of NP-completeness proofs).%
\footnote{Of course, 3LIN (i.e., the satisfiability of linear
equations (with three variables each) over $\GF(2)$) is easily
solvable in polynomial-time. Nevertheless, Bogdanov \etal~\cite{BOT}
use a reduction of 3LIN to 3-Colorability (via 3SAT)
that originates in the theory of NP-completeness
in order to reduce between the testing problems.}
%
Indeed, their reduction actually demonstrates that the set
of (3-colorable) graphs that are obtained by applying this reduction
to satisfiable 3LIN (mod~2) instances is hard to test
(i.e., requires linear query complexity in the bounded-degree graph model).%
\footnote{Like almost all reductions of this type, the analysis of the
reduction actually refers to the promise problem induced by the image
of the reduction (i.e., the image of both the yes- and no-instances).}
We mention that the resulting property contains only connected graphs,
which means that Theorem~\ref{pt-lb:thm} has some added value:
The fact that it applies to graphs with tiny connected components
is interesting, since testing properties of such graphs may
seem easy (or at least not extremely hard) at first thought.
\medskip

\BPF
Our starting point is a property $\Phi$ of (binary) strings
(equiv., Boolean functions)
that is recognizable in polynomial-time
but has a linear query complexity lower bound
(see, e.g.,~\cite[Sec.~7]{GKNR}).
This refers to a model in which one makes queries to bits
of the tested string, and the distance between strings is
the (relative) Hamming distance.
Such lower bounds were transported to the {\em dense graph model}
in~\cite[10.2.3]{GGR} (see also~\cite{GKNR}),
but~-- to the best of own knowledge~-- no such transportation
were performed before in the context of the bounded-degree graph model.
Using robustly self-ordered graphs of bounded degree,
we present such a transportation.

\Bct{\rm(From properties of strings to properties of bounded-degree graphs):}
\label{string2graph:con}
Suppose that $\{G_n=([n],E_n)\}_{n\in\N}$
is a family of robustly self-ordered graphs of maximum degree~$d-2$.
\BI
\item
For every $n\in\N$ and $s\in\bitset^n$,
we define the graph $G'_s=([3n],E'_s)$ such that
\begin{equation}\label{string2graph:eq}
E'_s = E_n\cup\{\{i,n+i\},\{i,2n+i\}:i\in[n]\}
            \cup\{\{n+i,2n+i\}:i\in[n]\wedge s_i=1\}
\end{equation}
That is, $G'_s$ consists of a copy of $G_n$ augmented
by $2n$ vertices such that vertex $i\in[n]$ forms
a triangle with $n+i$ and $2n+i$ if $s_i=1$,
and forms a wedge with $n+i$ and $2n+i$ otherwise.
\item
For a set of strings $\Phi$,
we define $\Pi=\bigcup_{n\in\N}\Pi_n$
as the set of all graphs that are isomorphic
to some graph $G'_s$ such that $s\in\Phi$;
that is,
\begin{equation}\label{string2graph:eq2}
\Pi_n=\{\pi(G'_s):s\in(\Phi\cap\bitset^n)\wedge\pi\in\Sym_{3n}\}
\end{equation}
where $\Sym_{3n}$ denote the set of all permutations over $[3n]$.
\EI
\Ect
We may assume, without loss of generality,
that $G_n$ has no isolated vertices.
Hence, given a graph of the form $\pi(G'_s)$,
the vertices of $G_n$ are easily identifiable
as having degree at least three
(since vertices outside $G_n$ have degree at most two).
The foregoing construction yields
a local reduction of $\Phi$ to $\Pi$,
where locality means that each query to $G'_s$ can be
answered by making a constant number of queries to $s$,
and the (standard) validity of the reduction is based
on the fact that $G_n$ is asymmetric.%
\footnote{Standard validity means that $s\in\Phi$
if and only if $G'_s\in\Pi$.
Evidently, $s\in\Phi$ is mapped to $G'_s\in\Pi$;
the asymmetry of $G_n$ is used to show that $s\not\in\Phi$
is mapped to $G'_s\not\in\Pi$, since $G'_s$ can not
be isomorphic to any graph $G'_w$ such that $w\neq s$.
This, by itself, does not mean that if $s$ is far from $\Phi$
then $G'_s$ is far from $\Pi$.}

In order to be useful towards proving lower bounds
on the query complexity of testing~$\Pi$, we need to show
that the foregoing reduction is ``distance preserving''
(i.e., strings that are far from $\Phi$
are transformed into graphs that are far from $\Pi$).
The hypothesis that $G_n$ is robustly self-ordered is pivotal
to showing that if the string $s$ is far from $\Phi$,
then the graph $G'_s$ is far from~$\Pi$.

\setcounter{techclm}{1}
\Bcm{\em(Preserving distances):}
\label{string2graph:clm}
If $s\in\bitset^n$ is $\e$-far from $\Phi$,
then the $3n$-vertex graph $G'_s$
{\rm(as defined in Construction~\ref{string2graph:con})}
is $\Omega(\e)$-far from $\Pi$.
\Ecm

\Bpf
We prove the contrapositive.
Suppose that $G'_s$ is $\delta$-close to $\Pi$.
Then, for some $r\in\Phi$ and a permutation $\pi:[3n]\to[3n]$,
it holds that $G'_s$ is $\delta$-close to $\pi(G'_r)$.
(The possible use of a non-trivial permutation arises from
the fact that $\Pi$ is closed under isomorphism.)
If $\pi(i)=i$ for every $i\in[n]$,
then $s$ must be $(3d\delta/2)$-close to $r$,
where $d$ is the degree bound (of the model),
since $s_i=1$ (resp., $r_i=1$) if and only if $i$ forms
a triangle with $n+i$ and $2n+i$ in $G'_s$ (resp., in $\pi(G'_r)=G'_r$).%
\footnote{Hence, $G'_s$ is $\delta$-close to $G'_r$
implies that $|\{i\!\in\![n]\!:\!s_i\neq r_i\}|\leq\delta\cdot3dn/2$,
which means that $s$ is $\frac{3\delta dn/2}{n}$-close to $r$.}
Unfortunately, the foregoing condition
(i.e., $\pi(i)=i$ for every $i\in[n]$)
need not hold in general.

In general, the hypothesis that $\pi(G'_r)$ is $\delta$-close to $G'_s$
implies that $\pi$ maps at most $3\delta dn/2$ vertices of $[n]$
to $\{n+1,\ldots,3n\}$. This is the case since each vertex of $[n]$
has degree at least three in $G'_r$, whereas the other vertices
have degree at most two in $G'_s$ (or in any other graph~$G'_{s'}$).
Hence, if $t=|\{i\!\in\![n]\!:\!\pi(i)\!\in\!\{n+1,\ldots,3n\}|$,
then $\pi(G'_r)$ and $G'_s$ differ on at least $t$ edges,
whereas the hypothesis is that the difference is at most $\delta\cdot3dn/2$.
It follows that $t\leq\delta\cdot3dn/2$.

Turning to the vertices $i\in[n]$ that $\pi$ maps to $[n]\setminus\{i\}$,
we upper-bound their number by $O(\delta d^2n)$,
since the difference between $\pi(G'_r)$ and $G'_s$
is at most $\delta\cdot3dn/2$,
whereas the hypothesis that $G_n$ is $c$-robustly self-ordered
implies that the difference between $\pi(G'_r)$ and $G'_s$
(or any other graph $G'_{w}$) is at least
$$\Delta =
     c\cdot |\{i\!\in\![n]\!:\!\pi(i)\neq i\}|
                     -d\cdot|\{i\!\in\![n]\!:\!\pi(i)\not\in[n]\}|.$$
(Compare Case~6 in the proof of Theorem~\ref{construction:thm}.)%
\footnote{Hence, $\Delta\leq\delta\cdot3dn/2$ implies that\\[-5ex]
\begin{eqnarray*}
|\{i\in[n]:\pi(i)\neq i\}|
&=& \frac{\Delta+d\cdot|\{i\in[n]:\pi(i)\not\in[n]\}|}{c} \\
&\leq& \frac{3\delta dn/2+d\cdot3\delta dn/2}{c}
\end{eqnarray*}\\[-4ex]
which is $O(\delta d^2n)$.}

Letting $I=\{i\!\in\![n]\!:\!\pi(i)\!=\!i\}|$,
observe that $D\eqdef|\{i\in I:r_i\neq s_i\}|\leq3\delta dn/2$,
since $r_i\neq s_i$ implies that, for every $i\in I$,
the subgraph induced by $\{i,n+i,2n+i\}$ is different in $\pi(G'_r)$ and $G'_s$
(i.e., it is a triangle in one graph and contains two edges in the other),
whereas by the hypothesis $\pi(G'_r)$ and $G'_s$ differ
on at most $\delta\cdot3dn/2$ edges.
Recalling that $|I|=n-O(\delta d^2n)$,
it follows that $|\{i\in[n]:r_i\neq s_i\}|\leq(n-|I|)+D=O(\delta d^2n)$.
Recalling that $d$ is a constant,
we infer that $s$ is $O(\delta)$-close to $r\in\Phi$,
and the claims follows.
\Epf

\mypar{Conclusion.}
Starting with Theorem~\ref{construction:thm}
(i.e., an efficient construction of robustly self-ordered graphs
of bounded degree),
using Construction~\ref{string2graph:con},
and applying Claim~\ref{string2graph:clm},
the theorem follows.
Specifically, we need to verify the following facts.
\BI
\item The set $\Pi$ is polynomial-time recognizable.

Given an $3n$-vertex graph $G'$, an adequate algorithm first tries
to identify and order the vertices of the corresponding graph $G_n$,
which means that it finds $s\in\bitset^n$ such that $G'$
is isomorphic to $G'_s$ (or determines that no such $s$ exists).
Having found $s$, the algorithm accepts if and only if $s\in\Phi_n$,
where $\Phi$ is polynomial-time recognizable by our starting hypothesis.

The first step relies on the hypothesis that $G_n$
can be constructed in polynomial-time, and proceeds as follows.
\BE
\item
Identifies a set of $n$ vertices, denoted $I$,
in $G'$ such that each vertex in $I$ has degree greater than 2,
rejecting if the number of such vertices is different from $n$.
\item
Finds the unique isomorphism between $G_n$
and subgraph of $G'$ induced by $I$,
rejecting if no such isomorphism is found.

(Here we rely on the fact that isomorphism between graphs
of bounded-degree can be found in polynomial-time~\cite{L}).
\item
The foregoing isomoprophism determines the ordering
of the vertices in $I$, which in turn determines $s$
(or indicates that $G'$ is not isomoprophic to any $G'_s$).
\EE

\item Testing $\Pi$ requires linear query complexity.

This is shown by reducing testing $\Phi$ to testing $\Pi$,
while recalling that testing $\Phi$ requires linear query complexity.
Given (proximity parameter $\e$ and)
oracle access to a string $s\in\bitset^n$,
we invoke the tester for $\Pi$ (with proximity parameter $\Omega(\e)$)
while emulating oracle access to $G'_s$ in a straightforward manner
(i.e., each query to $G'_s$ is answered by making
at most one query to $s$).
Recall that $s\in\Phi$ implies $G'_s\in\Pi$,
whereas by Claim~\ref{string2graph:clm}
if~$s$ is $\e$-far from $\Phi$
then $G'_s$ is $\Omega(\e)$-far from $\Pi$.
\EI
This completes the proof, since the $n$-vertex graphs
of Theorem~\ref{construction:thm}
have connected components of size $o(\log n)$.
\EPF

\paragraph{Digest: Reducing testing properties of strings
to testing graph properties.}
We wish to highlight the fact that the proof of Theorem~\ref{pt-lb:thm}
is based on a general reduction of testing any property $\Phi$
of strings to testing a corresponding (bounded-degree) graph property $\Pi$.
This reduction is described in Construction~\ref{string2graph:con}
and its validity is proved in Claim~\ref{string2graph:clm}.
Recall that, for any $n$, the graph property $\Pi$
consists of $3n$-vertex graphs (of bounded-degree)
that encode the different $n$-bit long strings in $\Phi$.
This reduction is local and preserves distances:
\BDes
\item[{\em Locality}:]
Each string $s\in\bitset^n$ is encoded by a graph $G'_s$
such that each query to $G'_s$ can be answered
by making at most one query to $s$.
\item[{\em Preserving distances}:]
If $s\in\Phi$ then $G'_s\in\Pi$,
whereas if $s$ is $\e$-far from $\Phi$
then $G'_s$ is $\Omega(\e)$-far from~$\Pi$.
\EDes
Recall that $G'_s$ consists of
a fixed robustly self-ordered $n$-vertex graph $G_n$
augmented by ($n$ two-vertex) gadgets that encode $s$.
Let us spell out the effect of this reduction.

\BCR[Implicit in the proof of Theorem~\ref{pt-lb:thm}]
\label{pt:reduction:cor}
For $\Phi$ and $\Pi$ as in Construction~\ref{string2graph:con},
let $Q_\Phi$ and $Q_\Pi$ denote the query complexities
of testing $\Phi$ and $\Pi$, respectively.
Then, $Q_\Phi(n,\e)\leq Q_\Pi(3n,\Omega(\e))$.
Likewise, letting $Q'_\Phi$ {\rm(resp., $Q'_\Pi$)} denote
the query complexity of tolerantly testing $\Phi$ {\rm(resp., $\Pi$)},
it holds that $Q'_\Phi(n,\eta,\e)\leq Q'_\Pi(3n,\eta/3,\Omega(\e))$.
\ECR
The tolerant testing part requires an additional justification.
Specifically, we observe that strings $s$ that are $\eta$-close to $\Phi$
yield graphs $G'_s$ that are $\eta/3$-close to $\Pi$.
This is the case because,
if the $n$-bit long strings $s$ and $r$ differ on $k$ bits,
then the $3n$-vertex graphs $G'_s$ and $G'_r$ differ on $k$ vertex pairs.


In preparation to proving the separation between the complexities
of testing and tolerant testing,
we show a reduction in the opposite direction.
This reduction holds provided that the robustly self-ordered graphs
used in the definition of $\Pi$ are locally reversed self-ordered
(see Definition~\ref{local-reversed-so:def}).

\BP[Reducing testing $\Pi$ to testing $\Phi$]
\label{pt:reverse-reduction:prop}
Suppose that the graphs used in Construction~\ref{string2graph:con}
are locally self-ordered and locally reversed self-ordered,
and let $\Phi,\Pi$ and $Q_\Phi,Q_\Pi$
be as in Corollary~\ref{pt:reduction:cor}.
Then, $Q_\Pi(3n,\e)\leq \poly(\log n)\cdot(Q_\Phi(n,2\e)+O(1/\e))$.
Furthermore, one-sided error probability is preserved.%
\footnote{A tester is said to have {\sf one-sided error probability}
if it always accepts objects that have the property.}
\EP
Recall that the hypothesis can be met by
using Corollary~\ref{local-reversed-so4rso:cor}.
\medskip

\BPF
Given oracle access to a graph $G'=([3n],E')$,
we first test that $G'$ is isomorphic to~$G'_s$,
for some $s\in\bitset^n$, and then invoke
the tester for $\Phi$ while providing it
with oracle access to $s$. Specifically,
when the latter tester queries the bit $i$,
we use the local reversed self-order algorithm
in order to locate the $i^\xth$ vertex of $G_n$ in $G'$,
and then determine the bit $s_i$ accordingly.
Details follow.

Let $V$ denote the set of vertices of the graph $G'=([3n],E')$
that have degree greater than~2 and neighbor two vertices
that have degree at most~2
and neighbor each other if they have degree~2.
Evidently, the vertices of $V$ are easy to identify by
querying $G'$ for their neighbors and their neighbors' neighbors.
Furthermore, $|V|\leq n$, since each vertex in $V$ has two
neighbors that are not connected to any other vertex in $V$,
and equality holds in case $G'\in\Pi$.
We try to find a (``pivot'') vertex $p\in V$ by picking
an arbitrary vertex in $G'$ and checking it and its neighbors.
If none of these is in $V$, then we reject.
Otherwise, we continue; we shall be using $p$ as an auxiliary input
in all (future) invocations of the local reversed self-ordering algorithm,
denoted $A$.

Intuitively, fixing the pivot $p$, allow to locate,
for each $i\in[n]$, the $i^\xth$ vertex of $G_n$ in $G'$;
that is, this vertex is $A^{G'}(p,i)$.
Indeed, this holds when $G'$ is in $\Pi$,
but all bets are off otherwise.
In particular, when $G'\not\in\Pi$,
the answer $A^{G'}(p,i)$ may not be a vertex in $V$,
or may equal $A^{G'}(p,j)$ for some $j\neq i$.
We avoid these cases by checking that $A^{G'}(p,i)\in V$
and that the vertex in $G_n$ that corresponds to $A^{G'}(p,i)$
is indeed indexed $i$ (by using the local self-ordering algorithm).
In particular, we define $A'(i)\eqdef A^{G'}(p,i)$
if both conditions hold, and let $A'(i)$ be undefined otherwise.
Hence, $i\mapsto A'(i)$ is always a bijection from a subset of $[n]$
to a subset of $V$. This strategy is detailed next.

Using the foregoing algorithm $A$ and the pivot $p\in V$,
we define $A'(i)=A^{G'}(p,i)$ if $A^{G'}(p,i)\in V$ and invoking
the local self-ordering algorithm on input $A^{G'}(p,i)$ yields~$i$.
Otherwise $A'(i)$ is undefined.
Hence, evaluating $A'$ amounts to evaluating $A$
as well as evaluating the local self-ordering algorithm.
Letting $I'\subseteq[n]$ denote the set of ``indices''
(i.e., vertices of $G_n$) on which $A'$ is defined,
we note that $A'$ is a bijection from $I'$
to $V'\eqdef\{A'(i)\!:\!i\in\!I'\}\subseteq V$,
and that $I'=[n]$ (and $V'=V$) if $G'\in\Pi$.
Hence, our first test is testing whether $I'=[n]$,
which is done by selecting at random $O(1/\e)$ elements of $[n]$,
and rejecting if $A'$ is undefined on any of them.
Otherwise, we proceed, while assuming that $|I'|\geq(1-0.1\e)\cdot n$.

Next, we test whether the subgraph of $G_n$ induced by $I'$
is isomorphic to the subgraph of $G'$ induced by $V'$,
{\em where the isomorphism is provided by $A'$}\/
(which maps $I'$ to $V'$).
In other words, we actually test whether the $A'$-image of
the subgraph of $G_n$ induced by $I'$
equals the subgraph of $G'$ induced by $V'$.
This can be done by sampling $O(1/\e)$ vertices of $G_n$
and comparing their neighbors to the neighbors
of the corresponding vertices in $G'$,
which are found by $A'$.
Specifically, for every sampled vertex $i\in[n]$,
we determine its set of neighbors $S_i$ in $G_n$,
obtain both $A'(i)$ and $A'(S_i)\eqdef\{A'(j)\!:\!j\!\in\!S_i\}$,
which are supposedly the corresponding vertices in $G'$,
and check whether $A'(S_i)$ is the set of neighbors of $A'(i)$ in $G'$.
We reject if $A'$ is undefined on any of these vertices
(i.e., on sampled vertices or their neighbors in $G_n$).
Needless to say,
we also reject if any of the foregoing neighborhood checks fails.
%

Assuming that we did not reject so far,
we may assume that $G'$ is $\e/2$-close to being isomorphic to some $G'_s$,
where the isomorphism is consistent with the inverse of $A'$.
At this point, we invoke the tester for $\Phi$, denoted $T$,
in order to test whether $s\in\Phi$.
This is done by providing~$T$ with oracle access to $s$ as follows.
When $T$ makes a query $i\in[n]$, we determine $A'(i)$,
and use our query access to $G'$ in order to determine
the two neighbors of $A'(i)$ that have degree at most~2
(and are either connected or have degree~1).
If this fails, we reject.
Otherwise, we answer~1 if and only if these two neighbors
are connected in $G'$.

To summarize, we employ three tests to $G'$:
An {\em initial test}\/ of the size $I'$
(which also includes finding a pivot $p\in V$),
an {\em equality test} between the $A'$-image of the subgraph
of $G'$ induced by $I'$ and the subgraph of $G_n$ induced by $V'$,
and an emulation of the testing of $\Phi$.
(In all tests, if we encounter an index in $[n]\setminus I'$,
we suspend the execution and reject.)
For simplicity and without loss of generality,
we may assume that $T$ is correct with high (constant) probability.

Note that if $G'\in\Pi$, then it holds that $G'=\pi(G'_s)$
for some $s\in\Phi$ and some permutation $\pi\in\Sym_{3n}$.
In this case, it holds that $|I'|=n$ and we always find a pivot $p\in V$.
Furthermore, $A'$ equals the restriction of $\pi$ to $[n]$,
the ``equality (between induced subgraphs) test'' always succeeds,
and the emulation of oracle access to $s$ is perfect.
Hence, we accept with high probability
(or always, if $T$ has one-sided error probability).

On the other hand, suppose that $G'$ is $\e$-far from $\Pi$.
If either $|I'|<(1-0.1\e)\cdot n$
or the subgraph of $G'$ induced by $V'$
is $0.1\e$-far from the $A'$-image of the subgraph of $G_n$ induced by $I'$,
then we reject with high probability due to one of the first two tests.
Otherwise, letting $\pi$ be an arbitrary bijection of $[3n]$ to $[3n]$
that extends $A'$, it follows that for some $s\in\bitset^n$
the graph~$G'$ is $0.2\e$-close to $\pi(G'_s)$,
since we may obtain $\pi(G'_s)$ from $G'$ by modifying
the neighborhood of $0.1n$ vertices in $I'$
as well as of the vertices in $[n]\setminus I'$.
Furthermore, for every $i\in[n]$ on which~$A'$ is defined,
it holds that $s_i=1$ if and only if the two neighbors
of $A'(i)$ that have degree at most~2 are connected.
By the hypothesis regarding $G'$,
the string $s$ must be $\frac{0.8\e\cdot3dn/2}{n}$-far from~$\Phi$,
and $A'(i)=\pi(i)$ whenever $A'$ is defined on $i\in[n]$.
It follows that either the emulation of $T$ was abruptly terminated
(leading to rejection)
or the answers provided to $T$ are according to $s$.
Hence, we reject with high probability.
\EPF

\paragraph{Digest: Tightly reducing testing properties of strings
to testing graph properties.}
In continuation to (the main part of) Corollary~\ref{pt:reduction:cor},
we highlight the fact that Construction~\ref{string2graph:con}
not only reduces testing the string property $\Phi$
to testing the graph property~$\Pi$,
but rather does so in a rather tight manner.
Specifically, for $\Phi,\Pi$ and $Q_\Phi,Q_\Pi$
as in Corollary~\ref{pt:reduction:cor},
it holds that $Q_\Phi(n,\e)$ and $Q_\Pi(\Theta(n),\Theta(\e))$
agree up to a $\poly(\log n)$ factor.
In other words, {\em for any property of strings $\Phi$,
there exists a property of bounded-degree graphs $\Pi$
such that the {\rm(}query and time{\rm)} complexity of testing $\Phi$
is reflected in the {\rm(}query and time{\rm)} complexity of testing $\Pi$},
where our notion of reflection allows for a polylogarithmic slackness.
Recall that the transformation of strings in $\Phi$ to graphs in $\Pi$
is (strongly/locally) efficient.

\subsection{Separating tolerant testing from testing}
Using Corollary~\ref{pt:reduction:cor}
and Proposition~\ref{pt:reverse-reduction:prop},
we transport the separation between tolerant testing and testing,
{from} the domain of testing strings,
where it has been established by~\cite{FF},
to the domain of testing graph properties in the bounded-degree graph model.

\BT[In the bounded-degree graph model,
tolerant testing is harder than testing]
\label{pt-tolerant:thm}
For any sufficiently large constant~$d$ and any constant $c\in(0,1)$,
there exists a graph property $\Pi$ such that testing $\Pi$
in the bounded-degree graph model {\rm(with degree bound $d$)}
has query complexity $O(\poly(\log n)/\e)$,
but tolerantly testing $\Pi$
has query complexity $\Omega(n^{\Omega(1-c)})$, provided that
the tolerance parameter is not smaller than $n^{-c}$.
Furthermore, $\Pi$ is efficiently recognizable.
\ET

\BPF
A small variant on the proof of~\cite[Thm.~1.3]{FF}
yields an efficiently recognizable set of strings $\Phi$
that is testable in $O(1/\e)$ queries but tolerantly testing it
requires $\Omega(n^{\Omega(1-c)})$ queries.%
\footnote{Basically, the construction of~\cite{FF} consists of
repeating some $m$-bit long string $\poly(m)$ times and augmenting
it with a PCP of Proximity (PCPP)~\cite{BGHSV,DR} of membership
in some polynomial-time recognizable set that is hard to test.
Essentially, the PCPP helps the tester,
but it may be totally useless (when corrupted)
in the tolerant testing setting.
While~\cite{FF} lets the PCPP occupy an $o(1/\log\log n)$
fraction of the final $n$-bit string, we let it occupy
just a $n^{-c}$ fraction (and use $m=n^{\Omega(1-c)}$).
This requires using a different PCPP than the one used in~\cite{FF};
e.g., using a strong PCPP with linear detection
probability~\cite[Def.~2.2]{DGG} will do,
and such a PCPP is available~\cite[Thm.~3.3]{DGG}.}
Using Construction~\ref{string2graph:con}
with graphs that are locally self-ordered
and locally reversed self-ordered
(as provided by Corollary~\ref{local-reversed-so4rso:cor}),
we obtain the desired graph property $\Pi$.
By Corollary~\ref{pt:reduction:cor} tolerantly testing $\Pi$
requires $\Omega(n^{\Omega(1)})$ queries,
whereas by Proposition~\ref{pt:reverse-reduction:prop}
(non-tolerant) testing~$\Pi$
has query complexity $\poly(\log n)\cdot O(1/\e)$.
The claim follows.
\EPF

\section{Random Regular Graphs are Robustly Self-Ordered}
\label{random-graphs:sec}
While Theorem~\ref{existence:thm} only asserts the
existence of robustly self-ordered $d$-regular graphs,
we next show that almost all $d$-regular graphs
are robustly self-ordered.
This extends work in probabilistic graph theory,
which proves a similar result for the weaker notion
of self-ordered (a.k.a asymmetric) graphs~\cite{B1,B2}.

\BT[Random $d$-regular graphs are robustly self-ordered]
\label{random-works:thm}
For any sufficiently large constant~$d$,
a random $2d$-regular $n$-vertex graph
is robustly self-ordered with probability $1-o(1)$.
\ET
Recall that, with very high probability, these graphs are expanders.
We mention that the proof of Theorem~\ref{existence:thm}
actually established that $n$-vertex graphs drawn from
a weird distribution (which has min-entropy $\Omega(n)$)
are robustly self-ordered with probability $1-o(1)$.
However, this is established by using the edge-coloring variant,
and requires employing the transformation presented
in Section~\ref{edge-colored:trans:sec}.
In contrast, the following proof works directly with the original
(uncolored) variant, and is completely self-contained.
\medskip

\BPF
The proof is quite similar to the proof Claim~\ref{random-colored:clm},
but it faces complications that were avoided in the prior proof
by using edge-colors and implicitly directed edges.
Specifically, for candidate permutations $\pi_1,\ldots,\pi_d:[n]\to[n]$
(to be used in the construction)
and all (non-trivial) permutations $\mu:[n]\to[n]$,
the proof of Claim~\ref{random-colored:clm} considered events
of the form $(\forall j\!\in\![d])\;\pi_j(i)=\mu(\pi_j(\imu(i)))$,
whereas here we shall consider events of the form
$\{\pi^b_j(i):j\!\in\![d]\,\&\,b\!\in\!\{\pm1\}\}
 =\{\mu(\pi_k^c(\imu(i))):k\!\in\![d]\,\&\,c\!\in\!\{\pm1\}\}$.
That is, rather than considering a sequence of equations that
refers to single $\pi_j$'s, we consider a sequence of equations
that refer to the set of all $\pi_j$'s, which were expressed
above in terms of equations between multi-sets.
These multi-set equations will be reduced to equations among sequences
by considering all possible ordering of these multi-sets.
This amounts to taking a union bound over all possible ordering
and results in a more complicated analysis (due to the $\pi_j^{-1}$'s)
and much more cumbersome notation.

To facilitate the proof,
we use the standard methodology (cf.~\cite[Apdx.~2]{E})
of first proving the result in the {\em random permutation model},
then transporting it to the {\em configuration model}\/
(by using a general result of~\cite{GJKW}),
and finally conditioning on the event that the generated graph
is simple (which occurs with positive constant probability).
Indeed, both models generate multi-graphs
that are not necessarily simple graphs
(i.e., these multi-graphs may have self-loops and parallel edges).
We also use the fact that the simple graphs that are
generated by the configuration model (for degree $d'$)
are uniformly distributed among all $d'$-regular graphs.

Recall that in the {\sf random permutation model}
a $2d$-regular $n$-vertex multi-graph is generated by selecting
uniformly and independently $d$ permutations $\pi_1,\ldots,\pi_d:[n]\to[n]$.
The multi-graph, denoted $G_{(\pi_1,\ldots,\pi_d)}$,
consists of the edge multi-set
$\bigcup_{j\in[d]}\{\{i,\pi_j(i)\}:i\in[n]\}$,
where the $2j^\xth$ (resp., $(2j-1)^\st$) neighbor of vertex $i$
is $\pi_j(i)$ (resp., $\pi_j^{-1}(i)$).
Note that this multi-graph may have self-loops (due to $\pi_j(i)=i$),
which contributed two units to the degree of a vertex,
as well as parallel edges
(due to $\pi_j(i)=\pi_k(i)$ for $j\neq k$
and $\pi_j(i)=\pi_k^{-1}(i)$ for any $j,k$).
We denote the $j^\xth$ neighbor of vertex $i$ by $g_j(i)$;
that is, $g_j(i)=\pi_{j/2}(i)$ if $j$ is even,
and $g_j(i)=\pi_{(j+1)/2}^{-1}(i)$ otherwise.

Consider an arbitrary permutation $\mu:[n]\to[n]$,
and let $T=\{i\!\in\![n]\!:\!\mu(i)\neq i\}$
be its set of non-fixed-point.
We shall show that,
with probability $1-\exp(-\Omega(d\cdot|T|\cdot\log n))$
over the choice of ${\ov\pi}=(\pi_1,\ldots,\pi_d)$,
the size of the symmetric difference
between $G_{\ov\pi}$ and $\mu(G_{\ov\pi})$ is $\Omega(|T|)$.
Note that this difference is (half) the sum over $i\in[n]$
of the size of the symmetric difference between the multi-set
of neighbors of vertex~$i$ in $G_{\ov\pi}$
and the multi-set of neighbors of vertex~$i$ in~$\mu(G_{\ov\pi})$.
We refer to the latter difference by the phrase
{\em the contribution of vertex $i$ to the difference
between $G_{\ov\pi}$ and $\mu(G_{\ov\pi})$}.

As a warm-up, we first show that each element of $T$
contributes a non-zero number of units to the difference
(between $G_{\ov\pi}$ and $\mu(G_{\ov\pi})$)
with probability $1-O(\poly(d)/n)^{d/6}$ over the choice of $\ov\pi$.
Recalling that $i\in T$ contributes to the difference
(between $G_{\ov\pi}$ and $\mu(G_{\ov\pi})$)
if the multi-sets of its neighbors in $G_{\ov\pi}$
and $\mu(G_{\ov\pi})$ differ,
it follows that $i\in T$ contributes to the difference if
{\em for every permutation $\sigma:[2d]\to[2d]$
there exists $j\in[2d]$
such that $g_j(i)\neq\mu(g_{\sigma(j)}(\imu(i)))$}.
Hence, the complementary event holds
(i.e., $i$ does not contribute to the difference)
if and only if there exists permutation $\sigma:[2d]\to[2d]$
such that for every $j\in[2d]$
it holds that $g_j(i)=\mu(g_{\sigma(j)}(\imu(i)))$.
Thus, the probability that $i$ does not contribute to the difference
is given by
\begin{eqnarray}
\lefteqn{\prob_{\ov\pi}
  \left[\exists\sigma\!\in\!\Sym_{2d}\;
        (\forall j\!\in\![2d])\;\;g_j(i)=\mu(g_{\sigma(j)}(\imu(i)))\right]}
\nonumber \\
&\leq& (2d)!\cdot\max_{\sigma\in\Sym_{2d}}
  \left\{\prob_{\ov\pi}\left[(\forall j\!\in\![2d])\;\;
            g_j(i)=\mu(g_{\sigma(j)}(\imu(i)))\right] \right\}.
\label{true-probability:eq}
\end{eqnarray}
Fixing $\sigma$ that maximizes the probability,
and denoting it $\sigma_i$,
consider any $J_i\subseteq[d]$ such that for the $j$'s in $J_i$
the multi-sets $\{j,\ceil{\sigma_i(2j)/2}\}$'s are disjoint
(i.e., $\{j,\ceil{\sigma_i(2j)/2}\}\cap\{k,\ceil{\sigma_i(2k)/2}\}=\emptyset$
for any $j\neq k\in J_i$).
Note that we may select $J_i$ such that $|J_i|\geq d/6$,
since taking $j$ to $J_i$ rules out taking (to $J_i$)
at most five other values (i.e., letting $v\eqdef\ceil{\sigma_i(2j)/2}$,
the value $k$ is ruled out if either $k\in\{j,v\}$
or $\sigma_i(2k)\in\{2j-1,2j,2v-1,2v\}$).
\ifnum\exclaims=0 
Then, \eqref{true-probability:eq} is (crudely\footnotemark)
\footnotetext{One may indeed analyze \eqref{true-probability:eq}
directly, and obtain a bound of $O(d/n)^{2d}$ by considering
all the $2d$ events and accounting for their small dependency.
On the other hand, we can obtain higher robustness parameter
by considering smaller sets $J_i$'s (say of size $d/4$),
which suffice for counting vertices that
contribute (say) $d/4$ units to the difference
between $G_{\ov\pi}$ and $\mu(G_{\ov\pi})$.\label{random-bd:fn}}
%
upper-bounded by
\else 
Using this feature of $J_i$, we prove~--

\setcounter{subsection}{1}

\Bcm{\em(Warm-up):}\footnote{One may obtain a better bound
of $O(d/n)^{2d}$ by analyzing \eqref{true-probability:eq} directly,
by considering all the $2d$ events and accounting for their small dependency.
On the other hand, we can obtain higher robustness parameter
by considering smaller sets $J_i$'s (say of size $d/12$),
which suffice for counting vertices that
contribute (say) $d/2$ units to the difference
between $G_{\ov\pi}$ and $\mu(G_{\ov\pi})$.\label{random-bd:fn}}
%
\label{random-works-warm-up:clm}
\eqref{true-probability:eq} is upper-bounded
by $(2d)^{2d}\cdot(3/n)^{|J_i|}$.
\Ecm

\Bpf
Recalling that $\sigma_i$ maximizes \eqref{true-probability:eq},
we upper-bound \eqref{true-probability:eq} by
\fi 
\begin{eqnarray}
\nonumber
\lefteqn{(2d)!\cdot
  \prob_{\ov\pi}\left[(\forall j\!\in\! J_i)\;\;
	    g_{2j}(i)=\mu(g_{\sigma_i(2j)}(\imu(i)))\right]} \\
&=& (2d)!\cdot \prod_{j\in J_i}
     \prob_{\pi_{j},\pi_{\ceil{\sigma_i(2j)/2}}}
          \left[g_{2j}(i)=\mu(g_{\sigma_i(2j)}(\imu(i))) \right]
\label{crude-bound:eq}
\end{eqnarray}
where the equality uses the disjointness of
the multi-sets $\{j,\ceil{\sigma_i(2j)/2}\}$
for the $j$'s in $J_i$.
Next, letting $\sigma'_{i}(2j)\eqdef{\ceil{\sigma_{i}(2j)/2}}$,
and $\sigma''_{i}(2j)\eqdef(-1)^{\sigma_{i}(2j)\bmod2}$
(so that $\sigma_{i}(2j)=2\sigma'_{i}(2j)-0.5(1-\sigma''_{i}(2j))$),
we upper-bound \eqref{crude-bound:eq} by
\begin{equation} \label{crude-bound:eq2}
(2d)!\cdot \prod_{j\in J_i}
	\prob_{\pi_{j},\pi_{\sigma'_i(2j)}}\left[
	\pi_{j}(i)=\mu(\pi^{\sigma''_i(2j)}_{\sigma'_i(2j)}(\imu(i)))
           \right]
\;<\; (2d)^{2d}\cdot(3/n)^{|J_i|},
\end{equation}
where $\prob_{\pi_j,\pi_j}[\cdot]$ stands for $\prob_{\pi_j}[\cdot]$
and $\pi^1$ stands for $\pi$,
while the inequality is justified by considering
the following three cases (w.r.t each $j\in J_i$).
\BE
\item If $k\eqdef\sigma'_i(2j)\neq j$
(equiv., $\sigma_i(2j)\not\in\{2j-1,2j\}$),
then, letting $b=\sigma''_i(2j)$,
the corresponding factor in the l.h.s of \eqref{crude-bound:eq2} is
$$\prob_{\pi_{j},\pi_{k}}\left[\pi_{j}(i) = \mu(\pi^b_k(\imu(i)))\right]$$
which equals $1/n$ by fixing $\pi_k$, letting $v=\mu(\pi^b_k(\imu(i)))$,
and using $\prob_{\pi_j}[\pi_j(i)\!=\!v]=1/n$.
\item If $\sigma_i(2j)=2j$,
then the corresponding factor in the l.h.s of \eqref{crude-bound:eq2} is
$$\prob_{\pi_{j}}\left[\pi_{j}(i) = \mu(\pi_j(\imu(i)))\right]$$
which is at most $1/(n-1)$ since $\mu(i)\neq i$;
specifically, fixing the value of $\pi_j(\imu(i))$,
and denoting this value by $v$,
leaves $\pi_j(i)$ uniformly distributed in $[n]\setminus\{v\}$,
which means that
$\prob_{\pi_j}[\pi_j(i)\!=\!\mu(v)\,|\,v\!=\!\pi_j(\imu(i))]\leq1/(n-1)$
(where equality holds if $\mu(v)\neq v$).
\item If $\sigma_i(2j)=2j-1$,
then the corresponding factor in the l.h.s of \eqref{crude-bound:eq2} is
$$\prob_{\pi_{j}}\left[\pi_{j}(i) = \mu(\pi_j^{-1}(\imu(i)))\right]$$
which is shown to be less than $3/n$.
In this case, we consider two sub-cases
depending on whether or not $\pi_j(i) = \imu(i)$,
while noting that the first case occurs with probability $1/n$ whereas
$\prob_{\pi_j}[\pi_{j}(i)=\mu(\pi_j^{-1}(\imu(i)))|\pi_j(i)\neq\imu(i)]
 \leq1/(n-1)$.
\EE
\ifnum\exclaims=1
Hence, each of the factors in the l.h.s of \eqref{crude-bound:eq2}
is upper-bounded by $3/n$, and the claim follows.
\Epf

\mypar{The general case.}
\fi
The same argument generalizes to any set $I\subseteq T$
such that $I\cap\mu(I)=\emptyset$.
In such a case we get
\ifnum\exclaims=0
\begin{eqnarray}
\lefteqn{\prob_{\ov\pi}\left[(\forall i\!\in\!I)\,
                    (\exists\sigma_i\!\in\!\Sym_{2d})\,
                    (\forall j\!\in\![2d])\;\;
                  g_j(i)=\mu(g_{\sigma_i(j)}(\imu(i)))\right]}
\nonumber \\
&\leq& (2d)!^{|I|}\cdot\max_{\sigma_1,\ldots,\sigma_n}
  \left\{\prob_{\ov\pi}\left[(\forall i\!\in\!I)\,(\forall j\!\in\![2d])
            \;\;g_j(i)=\mu(g_{\sigma_i(j)}(\imu(i)))\right] \right\}
\label{true-probability:eq2}  \\
&<& (2d)^{2d\cdot|I|}\cdot(2/(n-2(|I|-1)))^{|I|\cdot d/6},
\label{true-probability:eq3}
\end{eqnarray}
where the inequality uses the hypothesis that $I\cap\mu(I)=\emptyset$
and is justified analogously to the warm-up;
specifically, for every $i\in I=\{i_1,\ldots,i_m\}$,
\else 
\begin{eqnarray}
\lefteqn{\prob_{\ov\pi}\left[(\forall i\!\in\!I)\,
                    (\exists\sigma_i\!\in\!\Sym_{2d})\,
                    (\forall j\!\in\![2d])\;\;
                  g_j(i)=\mu(g_{\sigma_i(j)}(\imu(i)))\right]}
\nonumber \\
&\leq& (2d)!^{|I|}\cdot\max_{\sigma_1,\ldots,\sigma_n}
  \left\{\prob_{\ov\pi}\left[(\forall i\!\in\!I)\,(\forall j\!\in\![2d])
            \;\;g_j(i)=\mu(g_{\sigma_i(j)}(\imu(i)))\right] \right\}
\label{true-probability:eq2}
\end{eqnarray}
\Bcm{\em(Actual analysis):}
\label{random-works-actual:clm}
\eqref{true-probability:eq2} is upper-bounded by
\begin{equation} \label{true-probability:eq3}
(2d)^{2d\cdot|I|}\cdot(3/(n-2(|I|-1)))^{|I|\cdot d/6}.
\end{equation}
\Ecm

\Bpf
Fixing $\sigma_1,....,\sigma_n$
that maximize \eqref{true-probability:eq2},
we proceed as in the proof of Claim~\ref{random-works-warm-up:clm}.
For every $i\in I=\{i_1,\ldots,i_m\}$,
\fi
we fix a set $J_i$ of size at least $d/6$ such that
the multi-sets $\{j,\ceil{\sigma_i(2j)/2}\}$'s are disjoint,
and refer to the events $E_{j,k}(\pi_1,\ldots,\pi_{2d})$
that depend only on the value of $\pi_j$
and $\pi^{\sigma''_{i_k}\!(2j)}_{\sigma'_{i_k}\!(2j)}$
on the points $i_1,\ldots,i_{k-1}$ and $\imu(i_1),\ldots,\imu(i_{k-1})$,
respectively, where $\sigma'_{i}(2j)\eqdef{\ceil{\sigma_{i}(2j)/2}}$,
and $\sigma''_{i}(2j)\eqdef(-1)^{\sigma_{i}(2j)\bmod2}$
(as in the proof of Claim~\ref{random-works-warm-up:clm}).
Specifically, $E_{j,k}(\pi_1,\ldots,\pi_{2d})$ is the event
$$(\forall k'\!\in\![k-1])
             \;\;g_{2j}(i_{k'})=\mu(g_{\sigma_{i_{k'}}\!(2j)}(\imu(i_{k'})))$$
which can be written as
$$(\forall k'\!\in\![k-1]) \;\; \pi_j(i_{k'})
  =\mu(\pi^{\sigma''_{i_{k'}}\!(2j)}_{\sigma'_{i_{k'}}\!(2j)}(\imu(i_{k'}))).$$
Using the disjointness of the $\{j,\ceil{\sigma_i(2j)/2}\}$'s in $J_i$,
we upper-bound \eqref{true-probability:eq2} by

\begin{eqnarray}
\lefteqn{(2d)!^{m}\cdot\prod_{k\in[m]~} \prod_{~j\in J_{i_k}}
     \prob_{\pi_1,\ldots,\pi_{2d}}
          \left[g_{2j}(i_k)=\mu(g_{\sigma_{i_k}\!(2j)}(\imu(i_k)))\left|
                E_{j,k}(\pi_1,....,\pi_{2d}) \right.\right]}
\nonumber \\
&=& (2d)!^{m}\cdot\prod_{k\in[m]~} \prod_{~j\in J_{i_k}}
     \prob_{\pi_1,\ldots,\pi_{2d}}
          \left[\pi_{j}(i_k)
      =\mu(\pi^{\sigma''_{i_k}\!(2j)}_{\sigma'_{i_k}\!(2j)}
	  (\imu(i_k)))\left|E_{j,k}(\pi_1,....,\pi_{2d}) \right.\right] .
\label{true-probability:eq4}
\end{eqnarray}
Now, when analyzing the foregoing conditional probability
in \eqref{true-probability:eq4},
we consider two cases for each $k\in[m]$ and $j\in J_{i_k}$.
If $j\neq\sigma'_{i_k}\!(2j)$, then
we fix the value of each of these two permutations
(i.e., $\pi_j$ and $\pi_{\sigma'_{i_k}\!(2j)}$)
on the corresponding $k-1$ points that occur in the condition $E_{j,k}$,
and the value of these permutations on the $k^\xth$ points
(i.e., $i_k$ and $\imu(i_k)$) is restricted accordingly
(i.e., to the remaining $n-(k-1)$ values).
Otherwise (i.e., $j=\sigma'_{i_k}\!(2j)$),
we fix the value of $\pi_j$ on these $2(k-1)$ points.
Hence, the argument in the warm-up analysis applies with $n$
replaces by either $n-(k-1)$ or $n-2(k-1)$.
It follows that \eqref{true-probability:eq4} is upper-bounded by
$$(2d)!^{m}\cdot\prod_{k\in[m]~}(3/(n-2(m-1)))^{|J_{i_k}|}.$$
Using $|J_{i_k}|\geq d/6$ for every $k\in[m]$,
\ifnum\exclaims=1
the claim follows.
\Epf
\else
we have established the upper-bound asserted in \eqref{true-probability:eq3}.
\fi
\medskip

Recall that \eqref{true-probability:eq3} refers to
a fixed set $I\subseteq T$ such that $I\cap\mu(I)=\emptyset$,
and that it constitutes an upper bound
on the probability (over the choice of $\ov\pi$)
that, for each $i\in I$ there exists
a permutation $\sigma_i:[2d]\to[2d]$
such that $g_j(i)=\mu(g_{\sigma_i(j)}(\imu(i)))$
holds for all $j\in[2d]$.
This upper bound
(i.e., $(2d)^{2d\cdot|I|}\cdot(3/(n-2(|I|-1)))^{|I|\cdot d/6}$)
simplifies to $(2d)^{2d\cdot|I|}\cdot(5/n)^{|I|\cdot d/6}$,
provided that $|I|\leq\ceil{n/6}$.


Recalling that $t\eqdef|T|\in[n]$,
we shall upper-bound the probability (over the choice of $\ov\pi$)
that $T$ contains a $\ceil{t/2}$-subset $T'$
such that for each $i\in T'$ there exists
a permutation $\sigma_i:[2d]\to[2d]$
such that $g_j(i)=\mu(g_{\sigma_i(j)}(\imu(i)))$
holds for all $j\in[2d]$.
We do so by taking a union bound over all $\ceil{t/6}$-subsets $I$
such that $I\cap\mu(I)=\emptyset$
and for each $i\in I$ there exists a permutation $\sigma_i:[2d]\to[2d]$
such that $g_j(i)=\mu(g_{\sigma_i(j)}(\imu(i)))$ holds for all $j\in[2d]$.
(Note that such a $\ceil{t/6}$-subset $I$ exists
in each $\ceil{t/2}$-subset $T'$, and that $\ceil{t/6}<n/3$.)
Using the aforementioned simplified form of \eqref{true-probability:eq3},
we conclude that, with probability at most
$$\binom{t}{\ceil{t/6}}
   \cdot(2d)^{2d\cdot\ceil{t/6}}\cdot(5/n)^{\ceil{t/6}\cdot{d/6}}
   \;<\; 2^t\cdot(5\cdot(2d)^{12}/n)^{\ceil{t/6}\cdot d/6}
   \;=\; \exp(-\Omega(dt\log n))$$
over the choice of $\ov\pi$,
the set $T$ contains no $\ceil{t/6}$-subset $I$ as above.
This means that, with probability at most $\exp(-\Omega(d t\log n))$,
less than ${t/2}$ of the indices $i\in T$ contribute
a non-zero number of units to the difference
(between $G_{\ov \pi}$ and $\mu(G_{\ov \pi})$).

Letting $c'=1/2$ and
considering all (non-trivial) permutations $\mu:[n]\to[n]$,
we conclude that the probability, over the choice of $\ov\pi$,
that $G_{\ov\pi}$ is not $c'$-robustly self-ordered is at most
\begin{eqnarray*}
\sum_{t\in[n]}\binom{n}{t}\cdot\exp(-\Omega(d t\log n))
&=& \sum_{t\in[n]}\exp(-\Omega((d-O(1))\cdot t\log n)) \\
&=& \exp(-\Omega((d-O(1))\cdot \log n)),
\end{eqnarray*}
and the claim follows for the permutation model
(and for any sufficiently large $d$).

As stated upfront,
using the general result of~\cite[Thm.~1.3]{GJKW},
we infer that a uniformly distributed $2d$-regular $n$-vertex
multi-graph fails to be $c'$-robustly self-ordered with probability $o(1)$.
Lastly, recalling that such a $2d$-regular multi-graph is actually
a simple graph with probability $\exp(-((2d)^2-1)/4)$,
the theorem follows.
\EPF

\paragraph{Digest.}
The proof of Theorem~\ref{random-works:thm} is quite similar
to the proof Claim~\ref{random-colored:clm}, but it faces
two complications that were avoided in the prior proof
(by using edge-colors and implicitly directed edges).
Most importantly, the current proof has to handle equality
between multi-sets instead of equality between sequences.
This is done by considering all possible ordering of these multi-sets,
which amounts to taking a union bound over all possible ordering
and results in more complicated analysis and notation.
(Specifically, see the introduction of $\sigma_i$'s and $J_i$'s
and the three cases analyzed in the warm-up.)
In addition, since edges are defined by permutations over the vertex-set
rather than by perfect matching, we have to consider both the forward
and backward direction of each permutation,
which results in further complicating the analysis and the notation.
(Specifically, see the introduction of $\sigma'_i$'s and $\sigma''_i$'s
and the three cases analyzed in the warm-up.)

\paragraph{An alternative proof of Theorem 
\protect\ref{construction:thm}.}
We mention that combining an extension of Theorem~\ref{random-works:thm}
with some of the ideas underlying the proof of Theorem~\ref{construction:thm}
yields an alternative proof of Theorem~\ref{construction:thm}
(i.e., an alternative construction of robustly self-ordered
bounded-degree graphs).

\BR[An alternative construction
of $d$-regular robustly self-ordered graphs]
\label{random-works-alt:rem}
On input~$1^n$, we set $\ell=\frac{O(\log n)}{\log\log n}$,
and proceeds in three steps.
\BE
\item
Extending the proof of Theorem~\ref{random-works:thm},
we show that for all sufficiently large constant~$d$,
for any set $\cal G$ of $t=t(\ell)<n=\ell^{\Omega(\ell)}$
{\rm($2d$-regular)} $\ell$-vertex graphs,
with probability $1-o(1)$,
a random $2d$-regular $\ell$-vertex graph is both robustly self-ordered
and far from being isomorphic to any graph in $\cal G$.
Note that, with probability $1-o(1)$, such a graph is also expanding.

Here two $\ell$-vertex graphs are said to be {\sf far apart}
if they disagree on $\Omega(\ell)$ vertex-pairs.

The proof of Theorem~\ref{random-works:thm} is extended
by considering, for a random graph, the event that it is
either not robustly self-ordered or is not far from
an isomorphic copy of one of the $t$ {\rm(fixed)} graphs.
The later event
{\em(i.e., being close to isomorphic to one of these graphs)}
occurs with probability $o(t/n)$.
\item
Relying on Step~1, we find a sequence of $n/\ell$
robustly self-ordered $2d$-regular $\ell$-vertex graphs that are
expanding and pairwise far from being isomorphic to one another.

This is done by iteratively finding
robustly self-ordered $2d$-regular $\ell$-vertex expanding graphs
that are far from being isomorphic to all prior ones,
where scanning all possible graphs and checking the condition
can be done in time $n\cdot\ell^{d\ell/2}\cdot(\ell!)=\poly(n)$.
\item
Using the sequence of $n/\ell$ graphs found in Step~2,
we consider the $n$-vertex graph that consists of
these $\ell$-vertex graphs as its connected components,
and use parts of the proof of Theorem~\ref{construction:thm}
to show that this graph is robustly self-ordered.
Specifically, we only need to consider cases that
are analogous to Cases~2, 6 and~7.
The treatment of the analogous cases is slightly simpler
than in the proof of Theorem~\ref{construction:thm},
since the graphs are somewhat simpler.
\EE
Note that the resulting graphs are not locally constructable.
\ER

\part{The Case of Dense Graphs}
\label{dense:part}
Recall that when considering graphs of unbounded degree,
we ask whether we can obtain unbounded robustness parameters.
In particular, we are interested in $n$-vertex graphs
that are $\Omega(n)$-robustly self-ordered,
which means that they must have $\Omega(n^2)$ edges.

In Section~\ref{dense-basics:sec} we prove the existence
of $\Omega(n)$-robustly self-ordered $n$-vertex graphs,
and show that they imply $\Omega(1)$-robustly self-ordered
bounded-degree $O(n^2)$-vertex graphs.
In Section~\ref{dense+nmE:sec},
we reduce the construction of the former (dense) $n$-vertex graphs
to the construction of non-malleable two-source extractors
(with very mild parameters).
We actually show two reductions:
The first reduction (presented in Section~\ref{dense:rso1:sec})
requires the extractors to have an additional natural feature,
called quasi-orthogonality, and yields a construction
of such $n$-vertex graphs that runs in $\poly(n)$-time.
The second reduction (presented in Section~\ref{dense:rso2:sec})
does not make this additional requirement,
and yields an algorithm that computes the adjacency predicate
of such $n$-vertex graphs in $\poly(\log n)$-time.


In Section~\ref{dense-pt:sec} we demonstrate the applicability
of $\Omega(n)$-robustly self-ordered $n$-vertex graphs to property testing;
specifically, to proving lower bounds (on the query complexity)
for the dense graph testing model.
Lastly, in Section~\ref{inter-deg:sec},
we consider the construction of $\Omega(d(n))$-robustly
self-ordered $n$-vertex graphs of maximum degree $d(n)$,
for every $d:\N\to\N$ such that $d(n)\in[\Omega(1),n]$.

\section{Existence and Transformation to Bounded-Degree Graphs}
\label{dense-basics:sec}
It seems easier to prove that random $n$-vertex graphs
are $\Omega(n)$-robustly self-ordered
(see Proposition~\ref{dense:random:clm})
than to prove that random bounded-degree graphs
are $\Omega(1)$-robustly self-ordered
(or even just prove that such bounded-degree graphs exist).
In contrast, it seems harder
to construct $\Omega(n)$-robustly self-ordered $n$-vertex graphs than
to construct $\Omega(1)$-robustly self-ordered bounded-degree graphs.
In particular, we show that $\Omega(n)$-robustly
self-ordered $n$-vertex graphs can be easily transformed
into $O(n^2)$-vertex bounded-degree graphs that
are $\Omega(1)$-robustly self-ordered
(see Proposition~\ref{deg-reduction:clm}).
We stress that the construction of
robustly self-ordered bounded-degree graphs
that is obtained by combining the foregoing transformation
with Theorem~\ref{dense:ithm} is entirely different from
the constructions presented in the first part of the paper.

\paragraph{Random graphs are robustly self-ordered.}
We first show that, with very high probability,
a random $n$-vertex graph $G_n=([n],E_n)$,
where $E_n$ is a uniformly distributed subset of $\binom{[n]}{2}$,
is $\Omega(n)$-robustly self-ordered.

\BP[Robustness analysis of a random graph]
\label{dense:random:clm}
A random $n$-vertex graph $G_n=([n],E_n)$
is $\Omega(n)$-robustly self-ordered
with probability $1-\exp(-\Omega(n))$.
\EP
As stated above, the following proof is significantly easier
than the proof provided for the bounded-degree analogue
(i.e., Theorem~\ref{random-works:thm}).
\medskip

\BPF
For each (non-trivial) permutation $\mu:[n]\to[n]$,
letting $T\eqdef\{i\!\in\![n]\!:\!\mu(i)\neq i\}$
denote its (non-empty) set of non-fixed-points,
we show that, with probability $1-\exp(-\Omega(n\cdot|T|))$,
the size of the symmetric different between
a random $n$-vertex graph $G_n=([n],E_n)$
and $\mu(G_n)$ is $\Omega(n\cdot |T|)$.

For every $u,v\in[n]$ such that $u<v$,
let $\ee_{u,v}=\ee_{u,v}^\mu(G_n)$ represent the event that
{\em the pair $(\mu(u),\mu(v))$ contributes to
the symmetric difference between $G_n$ and $\mu(G_n)$};
that is, $\ee_{u,v}=1$ if exactly one of the edges $\{\mu(u),\mu(v)\}$
and $\{u,v\}$ is in $G_n$, since $\{u,v\}$ is an edge of $G_n$
if and only if $\{\mu(u),\mu(v)\}$ is an edge of $\mu(G_n)$.
Note that $\prob[\ee_{u,v}(G_n)\!=\!1]\,=\,1/2$
if $\{u,v\}\neq\{\mu(u),\mu(v)\}$,
because fixing the adjacency relation of the pair $(\mu(u),\mu(v))$
leaves the adjacency relation of pair $(u,v)$ totally random.
%
%
We shall prove that
\begin{equation}\label{T-claim:eq}
\prob_{G_n}
 \left[\sum_{u<v\in[n]}\ee_{u,v}^\mu(G_n)\,<\,\frac{n\cdot|T|}{20}\right]
\;=\; \exp(-\Omega(n\cdot|T|)).
\end{equation}
We prove \eqref{T-claim:eq} by identifying a set~$S$
of $\Omega(n\cdot|T|)$ vertex pairs such that
the random variables in $\{\ee_{u,v}^\mu(G_n):\{u,v\}\in S\}$
are totally independent and uniformly distributed in $\bitset$.
Specifically if~$S$ and $S'=\{\{\mu(u),\mu(v)\}:\{u,v\}\in S\}$
are disjoint, then fixing the adjacencies of the pairs in $S'$
leaves the adjacency of pairs in~$S$ totally random.

Fixing a $\ceil{|T|/3}$-subset $I\subseteq T$
such that $I\cap\mu(I)=\emptyset$,
let $J=[n]\setminus(I\cup\imu(I))$.
Note that $\mu(I\cup J)\subseteq[n]\setminus I$
and that $|J|=n-2\cdot\ceil{|T|/3})\geq (n/3)-2$.
Observe that, for every $(u,v)\in J\times I$,
it holds that $u\neq v$
and $\prob[\ee_{u,v}\!=\!1]\,=\,1/2$,
where the equality is due to $\{u,v\}\neq\{\mu(u),\mu(v)\}$,
which holds since $(u,v)\!\in\!J\times I$
but $\mu(u),\mu(v)\in[n]\setminus I$.
Furthermore, the events the correspond to the pairs in $J\times I$
are independent, because the sets $\{\{u,v\}\!:\!(u,v)\!\in\!J\times I\}$
and $\{\{\mu(u),\mu(v)\}\!:\!(u,v)\!\in\!J\times I\}$ are disjoint;
that is, $(u,v)\in J\times I$
implies $(\mu(u),\mu(v))\in([n]\setminus I)\times([n]\setminus I)$.
%
Hence (using $n\leq3(|J|+2)$ and $|T|\leq3|I|$
(as well as $3(|J|+2)\cdot3|I|<9.9\cdot|J|\cdot|I|$)),
the l.h.s. of \eqref{T-claim:eq} is upper-bounded by
\begin{eqnarray*}
\prob_{G_n}\left[\sum_{(u,v)\in J\times I}
    \ee_{u,v}^\mu(G_n)\,<\,\frac{3(|J|+2)\cdot3|I|}{20}\right]
&\leq& \prob_{G_n}\left[\sum_{(u,v)\in J\times I}
    \ee_{u,v}^\mu(G_n)\,<\,\frac{0.99\cdot|J|\cdot|I|}{2}\right] \\
&=& \exp(-\Omega(|J|\cdot|I|))
\end{eqnarray*}
which is $\exp(-\Omega(n\cdot|T|))$.
Having established \eqref{T-claim:eq},
the claim follows by a union bound
(over all non-trivial permutations $\mu:[n]\to[n]$);
specifically, denoting the set of non-trivial permutations by $P_n$,
we upper-bound the probability that $G_n$ is not $\frac{n}{20}$-robust by
\begin{eqnarray*}
\sum_{\mu\in P_n}
\lefteqn{\prob_{G_n}\left[
   \mbox{\rm $\mu$ violates the condition in \eqref{T-claim:eq}}\right]} \\
&\leq& \sum_{t\in[n]}\binom{n}{t}\cdot(t!)\cdot \exp(-\Omega(n\cdot t)) \\
&<& n\cdot\max_{t\in[n]}\{n^t\cdot \exp(-\Omega(n\cdot t))\} \\
&=& \exp(-\Omega(n))
\end{eqnarray*}
where $t$ represents the size of the set of non-fixed-points
(w.r.t $\mu$).
\EPF

\paragraph{Obtaining bounded-degree robustly self-ordered graphs.}
We next show how to transform $\Omega(n)$-robustly
self-ordered $n$-vertex graphs to $O(n^2)$-vertex bounded-degree graphs
that are $\Omega(1)$-robustly self-ordered.
Essentially, we show that the standard ``degree reduction via expanders''
technique works (when using a different color for the expanders' edges,
and then using gadgets to replace colored edges).
%
Specifically, we replace each vertex in $G_n=([n],E_n)$
by an $(n-1)$-vertex expander graph and connect each
of these vertices to at most one vertex in a different expander,
while coloring the edges of the expanders with~1,
and coloring the other edges by~2.
Actually, the vertex $v$ is replaced by
the vertex-set $C_v=\{\ang{v,u}:u\!\in\![n]\setminus\{v\}\}$
and in addition to the edges of the expander, colored~1,
we connect each vertex $\ang{v,u}\in C_v$ to the vertex $\ang{u,v}\in C_u$
and color this edge~2 if $\{u,v\}\in E_n$ and~0 otherwise.%
\footnote{This is equivalent to first converting $G_n$ into a $n$-vertex
clique while coloring an edge~2 if and only if it is in $E_n$.}
This yields an $n\cdot(n-1)$-vertex $O(1)$-regular graph, denoted $G'_n$,
coupled with an edge-coloring, denoted $\chi'$, which uses three colors.
We stress that
each vertex in $G'_n$ is incident to a single even-colored edge.
Using the hypothesis that $G_n$ is $\Omega(n)$-robustly self-ordered,
we prove that $(G'_n,\chi')$ is $\Omega(1)$-robustly self-ordered
(in the colored sense).

\BP[Robustness analysis of the degree reduction]
\label{deg-reduction:clm}
If $G_n$ is $\Omega(n)$-robustly self-ordered,
then $(G'_n,\chi')$ is $\Omega(1)$-robustly self-ordered
{\rm(in the colored sense of Definition~\ref{colored-robust:def})}.
\EP
Using Theorem~\ref{colored2std:clm} (after adding self-loops),
we obtain a $O(1)$-regular $O(n^2)$-vertex graph that
is $\Omega(1)$-robustly self-ordered (in the standard sense).
\medskip

\BPF
Denoting the vertex-set of $G'_n$ by $V=\bigcup_{v\in[n]}C_v$,
we consider an arbitrary (non-trivial) permutation $\mu':V\to V$,
and the corresponding set of non-fixed-points $T'$.
Intuitively, if $\mu'$ maps vertices of $C_v$ to several $C_w$'s,
then we get a proportional contribution to the difference
between~$G'_n$ and $\mu'(G'_n)$ by the (1-colored) edges of the expander.
Otherwise, $\mu'$ induces a permutation $\mu$ over the vertices of $G_n$,
and we get a corresponding contribution via the (2-colored) edges of $G_n$.
Lastly, non-identity mappings inside the individual $C_v$'s are charged
using the (even-colored) edges that connect different $C_v$'s,
while relying on the fact each vertex in $G'_n$ is incident to
a single even-colored edge.
Details follow.

For a permutation $\mu':V\to V$ as above,
let $\mu:[n]\to[n]$ be a permutation that maximizes
the (average over $v\in[n]$ of the) number of vertices in $C_v$
that are mapped by $\mu'$ to vertices in~$C_{\mu(v)}$;
that is, for every permutation $\nu:[n]\to[n]$, it holds that
\begin{equation}\label{mu:eqdef}
\left|\left\{\ang{v,u}\!\in\!V:\mu'(\ang{v,u})\in C_{\mu(v)}\right\}\right|
\;\geq\;
\left|\left\{\ang{v,u}\!\in\!V:\mu'(\ang{v,u})\in C_{\nu(v)}\right\}\right|.
\end{equation}
We consider the following three cases.

\BDes
\item[{\em Case 1}:] $\sum_{v\in[n]}|B_v|=\Omega(|T'|)$,
where $B_v\eqdef\{\ang{v,u}\!\in\!C_v:\mu'(\ang{v,u})\not\in C_{\mu(v)}\}$.

(This refers to the case that many vertices are mapped by $\mu'$
to an expander that is different from the one designated by $\mu$,
which represents the best possible mapping of whole expanders.)

Letting $C_{v,w}\eqdef\{\ang{v,u}:\mu'(\ang{v,u})\in C_w\}$,
we first observe that for every $v$ it holds that
$\max_{w\neq\mu(v)}\{|C_{v,w}|\}\leq\frac23\cdot(n-1)$,
because otherwise we reach a contradiction to the maximality of $\mu$
by defining $\nu(v)=w$ and $\nu(\imu(w))=\mu(v)$,
where $w$ is the element obtaining the maximum,
and $\nu(x)=\mu(x)$ otherwise.

Next, observe that there exists $W_v\subseteq[n]\setminus\{\mu(v)\}$
such that $B'_v=\bigcup_{w\in W_v}C_{v,w}$
satisfies both $|B'_v|\leq\frac23\cdot(n-1)$ and $|B'_v|\geq|B_v|/3$.
Now, consider the sets $B'_v$ and $C_v\setminus B'_v$:
On the one hand, in $\mu'(G'_n)$
there are $\Omega(|B'_v|)$ 1-colored edges
connecting $\mu'(B'_v)$ and $\mu'(C_v\setminus B'_v)$,
due to the subgraph of $\mu'(G'_n)$ induced by $\mu'(C_v)$
which equals subgraph of $G'_n$ induced by $C_v$
(which, in turn, is an expander).
On the other hand, in $G'_n$ there are no 1-colored edges
between $\mu'(B'_v)$ and $\mu'(C_v\setminus B'_v)$,
since $\mu'(B'_v)\subseteq\bigcup_{w\in W_v} C_w$
and $\mu'(C_v\setminus B'_v)\subseteq\bigcup_{w\in[n]\setminus W_v} C_w$.

We conclude that, in this case,
the difference between $G'_n$ and $\mu'(G_n)$
is $\sum_v\Omega(|B'_v|)=\sum_v\Omega(|B_v|)=\Omega(|T'|)$.

\item[{\em Case 2}:] $\sum_{v\in[n]:\mu(v)\neq v}|C'_v|=\Omega(|T'|)$,
where $C'_v\eqdef\{\ang{v,u}\!\in\!C_v:\mu'(\ang{v,u})\in C_{\mu(v)}\}$.

(This refers to the case that many vertices are mapped by $\mu'$
to an expander that is designated by $\mu$,
but this expander is not the one in which they reside
(i.e., $\mu$ has many non-fixed-points).)

Letting $\gamma>0$ be a constant such that $G_n$
is $\gamma\cdot n$-robustly self-ordered,
we may assume that
$\sum_{v\in[n]:\mu(v)\neq v}|C'_v|
 \geq(1-0.5\cdot\gamma)\cdot\sum_{v\in[n]:\mu(v)\neq v}|C_v|$,
since otherwise we are done by Case~1.

By the $\gamma n$-robust self-ordering of $G_n$,
the difference between $G_n$ and $\mu(G_n)$
is at least $\Delta\eqdef\gamma n\cdot|\{v\in[n]:\mu(v)\neq v\}|$.
Assuming, for a moment,
that $\mu'(C_v)=C_v$ for every $v$ such that $\mu(v)\neq v$,
the difference between $G'_n$ and $\mu'(G'_n)$ is $\Delta$,
where the difference is due to edges colored~2
(i.e., the edges inherited from $G_n$).
This amount is prorotional to the number of vertices
in the current case, since
$$\Delta\;=\;\frac{\gamma n}{n-1}\cdot\sum_{v:\mu(v)\neq v}|C_v|
 \;>\; \gamma\cdot\sum_{v:\mu(v)\neq v}|C_v|.$$
In general, $\mu'(C_v)=C_v$ may not hold for some $v$,
and in this case we may loss the contribution
of the 2-colored edges incident at vertices
in $\bigcup_{v\in[n]:\mu(v)\neq v}(C_v\setminus C'_v)$.
Recalling that (by our hypothesis) the size of this set
is at most $0.5\cdot\gamma\cdot\sum_{v:\mu(v)\neq v}|C_v|$,
we are left with a contribution of
at least $0.5\gamma\cdot\sum_{v:\mu(v)\neq v}|C'_v|$.

We conclude that, in this case,
the difference between $G'_n$ and $\mu'(G_n)$
is $\Omega(\sum_{v:\mu(v)\neq v}|C'_v|)=\Omega(|T'|)$.

%

\item[{\em Case 3}:] $\sum_{v\in[n]}|C''_v|=\Omega(|T'|)$, where
$C''_v
 \eqdef\{\ang{v,u}\!\in\!C_v:\mu'(\ang{v,u})\in C_v\setminus\{\ang{v,u}\}\}$.

(This refers to the case that many vertices are mapped by $\mu'$
to a different vertex in the same expander in which they reside.)%
\footnote{Note that
if $\ang{v,u}\in C_v$ is not mapped by $\mu'$ to $C_v$,
then either $\mu'(\ang{v,u})\not\in C_{\mu(v)}$ holds (i.e., Case~1)
or $\mu'(\ang{v,u})\in C_{\mu(v)}$ such that $\mu(v)\neq v$ (i.e., Case~2).
Hence, if $\ang{u,v}\in T'$ is not counted in Cases~1 and~2,
then it must be counted in Case~3.\label{deg-reduction:fn}}

(This case would have been easy to handle if the expanders used
on the $C_v$'s were robustly self-ordered.
Needless to say, we want to avoid such an assumption.
Instead, we rely on the fact that in $G'_n$ different
vertices in $C_v$ are connected to different $C_u$'s.)

We may assume that
$\sum_{v\in[n]}|C''_v|
 \geq2\cdot\sum_{v\in[n]}|\{\ang{v,u}\!\in\!C_v:\mu'(\ang{v,u})\not\in C_v\}|$,
since otherwise we are done by either Case~1 or Case~2
(see Footnote~\ref{deg-reduction:fn}).
Now, consider a generic $\ang{v,u}\in C''_v$,
and let $w\neq u$ be such that $\mu'(\ang{v,u})=\ang{v,w}$.
Then, in $\mu'(G'_n)$ an edge colored either~0 or~2
connects $\ang{v,w}=\mu'(\ang{v,u})$ to $\mu'(\ang{u,v})$,
since $\ang{v,u}$ and $\ang{u,v}$ are so connected in $G'_n$,
whereas in $G'_n$ an even-colored edge
connects $\ang{v,w}$ to $\ang{w,v}\in C_w$.
Recall, however, that $\ang{v,w}$ is incident
to a single even-colored edge.
We consider two sub-cases.
\BI
\item If $\mu'(\ang{u,v})\in C_u$,
then $\ang{v,w}$ contributes to the difference
between $\mu'(G'_n)$ and $G'_n$,
because in $\mu'(G'_n)$ vertex $\ang{v,w}$ is connected
(by its unique even-colored edge) to a vertex in $C_u$
whereas in $G'_n$ vertex $\ang{v,w}$ is connected
(by its unique even-colored edge) to a vertex in $C_w$.

(Recall that $w$ is uniquely determined by $\ang{v,u}\in C''_n$,
since $\mu'(\ang{v,u})=\ang{v,w}$, and so this contribution
can be charged to $\ang{v,u}$.)
\item If $\mu'(\ang{u,v})\not\in C_u$,
then $\ang{u,v}$ resides in the set
$\bigcup_{x\in[n]}\{\ang{x,y}\!\in\!C_x:\mu'(\ang{x,y})\not\in C_x\}$,
which (by the hypothesis) has size at most $0.5\cdot\sum_{v\in[n]}|C''_v|$
\EI
Hence, at least half of $\bigcup_{v\in[n]}C''_v$
appears in the first sub-case, which implies that,
in this case, the difference between $G'_n$ and $\mu'(G_n)$
is at least $\frac12\cdot\sum_{v\in[n]}|C''_v|=\Omega(|T'|)$.
\EDes
Hence, the difference between $G'_n$ and $\mu'(G_n)$ is $\Omega(|T'|)$.
\EPF

\section{Relation to Non-Malleable Two-Source Extractors}
\label{dense+nmE:sec}
For $n=2^\ell$,
we reduce the construction of $\Omega(n)$-robustly self-ordered
(dense) $n$-vertex graphs
to the construction of non-malleable two-source extractors
for $(\ell,\ell-O(1))$-sources.
Recall that a random variable $X$ is called an {\sf $(\ell,k)$-source}
if $X$ is distributed over $[2^\ell]$ and has min-entropy at least $k$
(i.e., $\prob[X\!=\!i]\leq2^{-k}$ for every $i\in[2^\ell]$).%
\footnote{Indeed, for the sake of simplicity (of our arguments),
we do not require that $\ell\in\N$,
but rather only that $2^\ell\in\N$;
consequently, we consider distributions over $[2^\ell]$
rather than over $\bitset^\ell$.}
A function ${\tt E}:[2^\ell]\times[2^\ell]\to\bitset^m$
is called a (standard) {\sf two-source $(k,\e)$-extractor}
if, for every two independent $(\ell,k)$-sources $X$ and $Y$,
it holds that ${\tt E}(X,Y)$ is $\e$-close to the uniform
distribution over $\bitset^m$, denoted $U_m$.
Our notion of a non-malleable two-source extractor, presented next,
is one of (the weakest of) the notions considered in~\cite{CG:nmE,CGL}.

\BD[Non-malleable two-source extractors~{\cite[Def.~1.3]{CGL}}]
\label{nmExtractor:def}
A function ${\nmE}:[2^\ell]\times[2^\ell]\to\bitset^m$
is called a {\sf non-malleable two-source $(k,\e)$-extractor}
if, for every two independent $(\ell,k)$-sources $X$ and $Y$,
and for every two functions $f,g:[2^\ell]\to[2^\ell]$
such that at least one of them has no fixed-points,
it holds that $(\nmE(X,Y),\nmE(f(X),g(Y)))$ is $\e$-close
to $(U_m,\nmE(f(X),g(Y))$; that is,
\begin{equation}\label{nm-extractor:eq}
\frac12\cdot\sum_{\alpha,\beta}
    \left|\prob[(\nmE(X,Y),\nmE(f(X),g(Y)))\!=\!(\alpha,\beta)]
          -2^{-m}\cdot\prob[\nmE(f(X),g(Y))\!=\!\beta]\right|
 \;\leq\;\e.
\end{equation}
The parameter $\e$ is called the {\sf error} of the extractor.
\ED
We shall be interested in the special case in which $f$ and $g$
are permutations. In this case, the foregoing condition
(i.e., \eqref{nm-extractor:eq})
can be replaced by requiring that $(\nmE(X,Y),\nmE(f(X),g(Y)))$
is $2\e$-close to the uniform distribution over $\bitset^{m+m}$.%
\footnote{In this case, $f(X)$ and $g(Y)$ have min-entropy at
least $k$, which implies that $\nmE(f(X),g(Y))$ is $\e$-close
to the uniform distribution over $\bitset^{m}$.}
Furthermore, we shall focus on non-malleable two-source $(k,\e)$-extractors
that output a single bit (i.e., $m=1$),
and in this case non-triviality mandates $\e<0.5$.
In general, we view $\e$ as a constant,
but view $\ell$ and $k$ as varying (or generic) parameters,
and focus on the case of $k=\ell-O(1)$.

Recall that constructions of non-malleable two-source $(k,\e)$-extractors
with much better parameters are known~\cite[Thm.~1]{CGL}.
In particular, these constructions support $k=\ell-\ell^{\Omega(1)}$,
negligible error (i.e., $\e=\exp(-\ell^{\Omega(1)})$),
and $m=\ell^{\Omega(1)}$.
We stress that, as is the norm in the context of randomness extraction,
the extracting function is computable in polynomial-time
(i.e., in $\poly(\ell)$-time).

We shall show that any non-malleable two-source $(\ell-O(1),0.49)$-extractor
(for sources over $[2^\ell]$)
yields $\Omega(2^\ell)$-robustly self-ordered $O(2^\ell)$-vertex graphs.
Actually, we shall show two such constructions:
The first construction runs in $\poly(2^\ell)$-time,
and the second construction provides strong constructability
(a.k.a local computability) as claimed in Theorem~\ref{dense:ithm}.
Both constructions use a similar underlying reasoning,
which is more transparent in the first construction.

\subsection{The first construction}
\label{dense:rso1:sec}
For the first construction, we need the extractor to satisfy
the following natural (and quite minimal) requirement,
which we call quasi-orthogonality.
We say that an extractor ${\nmE}:[2^\ell]\times[2^\ell]\to\bitset$
is {\sf quasi-orthogonal} (with error $\e$)
if the following conditions hold:
\BE
\item
{\em The residual function obtained from $\nmE$ by any fixing
of one of its two arguments is almost unbiased}:
For every $x\in[2^\ell]$ and every $\sigma\in\bitset$ it holds that
$|\{y\!\in\![2^\ell]:\nmE(x,y)\!=\!\sigma\}|\leq(0.5+\e)\cdot2^{\ell}$;
ditto for every $y\in[2^\ell]$
and the corresponding set $\{x\!\in\![2^\ell]:\nmE(x,y)\!=\!\sigma]\}$.
\item
{\em The residual functions obtained from $\nmE$ by any two different
fixings of one of its two arguments are almost uncorrelated}:
For every $\{x,x'\}\in\binom{[2^\ell]}{2}$ it holds that
$|\{y\!\in\![2^\ell]:\nmE(x,y)\!\neq\!\nmE(x',y)\}|\geq(0.5-\e)\cdot2^{\ell}$;
ditto for every $\{y,y'\}\in\binom{[2^\ell]}{2}$
and the corresponding set $\{x\!\in\![2^\ell]:\nmE(x,y)\!\neq\!\nmE(x,y')]\}$.
\EE
As shown in Proposition~\ref{nmE-made-nice:prop},
any non-malleable two-source $(k,\e)$-extractor
can be transformed (in $\poly(2^\ell)$-time)
to a quasi-orthogonal one
at a cost of a small degradation in the parameters
(i.e.,~$\e$ increases by an additive term of $O(2^{-(\ell-k)})$
and $2^\ell$ decreases by an additive term of $O(2^{k})$).
Note that $\poly(2^\ell)$-time is acceptable
when one aims at constructing $O(2^\ell)$-vertex graphs;
however, aiming at strong/local constructability
(as in Theorem~\ref{dense:ithm}),
we shall avoid such a transformation in the second construction
(presented in Section~\ref{dense:rso2:sec}).

\BP[Transforming non-malleable two-source extractors
into ones that are quasi-orthogonal]
\label{nmE-made-nice:prop}
For every $k\leq\ell-3$,
there exists a $\poly(2^\ell)$-time transformation
that given a non-malleable two-source $(k,\e)$-extractor
${\nmE}:[2^\ell]\times[2^\ell]\to\bitset$,
returns a non-malleable two-source $(k,\e')$-extractor
${\nmE'}:[n']\times[n']\to\bitset$ such that $n'\geq2^\ell-O(2^k)$
and $\nmE'$ is quasi-orthogonal with error $\e'=\e+O(2^k/n')$.
\EP

\BPF
Essentially, $\nmE'$ is obtained from $\nmE$ by simply discarding inputs
that violate the quasi-orthogonality conditions.
Letting $n=2^\ell$, first note that the number of $x$'s that violate
the first condition is at most $2^{k+1}$,
because otherwise we obtain a contradiction
to the hypothesis that $\nmE$ is a two-source $(k,\e)$-extractor
(by letting $X$ be uniform on the $x$'s that satisfy
$|\{y\!\in\![n]:\nmE(x,y)\!=\!\sigma\}|>(0.5+\e)\cdot n$
for either $\sigma=0$ or $\sigma=1$, and $Y$ be uniform on $[n]$).
Next, consider the
residual $(k,\e)$-extractor $\nmE_1:[n_1]\times[n_1]\to\bitset$,
where $n_1\geq n-2^{k+1}$, obtained by omitting the exceptional $x$'s.
Note that $\nmE_1$ satisfies the first quasi-orthogonality condition
with respect to the first argument with error $\e$.
Doing the same for the second argument yields
a residual $(k,\e)$-extractor $\nmE_2:[n_2]\times[n_2]\to\bitset$,
where $n_2\geq n_1-2^{k+1}$ and $\nmE_2$ satisfies
the first quasi-orthogonality condition (for both arguments)
with error $\e+\frac{2^{k+1}}{n_1}$.
Likewise, we claim that there are at most $2^{k}$
disjoint pairs $\{x,x'\}$'s that violate the second condition (i.e.,
$|\{y\!\in\![n_2]:\nmE_2(x,y)\!\neq\!\nmE_2(x',y)\}|\geq(0.5-\e)\cdot n_2$),
because otherwise we obtain a contradiction to the hypothesis
that $\nmE_2$ is a {\em non-malleable}\/ two-source $(k,\e)$-extractor
(by using a function $f$ that maps each such $x$ to its matched $x'$,
and the identity permutation~$g$).%
\footnote{Formally, denoting the exceptional pairs by $(x_i,x'_i)$,
where $i\in[2^k]$, we define $f$ such that $f(x_i)=x'_i$ for each $i\in[2^k]$,
and let $X$ be uniform on $\{x_i:i\!\in\![2^k]\}$
and $Y$ be uniform on $[n_2]$.
Then, $\prob[\nmE_2(X,Y)\!\neq\!\nmE_2(f(X),g(Y))]<0.5-\e$,
in contradiction to the hypothesis regarding $\nmE_2$.}
And, again, we consider a residual extractor obtained
by omitting the exceptional pairs.
Doing the same for the $y$'s, we obtained the desired extractor.
\EPF

\medskip
Recall that non-malleable two-source extractors with much stronger
parameters than we need (i.e., min-entropy $\ell-\ell^{\Omega(1)}$,
negligible error, and $\ell^{\Omega(1)}$ bits of output),
were constructed in~\cite[Thm.~1]{CGL},
but these extractors are not quasi-orthogonal.
Employing Proposition~\ref{nmE-made-nice:prop},
we obtain a quasi-orthogonal non-malleable
two-source $(\ell-4,0.1)$-extractor that can be used
in the construction of the following Theorem~\ref{nmE2RSO:prop}.
%
Essentially, the construction consists of a bipartite graph,
with $2^\ell$ vertices on each side, such that
the edges between the two sides are determined by the extractor.
In addition, we add a clique on one of the two sides
\ifnum\newfigs=1 
so that the two sides are (robustly) distinguishable
(see Figure~\ref{nmE2RSO:fig}).
\else
so that the two sides are (robustly) distinguishable.
\fi
We stress that the resulting $2^{\ell+1}$-vertex graph
is $\Omega(2^\ell)$-robustly self-ordered
as long as the non-malleable extractor
is quasi-orthogonal and works for very mild parameters; that is,
we only require error that is bounded away from~$1/2$
with respect to min-entropy $\ell-O(1)$.

\ifnum\newfigs=1 
\begin{figure}
\centerline{\mbox{\includegraphics[width=0.8\textwidth]{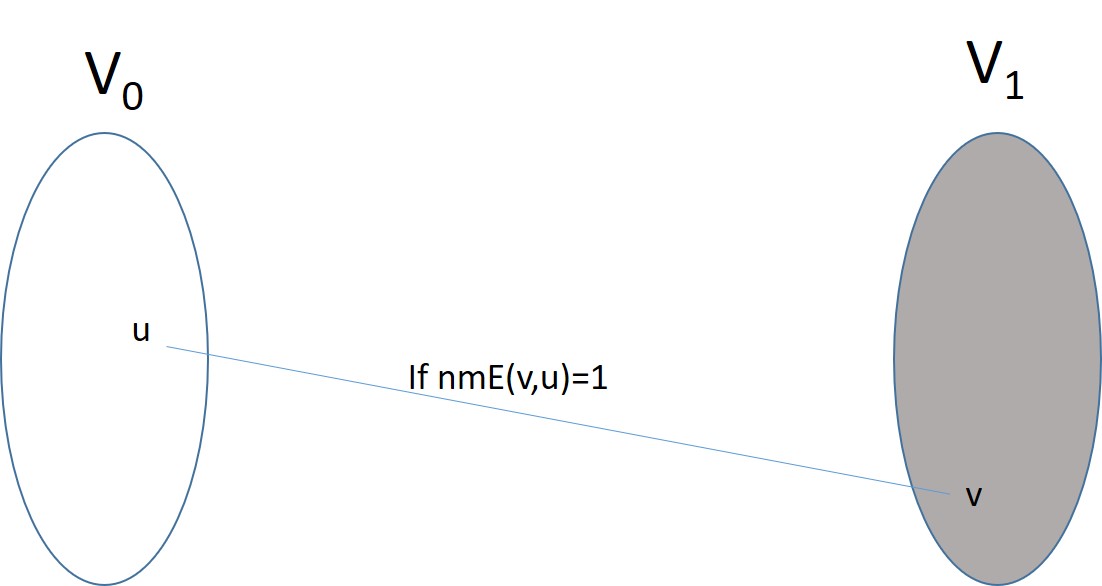}}}
\caption{Illustrating the construction of Theorem~\ref{nmE2RSO:prop}.}
\label{nmE2RSO:fig}
\end{figure}
\fi

\BT[Using a quasi-orthogonal non-malleable two-source extractor to
obtain a $\Omega(2^\ell)$-robustly self-ordered $O(2^\ell)$-vertex graph]
\label{nmE2RSO:prop}
For a constant $\e\in(0,0.5)$ varying $\ell\geq k$
such that $k\leq\ell-2+\log_2(0.5-\e)=\ell-O(1)$,
suppose that ${\nmE}:[2^\ell]\times[2^\ell]\to\bitset$
is a quasi-orthogonal {\em(with error $\e$)}
non-malleable two-source $(k,\e)$-extractor.
Then, the $2^{\ell+1}$-vertex graph $G=(V_1\cup V_0,E)$
such that $V_\sigma=\{\ang{\sigma,i}\!:\!i\!\in\![2^\ell]\}$
and
\begin{equation}\label{nmE2RSO:eq}
E\;=\;\left\{\{\ang{1,i},\ang{0,j}\}\!:\!\nmE(i,j)\!=\!1\right\}
\cup \binom{V_1}{2}
\end{equation}
is $\Omega(|V_1\cup V_0|)$-robustly self-ordered.
Furthermore, the claim holds even if
the non-malleability condition {\rm(i.e., \eqref{nm-extractor:eq})}
holds only for permutations $f$ and $g$ that have no fixed-points.
\ET
Indeed, the first set of edges, denoted $E'$, corresponds to
a bipartite graph between $V_1$ and~$V_0$ that is determined by $\nmE$,
and the second set corresponds to a $2^\ell$-vertex clique
(see Figure~\ref{nmE2RSO:fig}).
Note that the extraction parameters are extremely weak;
that is, the min-entropy may be very high (i.e., $k=\ell-O(1)$),
the error may be an arbitrary non-trivial constant (i.e., $\e<1/2$),
and we only extract one bit (i.e., $m=1$).
\medskip

\BPF
Let $V=V_1\cup V_0$,
and consider an arbitrary (non-trivial) permutation $\mu:V\to V$.
Intuitively, if $\mu$ maps a vertex of $V_1$ to $V_0$,
then the difference in degrees of vertices in the two sets
(caused by the clique edges) contributes
at least $(2^\ell-1)-2\e\cdot2^\ell$ units
to the symmetric difference between~$G$ and~$\mu(G)$,
where here we use the first quasi-orthogonality condition.
On the other hand,
if $\mu$ maps $\ang{1,i}\in V_1$ to $V_1\setminus\{\ang{1,i}\}$,
then the difference in the neighborhoods caused by the bipartite graph
contributes at least $(0.5-\e)\cdot2^{\ell}$ units to
the symmetric difference between~$G$ and~$\mu(G)$.
To prove this, we distinguish between the case that $\mu$
has relatively few non-fixed-points (in either $V_0$ or $V_1$),
which is analyzed using the second quasi-orthogonality condition,
and the case that $\mu$ has relatively many non-fixed-points
(in both $V_0$ and $V_1$),
which is analyzed using the non-malleability condition.
Details follow.

Let $T=\{v\!\in\!V\!:\!\mu(v)\neq v\}$ denote
the set of non-fixed-points of $\mu$.
Then, we consider two types of vertices:
Those that belong to the set
$T'=\bigcup_{\sigma\in\bitset}
  \{v\!\in\!V_\sigma\!:\!\mu(v)\not\in V_{\sigma}\}\subseteq T$
and those that belong to $T\setminus T'$.
The threshold for distinguishing these cases is set
to $K=(0.5-\e)\cdot2^{\ell-2}=\Omega(|V|)$.

\BDes
\item[{\em Case 1}:] $|T'|\geq K$.

(This refers to the case that many vertices are mapped by $\mu$
to the opposite side of the bipartite graph $(V,E')$,
where `many' means $\Omega(|V|)$.)

Each vertex in $T'$ contributes $(1-2\e)\cdot2^\ell-1$
units to the symmetric difference between~$G$ and~$\mu(G)$,
because the degree of each vertex in $V_1$
is at least $(2^{\ell}-1)+(0.5-\e)\cdot2^\ell$,
whereas the degree of each vertex in $V_0$
is at most $(0.5+\e)\cdot2^\ell$,
where we use the first quasi-orthogonality condition,
which implies that the number of bipartite edges incident
at each vertex is at least $(0.5-\e)\cdot2^\ell$
and at most $(0.5+\e)\cdot2^\ell$.

Hence, the symmetric difference between~$G$ and $\mu(G)$ is at least
$((1-2\e)\cdot2^\ell-1)\cdot|T'|=\Omega(|V|)\cdot|T'|$,
since $2^\ell=\Omega(|V|)$.
Using the case's hypothesis,
we have $|T'|=\Omega(|V|)=\Omega(|T|)$,
which means that in this case
the difference between~$G$ and $\mu(G)$ is $\Omega(|V|)\cdot|T|$.

We stress that the difference between~$G$ and $\mu(G)$
is at least $\Omega(|V|)\cdot|T'|$
also if the case hypothesis does not hold.

\item[{\em Case 2}:] $|T'|<K$.

(This refers to the case that few vertices are mapped by $\mu$
to the opposite side of the bipartite graph $(V,E')$,
where `few' means less than $K\leq|V|/20$ (assuming $\e\leq0.1$).)

For every $\sigma\in\bitset$,
let $V'_\sigma=V_\sigma\cap\imu(V_\sigma)$ and $T_\sigma=V'_\sigma\cap T$.
Indeed, $(T',T_0,T_1)$ is a three-way partition of $T$.
Note that the size of the symmetric difference between~$G$ and $\mu(G)$
is lower-bounded by
\begin{equation}\label{diff-vs-nmE:eq}
\left|\{(v,u)\in V'_1\times V'_0:\nmE(\mu(v),\mu(u))\neq\nmE(v,u)\}\right|,
\end{equation}
since, for any $(v,u)\in V'_1\times V'_0$,
it holds that $\mu(v)$ neighbors $\mu(u)$ in~$G$
if and only if $\nmE(\mu(v),\allowbreak\mu(u))=1$,
whereas $\mu(v)$ neighbors $\mu(u)$ in $\mu(G)$
if and only if $v$ neighbors $u$ in~$G$
which holds if and only if $\nmE(v,u)=1$.

We consider two sub-cases according to
whether or not $\min(|T_0|,|T_1|)$ is relatively large.
The threshold for distinguishing these sub-cases
is also set to $K=(0.5-\e)\cdot2^{\ell-2}$;
note that $K=\Omega(|V|)$ and $K\geq2^k$.

\BDes
\item[{\em Case 2.1}:] $\min(|T_0|,|T_1|)<K$.

In this case we shall use
the (second condition of) quasi-orthogonality of $\nmE$.

Suppose, without loss of generality, that $|T_0|\leq|T_1|$,
which implies $|T_0|<K$.
Then, the contribution of
each vertex $v\in T_1$ to \eqref{diff-vs-nmE:eq} equals
\begin{eqnarray*}
\lefteqn{|\{u\in V'_0:\nmE(\mu(v),\mu(u))\neq\nmE(v,u)\}|} \\
&\geq& |\{u\in V'_0:\nmE(\mu(v),u)\neq\nmE(v,u)\}| - |T_0| \\
&\geq& |\{u\in V_0:\nmE(\mu(v),u)\neq\nmE(v,u)\}|-|T'|-|T_0| \\
&\geq& (0.5-\e)\cdot2^\ell-2\cdot K \\
&=& (0.5-\e)\cdot2^{\ell-1}
\end{eqnarray*}
where the first inequality uses $\mu(u)=u$ for $u\in V'_0\setminus T_0$,
the second inequality uses $|V'_0|\geq|V_0|-|T'|$,
the third inequality uses $\mu(v)\neq v$
along with the (second condition of) quasi-orthogonality of $\nmE$
(and the hypotheses regarding $|T'|$ and $|T_0|$),
and the equality is due to $K=(0.5-\e)\cdot2^{\ell-2}$.

Hence, in this case,
the total contribution to \eqref{diff-vs-nmE:eq} is
$(0.5-\e)\cdot2^{\ell-1}\cdot|T_1|$,
which is $\Omega(|V|)\cdot(|T|-|T'|)$,
since $|T_1|\geq(|T|-|T'|)/2$.

\item[{\em Case 2.2}:] $\min(|T_0|,|T_1|)\geq K$.

In this case we shall use the non-malleable feature of $\nmE$.

Specifically, for each $\sigma\in\bitset$,
let $\mu_\sigma$ denote the restriction of $\mu$ to $T_\sigma$.
Essentially, using $K\geq2^k$,
the non-malleability condition of the $(k,\e)$-extractor $\nmE$ implies
$$\left|\left\{(i,j)\in T_0\times T_1:
            \nmE(i,j)\neq\nmE(\mu_0(i),\mu_1(j))\right\}\right|
  \geq(0.5-\e)\cdot|T_0|\cdot|T_1|.$$
This can be seen by
letting $X$ and $Y$ be uniform over $T_0$ and $T_1$, respectively.
(Also, for sake of formality, we extend $\mu_0$ and $\mu_1$
(which have no fixed-points)
to (arbitrary) derangements $f$ and $g$, respectively.)%
\footnote{Note that we may assume, w.l.o.g.,
that $|T_\sigma\cup\mu(T_\sigma)|\leq|V_\sigma|-2$.}
In this case, the non-malleability condition implies that
the distribution $(\nmE(X,Y),\nmE(f(X),g(Y)))$
is $\e$-close to $(U_1,\nmE(f(X),g(Y)))$,
which implies that $\prob[\nmE(X,Y)\neq\nmE(\mu_0(X),\mu_1(Y))]$
is at least $\prob[U_1\neq\nmE(\mu_0(X),\mu_1(Y))]-\e=0.5-\e$.

Hence, in this case,
the total contribution to \eqref{diff-vs-nmE:eq} is
$(0.5-\e)\cdot|T_0|\cdot|T_1|=\Omega(|V|)\cdot(|T|-|T'|)$,
where we use $\min(|T_0|,|T_1|)=\Omega(|V|)$
and $|T_0|+|T_1|=|T|-|T'|$.
\EDes
Hence, in both sub-cases, the difference between~$G$ and $\mu(G)$
is $\Omega(|V|)\cdot(|T|-|T'|)$.
\EDes
Recall that (by the last comment at Case~1)
the difference between~$G$ and $\mu(G)$
is $\Omega(|V|)\cdot|T'|$.
Combining this lower-bound with the conclusion of Case~2,
the difference between~$G$ and $\mu(G)$ is $\Omega(|V|)\cdot|T|$.
\EPF

\paragraph{Digest:}
Note that the quasi-orthogonality of $\nmE$ was used in Cases~1 and~2.1,
whereas the non-malleability of $\nmE$ (w.r.t derangements)
was used in Case~2.2. In particular, Case~1 only uses the
first condition of quasi-orthogonality, and does so in order to infer
that the degrees of all vertices in the bipartite graph
are approximately equal.
In Case~2.1 the second quasi-orthogonality condition
is used in order to assert that the neighborhoods of two
different vertices in $V_\sigma$ are significantly different.
This is useful only when the number of non-fixed-points in $V_{1-\sigma}$
is relatively small.
When the number of non-fixed-points in both $V_\sigma$'s is large
but few vertices are mapped to the other side (i.e., $|T'|\ll|T|$),
we only use Case~2.2, which does not refer to quasi-orthogonality at all.
Hence, we have the following~--

\BR[A special case of Theorem~\ref{nmE2RSO:prop}]
\label{nmE2RSO:rem}
For bipartite graphs $G=(V,E)$ such that $V=V_0\cup V_1$
and $E\subseteq V_0\times V_1$,
we consider the special case of robust self-ordering
that refers only to permutations $\mu:V\to V$
that are derangements that preserve the bipartition of $V$
{\rm(i.e., $\mu$ has no fixed-points and $\mu(V_0)=V_0$)}.%
\footnote{That is, the requirement regarding the symmetric difference
between~$G$ and $\mu(G)$ is made only for permutations $\mu$
that have no fixed-points and satisfy $\mu(V_0)=V_0$.}
%
In this case, assuming {\rm(only)} that $\nmE$ is
a non-malleable two-source $(\ell,\e)$-extractor
{\rm(i.e., the case of $k=\ell$)},
implies that, for any such $\mu$,
the size of the symmetric difference between~$G$ and $\mu(G)$
is $(0.5\pm\e)\cdot|V_0|\cdot|V_1|$.
Furthermore, the claim holds even if the non-malleability condition
holds only for permutations $f$ and $g$ that have no fixed-points
{\em(i.e., derangements)},
and the quasi-orthogonality condition is not necessary.
Note that the proof of Theorem~\ref{nmE2RSO:prop} simplifies,
since $T'=\emptyset$ and $T_\sigma=V_\sigma=V'_\sigma$ hold,
and the size of the symmetric difference between~$G$ and $\mu(G)$
equal the quantity in \eqref{diff-vs-nmE:eq}.
\ER
Interestingly, the special case of Theorem~\ref{nmE2RSO:prop}
asserted in Remark~\ref{nmE2RSO:rem} can be reversed in the
sense that a bipartite graph that is robustly self-ordered
in the foregoing restricted sense is actually
a non-malleable two-source $(\ell,0.5-\Omega(1))$-extractor
(w.r.t derangement).
In the following result $\e$ is an arbitrary constant (in $(0,0.5)$),
whereas~$G$ varies and $o(1)$ vanishes with~$|V|$.

\BP[A reversal of the special case of Theorem~\ref{nmE2RSO:prop}
(i.e., of Remark~\ref{nmE2RSO:rem})]
\label{reverse-nmE2RSO:prop}
Let $G=(V_0\cup V_1,E)$ be a bipartite graph
such that $|V_0|=|V_1|$ and $E\subseteq V_0\times V_1$.
Let $V=V_0\cup V_1$, and
suppose that for every derangement $\mu:V\to V$
such that $\mu(V_0)=V_0$ it holds that the
size of the symmetric difference between~$G$ and $\mu(G)$
is $(0.5\pm\e)\cdot|V_0|\cdot|V_1|$.
Then, $F:V_0\times V_1\to\bitset$
such that $F(x,y)=1$ if and only if $\{x,y\}\in E$
is a non-malleable two-source $(\ell,\e+{\sqrt{2\e}}+o(1))$-extractor
w.r.t derangement.%
\footnote{That is, the non-malleability condition
(i.e., \eqref{nm-extractor:eq}) is guaranteed
only for permutations $f$ and $g$ that have no fixed-points.}
\EP
Needless to say, the claim holds also if~$G$ is augmented
by complete graph on the vertex-set~$V_1$.
Note that we lose a $\sqrt{2\e}+o(1)$ term in the reversal.
\medskip


\BPF
Let $(f,g)$ and $(X,Y)$ be as in Definition~\ref{nmExtractor:def},
and note that in this case $X$ and $Y$ are independent distributions
that are each uniformly distributed on $[2^\ell]$.
Define $\mu:V\to V$ such that $\mu(z)=f(z)$ if $z\in V_0$
and $\mu(z)=g(z)$ otherwise, and note that $\mu$ is a derangement
that preserves the partition of $V$.
Recall that $(\mu(x),\mu(y))$ contributes to the symmetric difference
between~$G$ and $\mu(G)$ if and only if $F(\mu(x),\mu(y))\neq F(x,y)$,
since $\mu(x)$ is connected to $\mu(y)$ in~$\mu(G)$
if and only if $x$ is connected to $y$ in~$G$.
Hence, by the hypothesis, we have
\begin{equation}\label{reverse-nmE2RSO:eq}
\prob[F(X,Y)\neq F(\mu(X),\mu(Y))]=0.5\pm\e.
\end{equation}
Letting
$p_{\sigma,\tau}^\mu\eqdef\prob[(F(X,Y),F(\mu(X),\mu(Y)))\!=\!(\sigma,\tau)]$,
we have $p_{0,1}^\mu+p_{1,0}^\mu=0.5\pm\e$,
and using the fact that $(X,Y)$ and $(\mu(X),\mu(Y))$
are identically distributed we have $p_{1,0}^\mu=p_{0,1}^\mu$
(since $p_{1,1}^\mu+p_{1,0}^\mu=p_{1,1}^\mu+p_{0,1}^\mu$).
Hence, $p_{0,1}^\mu=0.25\pm0.5\e$.
Lastly, we show that $p_{1,1}^\mu+p_{1,0}^\mu=0.5\pm{\sqrt{\e/2}}+o(1)$,
and conclude that $p_{1,1}^\mu=0.25\pm(0.5\e+{\sqrt{\e/2}}+o(1))$;
it follows that $F$ is
a non-malleable (two-source) $(\ell,\e+{\sqrt{2\e}}+o(1))$-extractor.

To show that $p_{1,1}^\mu+p_{1,0}^\mu=0.5\pm{\sqrt{\e/2}}+o(1)$,
we first note that $p\eqdef p_{1,1}^\mu+p_{1,0}^\mu=\prob[F(X,Y)\!=\!1]$
is actually oblivious of $\mu$.
Hence, by considering a random derangement $\mu$ that preserves $V_0$
(i.e., $\mu(V_0)=V_0$), we observe that,
with overwhelmingly high probability (over the choice of $\mu$),
it holds that $\{(x,y)\!\in\!V_0\times V_1\!:\!F(x,y)\neq F(\mu(x),\mu(y))\}$
has size $(2p(1-p)\pm o(1))\cdot|V_0|\cdot|V_1|$.
Confronting this with \eqref{reverse-nmE2RSO:eq},
we infer that $p=0.5\pm({\sqrt{\e/2}}+o(1))$.
\EPF

\paragraph{Corollary.}
Combining Theorem~\ref{nmE2RSO:prop}
with the non-malleable two-source extractors of~\cite[Thm.~1]{CGL},
while using Proposition~\ref{nmE-made-nice:prop},
we obtain an efficient construction
of $\Omega(n)$-robustly self-ordered graphs
(alas not a strongly explicit (aka locally computable) one).

\BT[Constructing $\Omega(n)$-robustly self-ordered $n$-vertex graphs]
\label{dense-rso1:thm}
There exist an algorithm that, on input $n$,
works in $\poly(n)$-time and outputs an explicit description of
an $\Omega(n)$-robustly self-ordered $O(n)$-vertex graph.
Furthermore, each vertex in this graph has degree
at least $0.24\cdot n$ and at most $0.76\cdot n$.
\ET
The degree bounds follow by observing that the vertices
in the graph described in Theorem~\ref{nmE2RSO:prop}
have degree at least $(0.5-\e)\cdot n/2$
and at most $(1.5+\e)\cdot n/2$,
whereas~\cite[Thm.~1]{CGL} provides for $\e=o(1)$.

\subsection{The second construction}
\label{dense:rso2:sec}
Combining Theorem~\ref{nmE2RSO:prop}
with the non-malleable two-source extractors of~\cite[Thm.~1]{CGL},
while using Proposition~\ref{nmE-made-nice:prop},
we obtained an efficient construction
of $\Omega(n)$-robustly self-ordered $n$-vertex graphs
(see Theorem~\ref{dense-rso1:thm}).
Unfortunately, this construction is not locally computable
(as postulated in Theorem~\ref{dense:ithm}),
because the non-malleable two-source extractors of~\cite[Thm.~1]{CGL}
are not quasi-orthogonal and the transformation
of Proposition~\ref{nmE-made-nice:prop} runs in time
that is polynomial in the size of the resulting graph.

To avoid the foregoing transformation and prove Theorem~\ref{dense:ithm},
we employ a variant on the construction presented
in Theorem~\ref{nmE2RSO:prop}.
Rather than connecting two sets of vertices using
a bipartite graph that corresponds to
a {\em quasi-orthogonal}\/ non-malleable two-source extractor,
we connect three sets of vertices such that one pair
of vertex-sets is connected by a (not necessarily quasi-orthogonal)
non-malleable two-source extractor, whereas the other two pairs
are connected by bipartite graphs that are merely quasi-orthogonal.
In analogy to the definition of a quasi-orthogonal (two-source) extractor,
we say that a bipartite graph on the vertex-set $X\cup Y$
is {\sf quasi-orthogonal} (with error $\e$)
if the following two conditions hold regarding
its adjacency predicate $B:X\times Y\to\bitset$:
\BE
\item
{\em The degree of each vertex is approximately half
the number of the vertices on the other side}:
For each $x\in X$ (resp., $y\in Y$), it holds that
$|\{y\!\in\!Y\!:\!B(x,y)\!=\!1\}|=(0.5\pm\e)\cdot|Y|$
(resp., $|\{x\!\in\!X\!:\!B(x,y)\!=\!1\}|=(0.5\pm\e)\cdot|X|$).
\item
{\em For each pair of vertices on one side,
approximately half the vertices on the other side
neighbor exactly one of the vertices in the pair}:
For every $x\neq x'\in X$, it holds that
$|\{y\!\in\!Y\!:\!B(x,y)\!\neq\!B(x',y)\}|=(0.5\pm\e)\cdot|Y|$.
Similarly, for $y\neq y'\in Y$.
\EE
We note that inner-product (mod~2) extractor,
denoted $E_2:\bitset^\ell\times\bitset^\ell\to\bitset$,
corresponds to a quasi-orthogonal bipartite graph for
the case $X=Y=\bitset^\ell\setminus\{0^\ell\}$.
We will however need quasi-orthogonal bipartite graphs with
different-sized sides, which can be obtained by a simple variant.
Specifically, for the case of $X=\bitset^\ell\setminus\{0^\ell\}$
and $Y=\bitset^{\ell+2}\setminus\{0^{\ell+2}\}$,
we use the function $B(x,y)=E_2(G(x),y)$,
where $G:\bitset^\ell\to\bitset^{\ell+2}$
is a small-bias generator that satisfies $G(x)\neq0^{\ell+2}$
and $G(x)\neq G(x')$ for every $x\neq0^\ell$ and $x'\neq x$
(see Proposition~\ref{nice-bipartite:prop},
and note that $G(a,b,c,d)=(a,b,c,d,E_2(a,b),E_2(c,d))$ will do).
We stress that the foregoing construction is strongly explicit
(i.e., locally computable).

We shall also assume that the (bipartite graph corresponding to the)
non-malleable extractor ${\nmE}:[2^\ell-1]\times[2^\ell-1]\to\bitset$
has {\em linear degrees}\/ in the sense that for every $x$ it holds that
$|\{y\!\in\![2^\ell-1]\!:\!\nmE(x,y)\!=\!1\}|\geq\e'\cdot2^\ell$
for some constant $\e'>0$.
This can be enforced by starting with an arbitrary
non-malleable two-source $(k,\e')$-extractor
(e.g., the one of~\cite[Thm.~1]{CGL})
and resetting pairs in $m=\e'\cdot2^\ell$ fixed perfect matchings to~1
(i.e., for each $(x,y)$ in one of these matching, we reset $\nmE(x,y)\gets1$).%
\footnote{For example, we may use the matchings $\{(z,z+i):z\in[2^\ell-1]\}$
for $i\in[m]$, where addition is mod~$2^\ell-1$.
Furthermore, starting from an extractor that is defined
over $\ell$-bit strings, we may omit one of these strings
(and obtain an extractor defined over $[2^\ell-1]$).}
%
This increases the error of the extractor by an additive
term of $m/2^k=2^{\ell-k}\cdot\e'$, which we can afford
(e.g., $\e'=0.01$ and $k=\ell-4$,
yields extraction error $\e<0.2$).
We stress that this transformation preserves
polynomial-time computability of the extracting function.
%

\BT[Using a non-malleable two-source extractor with linear degrees to
obtain a $\Omega(2^\ell)$-robustly self-ordered $O(2^\ell)$-vertex graph]
%
\label{nmE2RSO-tri:thm}
For any constants $\e,\e'\in(0,0.5)$ and varying $k\leq\ell-4$,
where $\ell\in\N$,
suppose that ${\nmE}:[2^\ell-1]\times[2^\ell-1]\to\bitset$
is a non-malleable two-source $(k,\e)$-extractor
such that for every $x$ it holds that
$|\{y\!\in\![2^\ell-1]\!:\!\nmE(x,y)\!=\!1\}|>\e'\cdot2^\ell$.
Further suppose that $B:[2^\ell-1]\times[2^{\ell+2}-1]\to\bitset$
is quasi-orthogonal with error $0.1\cdot\e'$.
Then, the $(6\cdot2^{\ell}-3)$-vertex graph $G=(V_0\cup V_1\cup V_2,E)$
such that $V_\sigma=\{\ang{\sigma,i}\!:\!i\!\in\![2^{\ell_\sigma}-1]\}$,
where $\ell_0=\ell_1=\ell$ and $\ell_2=\ell+2$,
and
\begin{equation}\label{nmE2RSO-tri:eq}
E\;=\;\left\{\{\ang{1,i},\ang{0,j}\}\!:\!\nmE(i,j)\!=\!1\right\}
\cup\left\{\{
    \ang{\sigma,i},\ang{2,j}\}\!:\!B(i,j)\!=\!1,\sigma\in\bitset
       \right\}
\cup \binom{V_1}{2}
\cup \binom{V_2}{2}
\end{equation}
is $\Omega(|V|)$-robustly self-ordered, where $V=V_0\cup V_1\cup V_2$.
%
Furthermore, each vertex in this graph has degree
at least $0.3\cdot|V|$ and at most $0.9\cdot|V|$.
Moreover, the claim holds even if
the non-malleability condition {\rm(i.e., \eqref{nm-extractor:eq})}
holds only for permutations $f$ and $g$ that have no fixed-points.
\ET
Using the foregoing ingredients
(including the non-malleable extractor of~\cite[Thm.~1]{CGL}),
Theorem~\ref{dense:ithm} follows
(see also Remark~\ref{dense:rso:all-n's:rem}).
Looking at \eqref{nmE2RSO-tri:eq}, note that
the first set of edges corresponds to a bipartite
graph between $V_1$ and $V_0$ that is determined by $\nmE$,
the second set corresponds the bipartite graphs
between $V_\sigma$ (for $\sigma\in\bitset$) and $V_2$
that are determined by $B$,
and the other two sets correspond to cliques on $V_1$ and on $V_2$
(see Figure~\ref{nmE2RSO-tri:fig}).
\medskip

\ifnum\newfigs=1 
\begin{figure}
\centerline{\mbox{\includegraphics[width=0.8\textwidth]{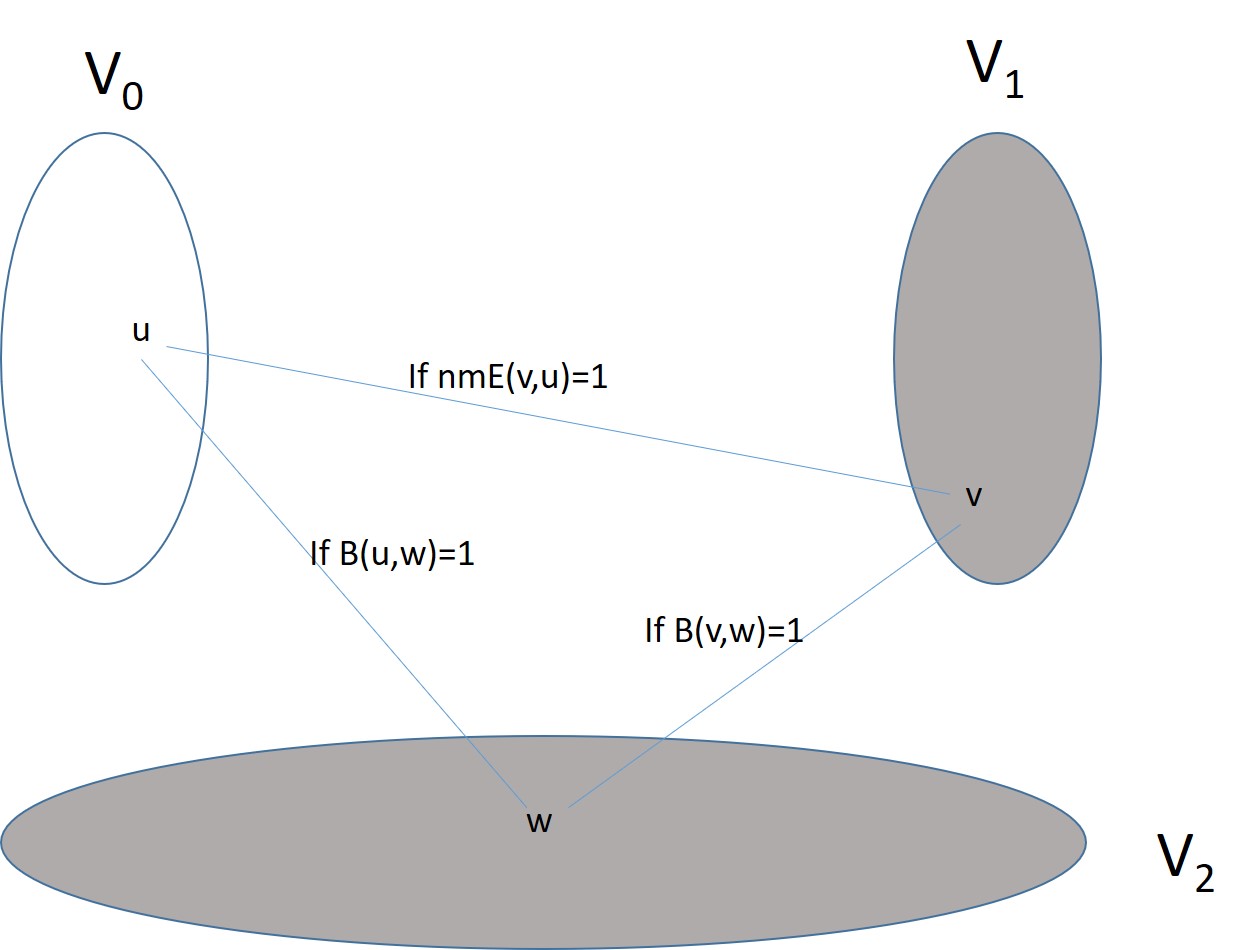}}}
\caption{Illustrating the construction of Theorem~\ref{nmE2RSO-tri:thm}.}
\label{nmE2RSO-tri:fig}
\end{figure}
\fi

\BPF
Recall that $V=V_0\cup V_1\cup V_2$, and
consider an arbitrary (non-trivial) permutation $\mu:V\to V$.
Intuitively, if $\mu$ maps a vertex of $V_0$ (or $V_1$) to $V_2$,
then the difference in the degrees of the vertices in the two sets
(caused by the $|V_2|$-clique edges) contributes $\Omega(|V|)$
units to the symmetric difference between~$G$ and $\mu(G)$,
where here we use the first quasi-orthogonality condition of $B$.
A similar argument, which uses the $V_1$-clique edges
and relies on the linear degrees of $\nmE$,
applies to a vertex of $V_\sigma$ mapped to $V_{1-\sigma}$
for any $\sigma\in\bitset$.
On the other hand, if for some $\sigma\in\{0,1,2\}$
the bijection $\mu$ maps $\ang{\sigma,i}\in V_\sigma$
to $V_\sigma\setminus\{\ang{\sigma,i}\}$,
then the difference in the neighborhoods caused by one
of the two relevant bipartite graphs contributes $\Omega(|V|)$
units to the symmetric difference between~$G$ and $\mu(G)$.
Here, we distinguishes between
the case that~$\mu$ has relatively few non-fixed-points
in either $V_0$ or $V_1$,
which is analyzed using the second quasi-orthogonality condition of $B$,
and the case that $\mu$ has relatively many non-fixed-points
in both $V_0$ and $V_1$,
which is analyzed using the non-malleability condition of $\nmE$.
Indeed, the structure of the proof is similar to the one
of Theorem~\ref{nmE2RSO:prop}, but the details are different
in many aspects, and so we provide them below.

Let $T=\{v\!\in\!V\!:\!\mu(v)\neq v\}$ denote
the set of non-fixed-points of $\mu$.
Then, we consider two types of vertices:
Those that belong to the set
$T'=\bigcup_{\sigma\in\{0,1,2\}}
  \{v\!\in\!V_\sigma\!:\!\mu(v)\not\in V_{\sigma}\}\subseteq T$
and those that belong to $T\setminus T'$.
The threshold for distinguishing these cases is set
to $K=(0.5-0.1\cdot\e')\cdot|V_0|/4=\Omega(|V|)$.%
\footnote{The threshold is set depending on the quasi-orthogonality
error of $B$. In the proof of Theorem~\ref{nmE2RSO:prop},
the threshold was set depending on the quasi-orthogonality
error of $\nmE$ (which equaled its extraction error).}
Recall that $\e$ denotes the extraction error of $\nmE$,
whereas $\e'$ is the fractional degree bound associated
with its linear degrees feature,
and $0.1\cdot\e'$ is the quasi-orthogonality error of~$B$.

\BDes
\item[{\em Case 1}:] $|T'|\geq K$.

(This refers to the case that many vertices are mapped by $\mu$
to a different part of the three-way partition $(V_0,V_1,V_2)$ of $V$,
where `many' means $\Omega(|V|)$.)

Each vertex in $T'$ contributes $\Omega(|V|)$ units
to the symmetric difference between~$G$ and $\mu(G)$,
because of the differences in the degrees of vertices
in the three parts. Specifically:
\BI
\item Vertices in $V_2$ have degree at least
$(|V_2|-1)+(0.5-0.1\e')\cdot(|V_0|+|V_1|)>(5-0.2\e')\cdot|V_0|-O(1)$,
where the first term is due to the clique edges
and the second term is due to the bipartite graphs
connecting $V_2$ to $V_0$ and to $V_1$
(and relies on the first quasi-orthogonality condition of $B$).
\item Vertices in $V_0$ have degree at most
$|V_1|+(0.5+0.1\e')\cdot|V_2|<(3+0.4\e')\cdot|V_0|+O(1)$,
where the first term is due to the edges (determined by $\nmE$)
connecting $V_0$ to $V_1$ and the second term is due to
the bipartite graph connecting $V_0$ to $V_2$.
\item
Vertices in $V_1$ have degree at least
$(|V_1|-1)+\e'\cdot|V_0|+(0.5-0.1\e')\cdot|V_2|>(3+0.6\e')\cdot|V_0|-O(1)$
and at most
$(|V_1|-1)+|V_0|+(0.5+0.1\e')\cdot|V_2| < (4+0.4\e')\cdot|V_0|$.
In both cases, the first term is due to clique edges,
the second term is due to the edges connecting $V_1$ to $V_0$
(as determined by $\nmE$), and the third term is due to
the edges connecting $V_1$ to $V_2$ (as determined by $B$).
The crucial fact is that the linear degrees of $\nmE$ provides
a non-trivial lower bound (of $\e'\cdot|V_0|$) on the second term.
\EI
Hence, the difference in the degrees of vertices in the different
parts is at least $0.2\e'\cdot|V_0|-O(1)$, where the minimum
is due to the difference between the degrees of vertices in $V_1$
and the degrees of vertices in $V_0$.

It follows that the symmetric difference between~$G$ and $\mu(G)$
is at least $(0.2\e'\cdot|V_0|-O(1))\cdot|T'|=\Omega(|V|)\cdot|T'|$,
since $|V_0|=\Omega(|V|)$ and $\e'=\Omega(1)$.
Using the case's hypothesis,
we have $|T'|=\Omega(|V|)=\Omega(|T|)$,
which means that in this case
the difference between~$G$ and $\mu(G)$ is $\Omega(|V|)\cdot|T|$.

We stress that the difference between~$G$ and $\mu(G)$
is at least $\Omega(|V|)\cdot|T'|$
also if the case hypothesis does not hold.

\item[{\em Case 2}:] $|T'|<K$.

(This refers to the case that few vertices are mapped by $\mu$
to a different part of the three-way partition $(V_0,V_1,V_2)$ of $V$.)

For every $\sigma\in\{0,1,2\}$,
let $V'_\sigma=V_\sigma\cap\imu(V_\sigma)$ and $T_\sigma=V'_\sigma\cap T$.
Indeed, $(T',T_0,T_1,T_2)$ is a four-way partition of $T$.
Note that the size of the symmetric difference between~$G$ and $\mu(G)$
is lower-bounded by
\begin{equation}\label{diff-vs-nmE-tri:eq}
\begin{array}{ll}
&\left|\{(v,u)\in V'_1\times V'_0:\nmE(\mu(v),\mu(u))\neq\nmE(v,u)\}\right| \\
&+\;\left|\{(v,u)\in V'_1\times V'_2:B(\mu(v),\mu(u))\neq B(v,u)\}\right| \\
&+\;\left|\{(v,u)\in V'_0\times V'_2:B(\mu(v),\mu(u))\neq B(v,u)\}\right|,
\end{array}
\end{equation}
since, for any $(v,u)\in V'_1\times V'_0$,
it holds that $\mu(v)$ neighbors $\mu(u)$ in~$G$
if and only if $\nmE(\mu(v),\allowbreak \mu(u))=1$,
whereas $\mu(v)$ neighbors $\mu(u)$ in $\mu(G)$
if and only if $v$ neighbors $u$ in~$G$
which holds if and only if $\nmE(v,u)=1$.
Ditto for the other two cases.

We consider two sub-cases according to
whether or not $\min(|T_0|,|T_1|)$ is relatively large.
The threshold for distinguishing these sub-cases is also set
to $K=(0.5-0.1\cdot\e')\cdot|V_0|/4$;
note that $K=\Omega(|V|)$ and $K>0.1\cdot|V_0|>2^{\ell-4}\geq2^k$.

\BDes
\item[{\em Case 2.1}:] $\min(|T_0|,|T_1|)<K$.

In this case we shall use the quasi-orthogonality of $B$.

Suppose, without loss of generality, that $|T_0|\leq|T_1|$,
which implies $|T_0|<K$. 

Depending on the relative sizes of $T_1$ and $T_2$,
we shall use either the quasi-orthogonal bipartite graph
between $V_1$ and $V_2$
or the quasi-orthogonal bipartite graph between $V_2$ and $V_0$.
\BE
\item
If $|T_1|>|T_2|$, then we consider
the quasi-orthogonal bipartite graph between $V_1$ and $V_2$.
The contribution of
each vertex $v\in T_1$ to \eqref{diff-vs-nmE-tri:eq} equals
\begin{eqnarray*}
\lefteqn{|\{u\in V'_2:B(\mu(v),\mu(u))\neq B(v,u)\}|} \\
&\geq& |\{u\in V'_2:B(\mu(v),u)\neq B(v,u)\}| - |T_2| \\
&>& |\{u\in V_2:B(\mu(v),u)\neq B(v,u)\}|-|T'|-|T_1| \\
&\geq& (0.5-0.1\cdot\e')\cdot|V_2|-K-|V_0| \\
&\geq& (4-0.25)\cdot(0.5-0.1\e')\cdot|V_0|-|V_0| \\
&>& 0.6\cdot|V_0|
%
\end{eqnarray*}
where the first inequality uses $\mu(u)=u$ for $u\in V'_2\setminus T_2$,
the second inequality uses $|V'_2|\geq|V_2|-|T'|$
and the hypothesis $|T_2|<|T_1|$,
the third inequality uses $\mu(v)\neq v$
along with the (second condition of) quasi-orthogonality of $B$
(and the hypotheses $|T'|<K$
and the fact that $|T_1|\leq|V_1|=|V_0|$),
the fourth inequality uses $|V_2|>4\cdot|V_0|$ and the definition of $K$,
and the last inequality uses $\e'<0.5$.

So the total contribution in this sub-case is
$|T_1|\cdot\Omega(|V|)\geq(|T|-|T'|)\cdot\Omega(|V|)$,
since $|T_1|\geq\max(|T_0|,|T_2|)$ and $|T_0|+|T_1|+|T_2|=|T|-|T'|$.
\item
If $|T_1|\leq|T_2|$, then
we consider the quasi-orthogonal bipartite graph between $V_2$ and $V_0$.
The contribution of
each vertex $v\in T_2$ to \eqref{diff-vs-nmE-tri:eq} equals
\begin{eqnarray*}
\lefteqn{|\{u\in V'_0:B(\mu(u),\mu(v))\neq B(u,v)\}|} \\
&\geq& |\{u\in V'_0:B(u,\mu(v))\neq B(u,v)\}| - |T_0| \\
&\geq& |\{u\in V_0:B(u,\mu(v))\neq B(u,v)\}|-|T'|-|T_0| \\
&\geq& (0.5-0.1\cdot\e')\cdot|V_0|-2\cdot K \\
&=& (0.5-0.1\cdot\e')\cdot|V_0|/2
\end{eqnarray*}
where the first inequality uses $\mu(u)=u$ for $u\in V'_0\setminus T_0$,
the second inequality uses $|V'_0|\geq|V_0|-|T'|$,
the third inequality uses $\mu(v)\neq v$
along with the (second condition of) quasi-orthogonality of $B$
(and the hypotheses regarding $|T'|$ and $|T_0|$),
and the equality is due to $K=(0.5-0.1\cdot\e')\cdot|V_0|/4$.

So the total contribution in this sub-case is
$|T_2|\cdot\Omega(|V|)\geq(|T|-|T'|)\cdot\Omega(|V|)$,
since $|T_2|\geq|T_1|\geq|T_0|$.
\EE
Hence, the total contribution (of Case~2.1)
to \eqref{diff-vs-nmE-tri:eq} is $\Omega(|V|)\cdot(|T|-|T'|)$.

\item[{\em Case 2.2}:] $\min(|T_0|,|T_1|)\geq K$.

In this case we shall use the non-malleable feature of $\nmE$.

Specifically, for each $\sigma\in\bitset$,
let $\mu_\sigma$ denote the restriction of $\mu$ to $T_\sigma$.
Essentially, using $K\geq2^k$,
the non-malleability condition of the $(k,\e)$-extractor $\nmE$ implies
$$\left|\left\{(i,j)\in T_0\times T_1:
            \nmE(i,j)\neq\nmE(\mu_0(i),\mu_1(j))\right\}\right|
  \geq(0.5-\e)\cdot|T_0|\cdot|T_1|.$$
As in (Case 2.2 of) the proof of Theorem~\ref{nmE2RSO:prop},
this can be seen by
letting $X$ and $Y$ be uniform over $T_0$ and $T_1$, respectively.%
\footnote{Recall that, formally, we also
extend $\mu_0$ and $\mu_1$ (which have no fixed-points)
to (arbitrary) derangements $f$ and $g$, respectively.
Then, the non-malleability condition implies
that $(\nmE(X,Y),\nmE(f(X),g(Y)))$ is $\e$-close to $(U_1,\nmE(f(X),g(Y)))$,
which implies that $\prob[\nmE(X,Y)\neq\nmE(\mu_0(X),\mu_1(Y))]$
is at least $\prob[U_1\neq\nmE(\mu_0(X),\mu_1(Y))]-\e=0.5-\e$.}
%

Hence, in this case,
the total contribution to \eqref{diff-vs-nmE-tri:eq} is
$(0.5-\e)\cdot|T_0|\cdot|T_1|=\Omega(|V|^2)$,
where we use $\min(|T_0|,|T_1|)=\Omega(|V|)$.
\EDes
Hence, in both sub-cases, the difference between~$G$ and $\mu(G)$
is $\Omega(|V|)\cdot(|T|-|T'|)$.
\EDes
Recall that (by the last comment at Case~1)
the difference between~$G$ and $\mu(G)$
is $\Omega(|V|)\cdot|T'|$.
Combining this lower-bound with the conclusion of Case~2,
the difference between~$G$ and $\mu(G)$ is $\Omega(|V|)\cdot|T|$.
As for the degree bounds,
note that each vertex has degree at most
$(|V_2|-1)+(0.5+0.1\e')\cdot(|V_0|+|V_1|)=\frac{5+0.2\e'}{6}\cdot|V|+O(1)$,
and at least $(0.5-0.1\e')\cdot|V_2|=\frac{2-0.4\e'}{6}\cdot|V|-O(1)$,
where maximum (resp., minimum) is obtained
by vertices in $V_2$ (resp., $V_0$).
\EPF

\paragraph{Digest:}
Compared to the construction used in Theorem~\ref{nmE2RSO:prop},
the construction in Theorem~\ref{nmE2RSO-tri:thm} decouples
the non-malleable feature from the quasi-orthogonality feature,
using non-malleable extractors for connecting one pair of vertex-sets
and quasi-orthogonal functions to connect the other two pairs.
The current analysis is slightly more complex because it has
to handle the fact that these features hold for different pairs.
Specifically, the quasi-orthogonality of $B$ is used in Cases~1 and~2.1,
whereas the non-malleability of $\nmE$ is used in Case~2.2.
In particular, Case~1 only uses the first condition of quasi-orthogonality,
and does so in order to infer that the degrees of all vertices
in the bipartite graph determined by $B$ are approximately equal.
In Case~2.1 the second quasi-orthogonality condition is used
in order to assert that the neighborhoods of two different vertices
in $V_\sigma$ (for every $\sigma\in\{0,1,2\}$) are significantly different.
This is useful only when the number of non-fixed-points
in the other side of the graph $B$ is relatively small.

In light of the key role that quasi-orthogonal unbalanced bipartite
graphs play in Theorem~\ref{nmE2RSO-tri:thm} and given their natural appeal,
it feel adequate to provide a general construction of these graphs,
which generalizes the construction outlined before
Theorem~\ref{nmE2RSO-tri:thm} (for the case of $\ell'=\ell+2$).

\BP[Quasi-orthogonal unbalanced bipartite graphs]
\label{nice-bipartite:prop}
For $\ell\leq\ell'$ and $S_\ell\eqdef\bitset^\ell\setminus\{0^\ell\}$
let $G:S_\ell\to S_{\ell'}$ be a small-bias generator with bias $\e$
such that $G(s)\neq G(s')$ for every $s\neq s'$,
and let $E_2$ denote the inner-product mod~2 function.
Then, the bipartite graph described by
the adjacency predicate $B:S_\ell\times S_{\ell'}\to\bitset$
such that $B(x,y)=E_2(G(x),y)$ is quasi-orthogonal
with error $\e$.
\EP
(Note that the hypothesis implies that $\e>1/|S_{\ell'}|$.
The definition of quasi-orthogonal bipartite graphs
appears before Theorem~\ref{nmE2RSO-tri:thm}.)
\medskip

\BPFS
Our starting point is the fact
that $E_2:S_{\ell'}\times S_{\ell'}\to\bitset$
is quasi-orthogonal with error $1/|S_{\ell'}|$.
The quasi-orthogonality feature of the first argument of $B$
follows as a special case of the corresponding feature of $E_2$.
Turning to fixings of the second argument of~$E_2$
and letting $X$ be uniform over $S_\ell$,
we observe that, for every $y\in S_{\ell'}$,
the bit $B(G(X),y)$ is a linear combination of the bits of $G(X)$,
and hence $\prob[B(G(X),y)\!=\!1]=0.5\pm\e$.
Similarly, for $y\neq y'$, it holds
that $B(G(X),y)\xor B(G(X),y')=B(G(X),y\xor y')$
is linear combination of the bits of~$G(X)$.
\EPFS

\BR[Obtaining $\Omega(n)$-robustly self-ordered $n$-vertex graphs,
for every~$n$]
\label{dense:rso:all-n's:rem}
Theorem~\ref{nmE2RSO-tri:thm} provides a construction
of $\Omega(n)$-robustly self-ordered $n$-vertex graphs,
for every $n$ of the form $6\cdot2^\ell-3$, where $\ell\in\N$.
A construction for every $n\in\N$ can be obtained
by using a few minor modifications.
\BI
\item
Rather than using $|V_2|=2^{\ell+2}-1=4\cdot(|V_0|+1)-3$,
we may use $|V_2|=n-2\cdot|V_0|$ such that $|V_0|=\Omega(n)$.
Specifically, we still use $|V_0|=2^\ell-1$,
for $\ell=\log_2n-\Theta(1)$,
along with $|V_2|\in[4\cdot|V_0|,10\cdot|V_0|]$.
Doing so requires decreasing the quasi-orthogonality error of $B$
to $0.04\e'$ so that $0.04\e'\cdot|V_2|\leq0.4\cdot|V_0|$ still holds.
\item
More importantly, we need a construction of a quasi-orthogonal bipartite
graph with an adjacency predicate $B:[2^\ell-1]\times[n']\to\bitset$
such that $n'=n-2\cdot(2^{\ell}-1)\geq 2n/3$.
The solution is to associated $[n']$ with an easily enumerable
small-bias space $S\subseteq\bitset^{\ell+4}\setminus\{0^{\ell+4}\}$
and use $B(x,y)=E_2(G(x),y)$, where $E_2$ and~$G$ are as
in Proposition~\ref{nice-bipartite:prop}.
Specifically, for $t=\log_2\log_2\ell$
and $D=\ceil{n'\cdot2^t/2^{\ell+4}}=\Theta(2^t)$,
we let~$S$ contain the $n'$ lexicographically first strings
in $S'\times\bitset^{\ell+4-t}$, where $S'$ is a small-bias sample
space of size $D$ over $\bitset^t$ that is found by exhaustive search.%
\footnote{Note that
for every $z=(z',z'')\in\bitset^{\ell+4}\setminus\{0^{\ell+4}\}$
and $Y=(Y',Y'')$ that is uniformly distributed over~$S$
such that $|z'|=|Y'|=t$ it holds that
$$\Exp[(-1)^{E_2(z,Y)}]
 =\Exp[(-1)^{E_2(z',Y')}]\cdot\Exp[(-1)^{E_2(z'',Y'')}]$$
where the absolute value of each of the factors is $o(1)$ if
the corresponding fixed string (i.e., $z'$ or $z''$) is non-zero.
Specifically, note that $Y'$ (resp., $Y''$) is $o(1)$-close to
being uniformly distributed over $S'$ (resp., $\bitset^{\ell+4-t}$).}
\EI
\ER


\subsection{Obtaining efficient self-ordering}
We say that a self-ordered graph $G=([n],E)$
is {\sf efficiently self-ordered} if there exists
a polynomial-time algorithm that,
given any graph $G'=(V',E')$ that is isomorphic to~$G$,
finds the unique bijection $\phi:V'\to[n]$ such that $\phi(G')=G$
(i.e., the unique isomorphism of $G'$ and~$G$).
Indeed, this isomorphism orders the vertices of $G'$
in accordance with the original (or target) graph~$G$.

Recall that in the case of bounded-degree graphs,
we relied on the existence of
a polynomial-time isomorphism test (see~\cite{L})
for efficiently self-ordering the robustly self-ordered graphs
that we constructed.
We cannot do so in the dense graph case, since a general
polynomial-time isomorphism test is not known (see~\cite{B}).
Instead, we augment the construction asserted in Theorem~\ref{dense:ithm}
so to obtain dense $\Omega(n)$-robustly self-ordered graphs
that are efficiently self-ordered.%
\footnote{Unlike in the bounded degree case (see Section~\ref{local-so:sec}),
at the time of writing this paper, we did not know
how to construct $\Omega(n)$-robustly self-ordered graphs
that support {\em local}\/ self-ordering,
but such constructions were presented in subsequent works.
Specifically, $\Omega(n)$-robustly self-ordered graphs with
information-theoretically local self-ordering were presented
in our follow-up work~\cite{GW:na-vs-ad}, and a construction
that supports (full-fledged) local self-ordering was
presented by the first author~\cite{G:rso-lso}.
We mention that the latter work provides a general study of
the relationship between robust self-ordering and local self-ordering
(in both regimes).}

%

\BT[Strengthening Theorem~\ref{dense:ithm}]
\label{dense:effective:thm}
There exist an infinite family
of dense $\Omega(n)$-robustly self-ordered graphs $\{G_n\}_{n\in\N}$
and a polynomial-time algorithm that,
given $n\in\N$ and a pair of vertices $u,v\in[n]$
in the $n$-vertex graph $G_n$,
determines whether or not $u$ is adjacent to~$v$ in~$G_n$.
Furthermore, these graphs are efficiently self-ordered,
and the degrees of vertices in $G_n$ reside in $[0.06n,0.73n]$.
\ET

\BPF
Our starting point is the construction of $m$-vertex graphs
that are $\Omega(m)$-robustly self-ordered
(see Theorem~\ref{dense:ithm}, which uses Theorem~\ref{nmE2RSO-tri:thm}).
Recall that the vertices in these graphs have degree that ranges
between $0.3\cdot m$ and $0.9\cdot m$ (see Theorem~\ref{nmE2RSO-tri:thm}).

The idea is to use two such graphs, $G_1$ and $G_2$,
one with $m$ vertices and the other with $4\cdot m$ vertices,
where $m=n/5$, and connect them in a way that assists finding
the ordering of vertices in each of these two graphs.
Specifically, we designate a set, denoted $S_1$,
of $s\eqdef2{\sqrt{\log_2n}}$ vertices in $G_1=([m],E_1)$,
and a set, denoted $S_2$,
of $\ell\eqdef\binom{s}{2}\in[1.5\cdot\log_2n,2\cdot\log_2n]$
vertices in $G_2=(\{m+1,\ldots,5m\},E_2)$,
and use them as follows (see Figure~\ref{dense:effective:fig}):

\ifnum\newfigs=1 
\begin{figure}
\centerline{\mbox{\includegraphics[width=0.8\textwidth]{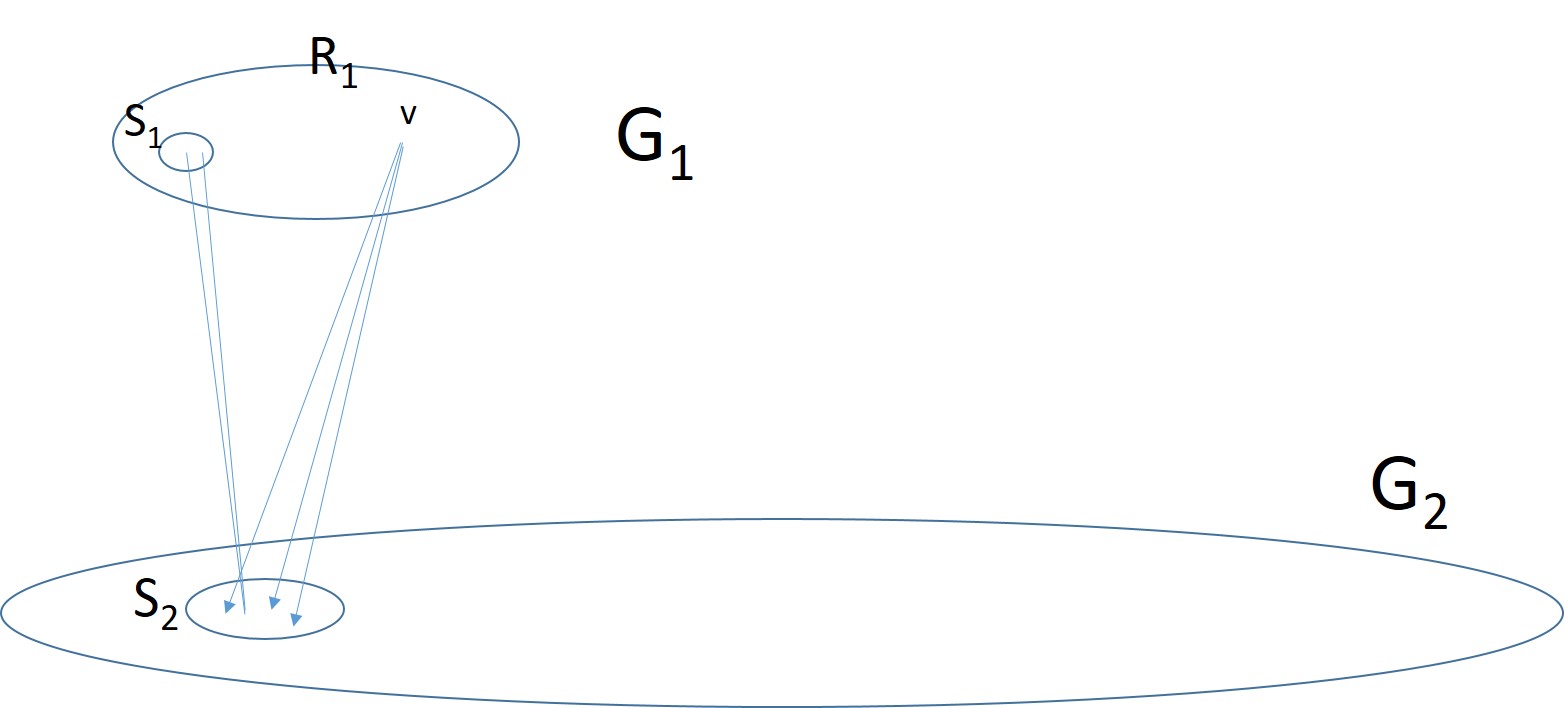}}}
\caption{The construction of Theorem~\ref{dense:effective:thm}.}
\label{dense:effective:fig}
\end{figure}
\fi

\BI
\item
Connect each vertex in $S_2$ to two different vertices in $S_1$,
while noting that each vertex in~$S_1$ is connected to $2\ell/s=o(\ell)$
vertices of $S_2$.
\item
Connect each vertex in $R_1\eqdef[m]\setminus S_1$
to a different set of neighbors in $S_2$ such that
each vertex in $R_1$ has at least $\ell/2$ neighbors in $S_2$.
(Note that $\binom{\ell}{\ell/2}>n$.)
\item
Connect each vertex in $R_2\eqdef\{m+1,\ldots,5m\}\setminus S_2$
to a different set of neighbors in $R_1$ such that
each vertex in $R_2$ has two neighbors in $R_1$
and each vertex in $R_1$ has at most eight neighbors in $R_2$.
\EI
Denote the resulting graph by $G=([n],E)$,
and note that the vertices of~$G_1$ have degree
at most $0.9\cdot m+\ell$, whereas
the vertices of $G_2$ have degree at least $0.3\cdot 4m$.
Given an isomorphic copy of the~$G$,
we can find the unique isomorphism (i.e., its ordering) as follows:

\BE
\item
Identify the vertices that belong to~$G_1$
by virtue of their lower degree.
\item
Identify the set $S_1$ as the set of vertices that belong to~$G_1$
and have $2\ell/s=o(\ell)$ neighbors in~$G_2$.

(Recall that each vertex in $R_1$ has at least $\ell/2$ neighbors in $S_2$.)
\item
Identify the set $S_2$ as the set of vertices that belong to $G_2$
and have (two) neighbors in $S_1$.
\item
For each possible ordering of $S_1$,
order the vertices of $S_2$ by their neighborhood in $S_1$, and
order the vertices of $R_1$ according to their neighborhood in $S_2$.

(Recall that the index of each $s\in S_2$
is determined by its (two) neighbors in $S_1$,
whereas the index of each $v\in R_1$
is determined by its neighbors in $S_2$.)

If the resulting ordering (of $S_1\cup R_1$) yields
an isomorphism to~$G_1$, them continue.
Otherwise, try the next ordering of $S_1$.
\item
Order the vertices of $R_2$
according to their neighborhood in $R_1$.
\EE
Note that by the asymmetry of~$G_1$,
there exists a unique ordering of its vertices,
and a unique ordering of $S_1$ that fits it
and leads the procedure to successful termination.
One the other hand,
the number of possible ordering of $S_1$ is $s!=n^{o(1)}$,
which means that the procedure is efficient.

It is left to show that the graph~$G$
is $\Omega(n)$-robustly self-ordered.
Let $\gamma\in(0,1]$ be a constant such that that~$G_1$ (resp., $G_2$)
is $\gamma\cdot m$-robustly self-ordered
(resp., $\gamma\cdot 4m$-robustly self-ordered).
Then, fixing an arbitrary permutation $\mu:[n]\to[n]$,
and letting $T=\{v\in[n]:\mu(v)\neq v\}$,
we consider the following cases.

\BDes
\item[{\em Case 1}:]
	$t\eqdef|\{v\in[m]:\mu(v)\not\in[m]\}|>\gamma\cdot|T|/10$.

In this case, we get a contribution of $t\cdot\Omega(m)=\Omega(m\cdot|T|)$
units to the symmetric difference between~$G$ and $\mu(G)$,
because of the difference in degree between vertices in $[m]$
and outside $[m]$.
(Recall that the former have degree at most $0.9\cdot m+\ell<m$,
whereas the latter have degree at least $0.3\cdot 4m=1.2\cdot m$.)

\item[{\em Case 2}:]
	$t=|\{v\in[m]:\mu(v)\not\in[m]\}|\leq\gamma\cdot|T|/10$.

In this case, at least $(1-0.1\gamma)\cdot|T|$ vertices in $T$
are mapped by $\mu$ to the side in which they belong
(i.e., each of these vertices $v$
satisfies $v\in[m]$ if and only if $\mu(v)\in[m]$).
Let $T_1\eqdef\{v\!\in\!T\cap[m]\!:\!\mu(v)\in[m]\}$
and $T_2\eqdef\{v\!\in\!T\setminus[m]\!:\!\mu(v)\not\in[m]\}$.
Then, the vertices in $T_1$ contribute
at least $|T_1|\cdot\gamma\cdot m-t\cdot m$
units to the symmetric difference between~$G$ and $\mu(G)$,
where the negative term is due to possible change
in the incidence with vertices that did not maintain their side.
Similarly, the vertices in $T_2$ contribute
at least $|T_2|\cdot\gamma\cdot 4m-t\cdot4m$
units to the symmetric difference.
Hence, it total, we get a contribution of
at least $(|T|-2t)\cdot\gamma\cdot m-t\cdot5m=\Omega(m\cdot|T|)$.
\EDes
The main claims follows.
As for the degree bounds,
note that the degree of each vertex in~$G_1$ is at least $0.3m=0.06n$,
whereas the degree of each vertex in $G_2$ is at most $0.9\cdot4m+s<0.73n$.
\EPF

\paragraph{Digest.}
The $n$-vertex graph constructed in the proof
of Theorem~\ref{dense:effective:thm}
is proved to be $\Omega(n)$-robustly self-ordered
by implicitly using the following claim.

\BCM[Combining two $\Omega(n)$-robustly self-ordered graphs]
\label{dense:bipartite:clm}
For $i\in\{1,2\}$,
let $G_i=(V_i,E_i)$ be an $\Omega(n)$-robustly self-ordered graph,
and consider a graph $G=(V_1\cup V_2,E_1\cup E_2\cup E)$
such that~$E$ contains edges with a single vertex in each $V_i$;
that is, $G$ consists of~$G_1$ and $G_2$
and an arbitrary bipartite graph that connects them.
If the maximum degree in~$G$ of each vertex in $V_1$
is smaller by an $\Omega(n)$ term from
the minimum degree of each vertex in $V_2$,
then~$G$ is $\Omega(n)$-robustly self-ordered.
\ECM
Indeed, Claim~\ref{dense:bipartite:clm} is analogous
to Claim~\ref{bd:bipartite:clm} (which refers to bounded-degree graphs).

We also comment that $\Omega(n)$-robustly self-ordered graph
maintain this feature also when $o(n)$ edges are added
(and/or removed) from the incidence of each vertex.
A related statement appears as Theorem~\ref{superimposing:thm}.



\section{Application to Testing Dense Graph Properties}
\label{dense-pt:sec}
In Section~\ref{pt:sec}, we demonstrated the applicability
of robustly self-ordered bounded-degree graphs to the study
of testing graph properties in the bounded-degree graph model.
In the current section, we provide an analogous demonstration
for the regime of dense graphs.
Hence, we refer to testing graph properties in the dense graph model,
which was introduced in~\cite{GGR} and is surveyed in~\cite[Chap.~8]{G:pt}.
In this model, graphs are represented by their adjacency predicate,
and distances are measured as the ratio of the number of
differing adjacencies to the maximal number of edges.

\paragraph{Background.}
We represent a graph $G=([n],E)$,
by the adjacency predicate $g:[n]\times[n]\to\bitset$
such that $g(u,v)=1$ if and only if $\{u,v\}\in E$,
and  oracle access to a graph means oracle access to
its adjacency predicate (equiv., adjacency matrix).
The distance between the graphs $G=([n],E)$ and $G'=([n],E')$
is defined as the fraction of entries (in the adjacency matrix)
on which the two graphs disagree.

\BD[Testing graph properties in the dense graph model]
\label{dense:test-dense.def}
A {\sf tester} for a graph property $\Pi$
is a probabilistic oracle machine that, on input parameters $n$ and $\e$,
and oracle access to an $n$-vertex graph $G=([n],E)$
outputs a binary verdict that satisfies the following two conditions.
\BE
\item
If $G\in\Pi$, then the tester accepts with probability at least~$2/3$.
\item
If~$G$ is $\e$-far from $\Pi$,
then the tester accepts with probability at most~$1/3$,
where~$G$ is {\sf $\e$-far} from $\Pi$ if for every $n$-vertex
graph $G'=([n],E')\in\Pi$ the adjacency matrices of~$G$ and $G'$
disagree on more than $\e\cdot n^2$ entries.
\EE
\ED
The {\sf query complexity} of a tester for $\Pi$ is a function
(of the parameters $n$ and $\e$)
that represents the number of queries made by the tester
on the worst-case $n$-vertex graph,
when given the proximity parameter $\e$.

\paragraph{Our result.}
We present a general reduction of testing any property $\Phi$
of (bit) strings to testing a corresponding graph property $\Pi$.
Loosely speaking, $n$-bit long strings will be encoded
as part of an $O({\sqrt n})$-vertex graph,
which is constructed using $\Omega({\sqrt n})$-robustly
self-ordered $\Theta({\sqrt n})$-vertex graphs.
This reduction is described in Construction~\ref{dense:string2graph:con}
and its validity is proved in Lemma~\ref{dense:string2graph:clm}.
Denoting the query complexities of $\Phi$ and $\Pi$
by $Q_\Phi$ and $Q_\Pi$, respectively,
we get $Q_\Phi(n,\e)\leq Q_\Pi(O(n^{1/2}),\Omega(\e))$.
Thus, lower bounds on the query complexity of testing $\Phi$,
which is a property of ``ordered objects'' (i.e., bit strings),
imply lower bounds on the query complexity of testing $\Pi$,
which is a property of ``unordered objects'' (i.e., graphs).


Our starting point is the construction of $m$-vertex graphs
that are $\Omega(m)$-robustly self-ordered.
Actually, wishing $\Pi$ to preserve the computational complexity of $\Phi$,
we use a construction of graphs that are efficiently self-ordered,
as provided by Theorem~\ref{dense:effective:thm}.
Recall that the vertices in these graphs have degree that ranges
between $0.06\cdot m$ and $0.73\cdot m$.

The idea is to use two such graphs, $G_1$ and $G_2$,
one with $m$ vertices and the other with $49\cdot m$ vertices,
where $m={\sqrt n}$,
and encode an $n$-bit string in the connection between them.
Specifically, we view the latter string as a $m$-by-$m$ matrix,
denoted $(s_{i,j})_{i,j\in[m]}$,
and connect the $i^\xth$ vertex of~$G_1$
to the $j^\xth$ vertex of $G_2$
if and only if $s_{i,j}=1$.

\BCT[From properties of strings to properties of dense graphs]
\label{dense:string2graph:con}
Suppose that $\{G_m=([m],E_m)\}_{m\in\N}$
is a family of $\Omega(m)$-robustly self-ordered $m$-vertex graphs.
For every $n\in\N$, we let $m={\sqrt n}$,
and proceed as follows.
\BI
\item
For every $s\in\bitset^n$ views
as $(s_{i,j})_{i,j\in[m]}\in\bitset^{{m}\times{m}}$,
we define the graph $G'_s=([50m],E'_s)$ such that
\begin{equation}\label{dense:string2graph:eq}
E'_s = E_m\cup \{\{m+i,m+j\}:\{i,j\}\in E_{49m}\}
            \cup\{\{i,m+j\}:i,j\in[m]\wedge s_{i,j}=1\}
\end{equation}
That is, $G'_s$ consists of a copy of $G_m$ and a copy of $G_{49m}$
that are connected by a bipartite graph that is determined by $s$.
\item
For a set of strings $\Phi$,
we define $\Pi=\bigcup_{n\in\N}\Pi_n$
as the set of all graphs that are isomorphic
to some graph $G'_s$ such that $s\in\Phi$;
that is,
\begin{equation}\label{dense:string2graph:eq2}
\Pi_n=\{\pi(G'_s):s\in(\Phi\cap\bitset^n)\wedge\pi\in\Sym_{50m}\}
\end{equation}
where $\Sym_{50m}$ denote the set of all permutations over $[50m]$.
\EI
\ECT
Note that, given a graph of the form $\pi(G'_s)$,
the vertices of $G_m$ are easily identifiable
(as having degree at most $0.73m+m<1.8m$).%
\footnote{In contrast, the vertices of $G_{49m}$
have degree at least $0.06\cdot49m>2.9m$.}
%
The foregoing construction yields
a local reduction of $\Phi$ to $\Pi$,
where locality means that each query to $G'_s$ can be
answered by making a constant number of queries to $s$.
The (standard) validity of the reduction
(i.e., $s\in\Phi$ if and only if $G'_s\in\Pi$)
is based on the fact that $G_m$ and $G_{49m}$ are asymmetric.

In order to be useful towards proving lower bounds
on the query complexity of testing~$\Pi$, we need to show
that the foregoing reduction is ``distance preserving''
(i.e., strings that are far from $\Phi$
are transformed into graphs that are far from $\Pi$).
The hypothesis that $G_m$ and $G_{49m}$ are $\Omega(m)$-robustly self-ordered
is pivotal to showing that if the string $s$ is far from $\Phi$,
then the graph $G'_s$ is far from $\Pi$.

\BL[Preserving distances]
\label{dense:string2graph:clm}
If $s\in\bitset^n$ is $\e$-far from $\Phi$,
then the $50m$-vertex graph $G'_s$
{\rm(as defined in Construction~\ref{dense:string2graph:con})}
is $\Omega(\e)$-far from~$\Pi$.
\EL

\BPF
We prove the contrapositive.
Suppose that $G'_s$ is $\delta$-close to $\Pi$.
Then, for some $r\in\Phi$ and a permutation $\pi:[50m]\to[50m]$,
it holds that $G'_s$ is $\delta$-close to $\pi(G'_r)$,
which means that these two graphs differ
on at most $\delta\cdot(50m)^2$ vertex pairs.
If $\pi(i)=i$ for every $i\in[2m]$,
then $s$ must be $O(\delta)$-close to $r$,
since $s_{i,j}=1$ (resp., $r_{i,j}=1$)
if and only if $i$ is connected to $m+j$
in $G'_s$ (resp., in $\pi(G'_r)=G'_r$).%
\footnote{Hence, $G'_s$ is $\delta$-close to $G'_r$ implies that
$|\{i,j\!\in\![n]\!:\!s_{i,j}\neq r_{i,j}\}|\leq\delta\cdot(50m)^2$,
which means that $s$ is $\frac{(50m)^2\delta}{n}$-close to $r$.
(Recall that $m={\sqrt n}$.)}
Unfortunately, the foregoing condition
(i.e., $\pi(i)=i$ for every $i\in[2m]$)
need not hold in general.

In general, the hypothesis that $\pi(G'_r)$ is $\delta$-close to $G'_s$
implies that $\pi$ maps at most $O(\delta m)$ vertices of $[m]$
to $\{m+1,\ldots,2m\}$, and maps to $[m]$ at most $O(\delta m)$
vertices that are outside it.
This is the case because each vertex of $[m]$
has degree smaller than $0.73m+m<1.8m$,
whereas the other vertices have degree at least $0.06\cdot49m>2.9m$.


Turning to the vertices $i\in[m]$ that $\pi$ maps to $[m]\setminus\{i\}$,
we upper-bound their number by $O(\delta m)$,
since the difference between $\pi(G'_r)$ and $G'_s$
is at most $\delta\cdot(50m)^2$,
whereas the hypothesis that $G_m$ is $c\cdot m$-robustly self-ordered
implies that the difference between $\pi(G'_r)$ and $G'_s$
(or any other graph $G'_{w}$) is at least
$$\Delta = c\cdot m\cdot|\{i\!\in\![m]\!:\!\pi(i)\neq i\}|
                     -m\cdot|\{i\!\in\![m]\!:\!\pi(i)\not\in[n]\}|.$$
(Hence,
$|\{i\!\in\![m]\!:\!\pi(i)\neq i\}|
 \leq \frac{\Delta+m\cdot O(\delta m)}{cm} = O(\delta m)$.)
The same considerations apply to the vertices $i\in\{m+1,\ldots,2m\}$
that $\pi$ maps to $\{m+1,\ldots,2m\}\setminus\{i\}$;
their number is also upper-bounded by $O(\delta m)$.

For every $k\in\{1,2\}$,
letting $I_k=\{i\!\in\![m]\!:\!\pi((k-1)\cdot m+i)\!=\!(k-1)\cdot m+i\}$,
observe that
$D\eqdef|\{(i,j)\in I_0\times I_1:r_{i,j}\neq s_{i,j}\}|
  \leq\delta\cdot(50m)^2$,
since $r_{i,j}\neq s_{i,j}$ implies that $\pi(G'_r)$ and $G'_s$
differ on the vertex-pair $(i,m+j)$.
Recalling that $m-|I_k|=O(\delta m)$, it follows that
$$|\{(i,j)\in[m]:r_{i,j}\neq s_{i,j}\}|
 \leq((m-|I_1|)-(m-|I_2|))\cdot m+D=O(\delta m^2).$$
Hence, $s$ is $O(\delta)$-close to $r\in\Phi$,
and the claims follows.
\EPF

\section{The Case of Intermediate Degree Bounds}
\label{inter-deg:sec}
While Section~\ref{edge-colored:sec}--\ref{random-graphs:sec}
study bounded-degree graphs
and Sections~\ref{dense-basics:sec}--\ref{dense-pt:sec}
study dense graphs (i.e., constant edge density),
in this section we shall consider graphs of intermediate degree bounds.
That is, for every $d:\N\to\N$ such that $d(n)\in[\Omega(1),n]$,
we consider $n$-vertex graphs of degree bound $d(n)$.
In this case, the best robustness we can hope for is $\Omega(d(n))$,
and we shall actually achieve it for all functions $d$.

\BT[Robustly self-ordered graphs for intermediate degree bounds]
\label{all-varying-degree:thm}
For every $d:\N\to\N$ such that $d(n)$ is computable in $\poly(n)$-time,
there exists an efficiently constructable family
of graphs $\{G_n\}_{n\in\N}$ such that $G_n$ has maximal
degree $d(n)$ and is $\Omega(d(n))$-robustly self-ordered.
\ET
We prove Theorem~\ref{all-varying-degree:thm} in three parts,
each covering a different range of degree-bounds (i.e., $d(n)$'s).
Most of the range (i.e., $d(n)=\Omega(\log n)^{0.5}$)
is covered by Theorem~\ref{varying-degree:thm},
whereas Theorem~\ref{small-degree:thm} handles small degree-bounds
(i.e., $d(n)=O(\log n)^{0.499}$) and Theorem~\ref{degree-gap:thm}
handles the degree-bounds that are in-between.
One ingredient in the proof of Theorem~\ref{degree-gap:thm}
is a transformation of graphs that makes them expanding,
while preserving their degree and robustness parameters
up to a constant factor.
This transformation, which is a special case
of Theorem~\ref{superimposing:thm}, is of independent interest.

\BT[Robustly self-ordered graphs for large degree bounds]
\label{varying-degree:thm} \label{VARYING-DEGREE:THM}
For every $d:\N\to\N$ such that $d(n)\geq O({\sqrt{\log n}})$
is computable in $\poly(n)$-time,
there exists an efficiently constructable family
of graphs $\{G_n\}_{n\in\N}$ such that $G_n$ has maximal
degree $0.79\cdot d(n)$ and is $\Omega(d(n))$-robustly self-ordered.
Furthermore, the minimal vertex degree is $0.02\cdot d(n)$.
\ET
The graphs will consist of connected components of size $d(n)$,
and in this case $d(n)=\Omega({\sqrt{\log n}})$ is necessary,
since these connected components must be different.
\medskip

\BPFS
We combine ideas from Construction~\ref{dense:string2graph:con}
with elements of the proof of Theorem~\ref{construction:thm}.
Specifically, as in Construction~\ref{dense:string2graph:con},
we shall use constructions of $m$-vertex and $9m$-vertex graphs
that are $\Omega(m)$-robustly self-ordered,
but here we set $m=d(n)/10$ (rather than $m={\sqrt n}$)
and use $n/d(n)$ different $d(n)$-vertex graphs
that are based on the foregoing two graphs.
As in the proof of Theorem~\ref{construction:thm},
these ($d(n)$-vertex) graphs will be far from being isomorphic
to one another and will form the connected components
of the final $n$-vertex graph.

Our starting point is the construction of $m$-vertex graphs
that are $\Omega(m)$-robustly self-ordered.
Specifically, we may use Theorem~\ref{dense-rso1:thm}
and note that in this case the vertices in these $m$-vertex graph
have degree that ranges between $0.24\cdot m$ and $0.76\cdot m$.
Furthermore, these graphs have extremely high conductance;
that is, in each of these graphs,
the number of edges crossing each cut (in the graph)
is at least $\Omega(m)$ times
the number of vertices in the smaller side (of the cut).

The idea is to use two such graphs, $G_1$ and $G_2$,
one with $m\eqdef0.1\cdot d(n)$ vertices
and the other with $0.9\cdot d(n)=9\cdot m$ vertices,
and connect them in various ways as done in Section~\ref{step2:sec}.
Specifically, using an error correcting code
with constant rate and constant relative distance and weight,
denoted $C:[2^k]\to\bitset^{m^2}$, we obtain a collection
of $2^k\geq n/d(n)$ strongly connected $d(n)$-vertex graphs
such that the $i^\xth$ graph consists of copies of~$G_1$ and $G_2$
that are connected according to the codeword $C(i)$;
more specifically, we use the codeword $C(i)$
(viewed as an $m$-by-$m$ matrix)
in order to determine the connections between the vertices of~$G_1$
and the first $0.1\cdot d(n)$ vertices of $G_2$.
The final $n$-vertex graph, denoted~$G$,
consists of $n/d(n)$ connected components that are
the first $n/d(n)$ graphs in this collection.%
\footnote{Note that we used $2^k\geq n/d(n)$ and $m^2=O(k)$,
where $m=0.1\cdot d(n)>{\sqrt k}$.
This setting allows for handling any $d(n)\geq O({\sqrt{\log n}})$.}

The analysis adapts the analysis of the construction
presented in the proof of Theorem~\ref{construction:thm}.
Towards this analysis, we let $G_j^{(i)}$ denote the $i^\xth$
copy of $G_j$; that is, the copy of $G_j$ that is part of
the $i^\xth$ connected component of~$G$.
Hence, for each $i\in[n/d(n)]$,
the $i^\xth$ connected component of~$G$ is isomorphic to
a graph that consists of copies
of $G_1=([m],E_1)$ and $G_2=(\{m+1,\ldots,10m\},E_2)$
such that for every $u,v\in[m]$
the vertex $u$ (of $G_1^{(i)}$)
is connected to the vertex $m+v$ (of $G_2^{(i)}$)
if and only if $C(i)_{u,v}=1$.
Loosely speaking, considering an arbitrary permutation $\mu:[n]\to[n]$,
we proceed as follows.%
\footnote{These cases are analogous to the cases treated
in the proof of Theorem~\ref{construction:thm},
with the difference that we merged Cases~2\&3 (resp., Cases~4\&5)
into our second (resp., third) case.}
\BI
\item 
The discrepancy between the degrees of vertices
in copies of~$G_1$ and $G_2$
(i.e., degree smaller than $0.76m+m$ versus degree at least $0.24\cdot9m$)
implies that each vertex that resides in a copy of~$G_1$
and is mapped by $\mu$ to a copy of $G_2$
yields a contribution of $\Omega(d(n))$ units
to the symmetric difference between~$G$ and $\mu(G)$.
\item 
Let $\si'(i)$ (resp., $\si''(i)$) denote the index of
the connected component to which $\mu$ maps a plurality of the
vertices that reside in~$G_1^{(i)}$ (resp., of $G_2^{(i)}$).
Then, the extremely high conductance of~$G_1$ (resp., $G_2$) implies
that the vertices that resides in $G_1^{(i)}$ (resp., of $G_2^{(i)}$)
and are mapped by $\mu$ to a connected component
different from $\si'(i)$ (resp., $\si''(i)$)
yields an average contribution of $\Omega(d(n))$ units
per each of these vertices.
\item 
The lower bound on the weight of the codewords of $C$,
which implies that there are $\Omega(d(n)^2)$ edges
between $G_1^{(i)}$ and $G_2^{(i)}$,
implies that every $i$ such that $\si'(i)\neq\si''(i)$
yields a contribution of $\Omega(d(n)^2)$ units,
since there are no edges between $G_1^{(\si'(i))}$ and $G_2^{(\si''(i))}$.
Here we assume that few vertices fell to the previous case
(i.e., are mapped by $\mu$ in disagreement with the relevant plurality vote);
analogously to the proof of Theorem~\ref{construction:thm},
each of these few exceptional vertices reduces the contribution
by at most $d(n)$ units.
\item 
The $\Omega(d(n))$-robust self-ordering of~$G_1$ (resp., $G_2$)
implies that each vertex that reside in~$G_1^{(i)}$ (resp., of $G_2^{(i)}$)
and is mapped by $\mu$ to a different location in $G_1^{(\si'(i))}$
(resp., in $G_2^{(\si''(i))}$) yields a contribution of $\Omega(d(n))$ units.
Again, this assumes that few vertices fell to the penultimate case,
whereas each of these few vertices reduces the contribution by one unit
(per each vertex in the current case).
\item 
The distance between the codewords of $C$ implies
that every $i$ such that $\si'(i)=\si''(i)\neq i$
yields a contribution of $\Omega(d(n)^2)$,
where we assume that few vertices fell to the previous cases.
\EI
As in the proof of Theorem~\ref{construction:thm},
there may be a double counting across the different cases,
but this only means that we overestimate the contribution
by a constant factor.
Overall the size of the symmetric difference is $\Omega(d(n))$
times the number of non-fixed-points of $\mu$.
\EPFS

\paragraph{Handling small degree bounds.}
Theorem~\ref{varying-degree:thm} is applicable only
for degree bounds that are at least $O(\log n)^{0.5}$.
A different construction allows handling degree bounds
up to $O(\log n)^{0.499}$, which leaves a small gap
(which we shall close in Theorem~\ref{degree-gap:thm}).

\BT[Robustly self-ordered graphs for small degree bounds]
\label{small-degree:thm} \label{SMALL-DEGREE:THM}
For every constant $\e>0$, and every $d:\N\to\N$
such that $d(n)\in[\Omega(1),(\log n)^{0.5-\e}]$
is computable in $\poly(n)$-time,
there exists an efficiently constructable family
of graphs $\{G_n\}_{n\in\N}$ such that $G_n$ has maximal
degree $d(n)$ and is $\Omega(d(n))$-robustly self-ordered.
\ET
In this case, the graphs will consist of connected components
of size $\frac{\Theta(\log n)}{d(n)\cdot\log\log n} > d(n)$.
\medskip

\BPFS
Setting
$m(n)\eqdef\frac{\Theta(\log n)}{d(n)\cdot\log\log n}
  > d(n)\cdot(\log n)^{\e}$,
we proceed in two main steps.

The first main step amounts to showing that,
with probability at least $1-\exp(-\Omega(d(n)\cdot\log m(n))=1-o(1)$,
a $d(n)$-regular $m(n)$-vertex multi-graph
(generated by the random permutation model)
is $\Omega(d(n))$-robustly self-ordered and expanding.
This is a tightening of Theorem~\ref{random-works:thm}
(where $m(n)=n$ and $d(n)=O(1)$)
and it is proved by observing that the original proof (as is)
extends to a varying degree bound
and that higher robustness can obtained as indicated
in Footnote~\ref{random-bd:fn}.
Specifically, we can show that each non-fixed-point $i$
contributes at least $d(n)/2$ units (rather than one)
to the symmetric difference,
by considering smaller sets of indices $j\in[2d(n)]$
for which equalities hold
(i.e., we use $|J_i|\geq d(n)/12$ instead of $|J_i|\geq d(n)/6$).
The second main step consists of mimicking
the construction outlined in Remark~\ref{random-works-alt:rem};
specifically, we use the following three steps.

\BE
\item
First, we extend the argument of the foregoing main step
and show that, for any set $\cal G$ of $t<n$ multi-graphs
(which is each $d(n)$-regular and has $m(n)$ vertices),
with probability at least
$1-\exp(-\Omega(d(n)\cdot\log m(n))
  -t\cdot\exp(\Omega(d(n)\cdot m(n)\cdot\log(m(n)/d(n))))
  = 1-o(1)$,
a random $d(n)$-regular $m(n)$-vertex multi-graph (as generated above)
is both $\Omega(d(n))$-robustly self-ordered (and expanding)
and far from being isomorphic to any multi-graph in $\cal G$.
Here two $d(n)$-regular $m(n)$-vertex multi-graphs
are said to be {\sf far apart}
if they disagree on $\Omega(d(n)\cdot m(n))$ vertex-pairs.
As in Step~1 of Remark~\ref{random-works-alt:rem},
this amounts to obserting that the probability that
such a random multi-graph is close to being isomorphic to
a fixed multi-graph is at most
$\exp(-\Omega(d(n)\cdot m(n)\cdot\log(m(n)/d(n))))=o(1/n^2)$,
where the last inequality is due to the setting of $m(n)$.)%
\footnote{For starters, the probability that an edge that appears
in the fixed multi-graph appears in the random graph is $d(n)/m(n)$.
Intuitively, these events are sufficiently independent so to prove
the claim; for example, we may consider the neighborhoods
of the first $m(n)/2$ vertices in the random graph,
and use an iterative process in which their incidences
are determined at random conditioned on all prior choices.}


(Note that this multi-graph may have parallel edges and self-loops,
but their number can be upper-bounded with high probability.
Specifically, for $t=1/\e$,
with probability at least $1-O(d(n)^t/m(n)^{t-1})$,
no vertex has $t$ (or more) self-loops
and no vertex is incident to $t+1$ (or more) parallel edges.
Hence, omitting all self-loops and all parallel edges leaves us
with a simple graph that is both $\Omega(d(n))$-robustly self-ordered
(and expanding) and far from being isomorphic to any graph in $\cal G$.)

\item
Next, using Step~1, we show that one can construct
in $\poly(n)$-time a collection of $n/m(n)$ graphs such that
each graph is $d(n)$-regular, has $m(n)$ vertices,
is $\Omega(d(n))$-robustly self-ordered and expanding,
and the graphs are pairwise far from being isomorphic to one another.

As in Step~2 of Remark~\ref{random-works-alt:rem},
this is done by iteratively finding $\Omega(d(n))$-robustly
self-ordered $d(n)$-regular $m(n)$-vertex expanding graphs
that are far from being isomorphic to all prior ones,
while relying on the fact that $m(n)^{d(n)\cdot m(n)}=\poly(n)$
(by the setting of $m(n)$).

\item
Lastly, we use the graphs constructed in Step~2 as connected components
of an $n$-vertex graph, and obtain the desired graph.
\EE
Note that we have used $m(n) > (\log n)^{\e}\cdot d(n)$
and $d(n)\cdot m(n)\cdot \log(m(n)/d(n)) = \Theta(\log n)$,
which is possible if (and only if) $d(n) \leq (\log n)^{0.5-\Theta(\e)}$.
\EPFS


\paragraph{Obtaining strongly connected graphs.}
The graphs constructed in the proofs of
Theorems~\ref{varying-degree:thm} and~\ref{small-degree:thm}
consists of many small connected components;
specifically, we obtain $n$-vertex graphs of maximum degree $d(n)$
with connected components of size $\max(O(d(n)),o(\log n))$
that are $\Omega(d(n))$-robustly self-ordered.
We point out that the latter graphs can be transformed
into ones with asymptotically maximal expansion
(under any reasonable definition of this term),
while preserving their maximal degree and robustness parameter
(up to a constant factor).
This is a consequence of the following general transformation
(applied to a $\Omega(d(n))$-robustly self-ordered graph~$G$
of maximal degree $d(n)$
and a $d'(n)$-regular graph $G'$ of maximal expansion).

\BT[The effect of super-imposing two graphs]
\label{superimposing:thm}
For every $d,d':\N\to\N$ and $\rho:\N\to\R$,
let~$G$ and $G'$ be $n$-vertex graphs such that~$G$
is $\rho(n)$-robustly self-ordered and has maximum degree $d(n)$,
and $G'$ has maximum degree $d'(n)$.
Then, the graph obtained by super-imposing~$G$ and $G'$
is $(\rho(n)-d'(n))$-robustly self-ordered
and has maximum degree $d(n)+d'(n)$.
\ET
Note that Theorem~\ref{superimposing:thm} is not applicable
to the constructions of bounded-degree graphs
obtained in the first part of this paper,
because their robustness parameter was a constant smaller than~1.
(This is due mostly to Construction~\ref{colored2standard:ct},
but also occurs in the proof of Theorem~\ref{construction:thm}.)%
\footnote{In contrast, the construction of Theorem~\ref{small-degree:thm},
which builds upon the proof of Theorem~\ref{random-works:thm},
does yield $\Omega(d)$-robustly self-ordered graphs
of maximum degree $d$, for sufficiently large constant~$d$.}
A typical application of Theorem~\ref{superimposing:thm}
may use $d'(n)=\rho(n)/2\geq3$.
(Recall that $\rho(n)\leq d(n)$ always holds.)
\medskip

\BPF
Fixing any permutation $\mu$ of the vertex set,
note that the contribution of each non-fixed-point of $\mu$
to the symmetric difference between $G\cup G'$ and $\mu(G\cup G')$
may decrease by at most $d'(n)$ units due to $G'$.
\EPF

\paragraph{Closing the gap
between Theorems~\ref{varying-degree:thm} and~\ref{small-degree:thm}.}
Recall that these theorems left few bounding functions untreated;
essentially, these were functions $d:\N\to\N$
such that $d(n)\in[(\log n)^{0.499},O(\log n)^{0.5}]$.
We close this gap now.

\BT[Robustly self-ordered graphs for the remaining degree bounds]
\label{degree-gap:thm}
For every $d:\N\to\N$ such that $d(n)\in[(\log n)^{1/3},(\log n)^{2/3}]$
is computable in $\poly(n)$-time,
there exists an efficiently constructable family
of graphs $\{G_n\}_{n\in\N}$ such that $G_n$ has maximal
degree $d(n)$ and is $\Omega(d(n))$-robustly self-ordered.
\ET
In this case, the graphs will consist of connected components
of size $O(\log n)$.
\medskip

\BPFS
We apply the proof strategy of Theorem~\ref{varying-degree:thm},
while setting $\ell=\log_2n$ and using as building blocks
$\ell$-vertex $\Omega(d(n))$-robustly self-ordered graphs
of degree bound $d(n)/2$
obtained from Theorem~\ref{varying-degree:thm} itself
(while relying on $d(n)\in[\omega(\log\ell)^{1/2},o(\ell)]$).
Actually, we shall use two versions
of the foregoing $\ell$-vertex graphs
that have sufficiently different degree ranges
and transform them to obtain maximal expansion
(using Theorem~\ref{superimposing:thm}).
Furthermore, we shall connect pairs of copies of these graphs
by subgraphs of a fixed $d(n)/2$-regular bipartite graph
(rather than by subgraphs of the complete $\ell$-by-$\ell$ bipartite graph).
As in the proof of Theorem~\ref{varying-degree:thm},
using sufficiently different bipartite subgraphs,
we shall obtain $2\ell$-vertex graphs
that are far from being isomorphic to one another.
Our final $n$-vertex graph will consist of these $2\ell$-vertex graphs
as its connected components. Details follow.


Our starting point is a construction of $\ell$-vertex graphs that,
for some constant $\gamma\in(0,0.01)$,
are $\gamma\cdot d(n)$-robustly self-ordered
and have maximum degree $0.79\cdot d(n)$
and minimum degree $0.02\cdot d(n)$.
Such graphs are obtained by Theorem~\ref{varying-degree:thm},
while relying on $d(n)\geq\ell^{1/3}=\omega(\log\ell)$ and $d(n)<\ell$.
Using Theorem~\ref{superimposing:thm}
(with $d'(n)=\gamma\cdot d(n)/2$)
we transform these graphs to ones of maximum degree $0.8d(n)$
and asymptotically maximal conductance
(i.e., in each of these graphs, the number of edges crossing
each cut (in the graph) is at least $\Omega(d(n))$ times
the number of vertices in the smaller side (of the cut)).
Note that the resulting graph, denoted~$G_1$,
is $\Omega(d(n))$-robustly self-ordered
and has minimum degree $0.02\cdot d(n)$.
Using the same process with degree bound $d_2(n)=0.01\cdot d(n)$,
we obtain an analogous graph, denoted $G_2$,
that is also $\Omega(d(n))$-robustly self-ordered
and has maximum degree $0.8\cdot d_2(n)<0.01\cdot d(n)$.
We stress that
both~$G_1$ and $G_2$ have asymptotically maximal conductance
(w.r.t degree bound $\Theta(d(n))$).

Next, we connect~$G_1$ and $G_2$ in various ways
so to obtain $n/2\ell$ graphs that are far from
being isomorphic to one another.
This is done by a small variation on the construction
used in the proof of Theorem~\ref{varying-degree:thm}.
Specifically, we fix $d(n)/200$ disjoint perfect matchings
between the vertices of~$G_1$ and the vertices $G_2$,
and use the error correcting code to determine which
of the corresponding $\ell\cdot d(n)/200=\omega(\log n)$ edges
to include in the graph.
More specifically, using an error correcting code
with constant rate and constant relative distance and weight,
denoted $C:[2^k]\to\bitset^{\ell\cdot d(n)/200}$,
we obtain a collection of $n/2\ell<2^k$
strongly connected $2\ell$-vertex graphs
such that the $i^\xth$ graph consists of copies of~$G_1$ and $G_2$
that are connected according to the codeword $C(i)$;
that is, the $(j,v)^\xth$ bit of the codeword $C(i)$
(viewed as an $d(n)/200$-by-$\ell$ matrix)
determines whether the edge of the $j^\xth$ matching that is incident
at vertex $v\in[\ell]$ (of~$G_1$) is included in the $i^\xth$ graph.
The final $n$-vertex graph, denoted~$G$,
consists of these $n/2\ell$ graphs as its connected components.

The analysis is almost identical to the analysis provided
in the proof of Theorem~\ref{varying-degree:thm},
since the key facts used there hold here too
(although the construction is somewhat different).
The key facts are that the degrees of vertices in~$G_1$ and $G_2$
differ in $\Omega(d(n))$ units,
that the relative conductance of the connected components is $\Omega(d(n))$,
that~$G_1$ and $G_2$ are both $\Omega(d(n))$-robustly self-ordered,
and that the bipartite graphs (used in the different connected components)
are far away from one another.
\EPFS


\section*{Acknowledgements}
\label{ack:ack}
We are grateful to Eshan Chattopadhyay
for discussions regarding non-malleable two-source extractors,
and to Dana Ron for discussions regarding tolerant testing.
We are deeply indebted to Jyun-Jie Liao for pointing out
an error in the proof of~\cite[Lem.~9.7]{GW:rso-vo}.
We are also grateful to the reviewers of this journal
for their extremely careful reading of the paper
and for their numerous comments.
\addcontentsline{toc}{section}{Acknowledgements}

Oded Goldreich is a Meyer W.~Weisgal Professor at
the Faculty of Mathematics and Computer Science
of the Weizmann Institute of Science, Rehovot, {\sc Israel}.
His research was partially supported
by the Israel Science Foundation (grant No.~1041/18) and
by the European Research Council (ERC) under the European Union's
Horizon 2020 research and innovation programme (grant agreement No.~819702).
Avi Wigderson is the Herbert Maass Professor at the School of Mathematics
of the Institute for Advanced Study in Princeton, NJ 08540, {\sc USA}.
His research was partially supported by NSF grant CCF-1900460.

\addcontentsline{toc}{section}{References}

\printbibliography

\end{document}